\documentclass[a4paper,12pt]{article}
\usepackage{setspace}
\usepackage[T1]{fontenc}
\usepackage{times}
\usepackage[english]{babel}
\usepackage[utf8]{inputenc}
\usepackage{dtklogos}
\usepackage{wallpaper}
\usepackage[absolute]{textpos}
\usepackage[top=2cm, bottom=2.5cm, left=3cm, right=3cm]{geometry}
\usepackage{appendix}
\usepackage[nottoc]{tocbibind}
\usepackage[colorlinks=true,
            linkcolor=black,
            urlcolor=blue,
            citecolor=black]{hyperref}
\usepackage{hyperref}
\usepackage{sectsty}
\sectionfont{\fontsize{14}{15}\selectfont}
\subsectionfont{\fontsize{12}{15}\selectfont}
\subsubsectionfont{\fontsize{12}{15}\selectfont}

\usepackage{csquotes} 

\usepackage{float} 

\usepackage{algorithm}  
\usepackage{algpseudocode} 

\newsavebox{\mybox}
\newlength{\mydepth}
\newlength{\myheight}
\newenvironment{sidebar}%
{\begin{lrbox}{\mybox}\begin{minipage}{\textwidth}}%
{\end{minipage}\end{lrbox}%
 \settodepth{\mydepth}{\usebox{\mybox}}%
 \settoheight{\myheight}{\usebox{\mybox}}%
 \addtolength{\myheight}{\mydepth}%
 \noindent\makebox[0pt]{\hspace{-20pt}\rule[-\mydepth]{1pt}{\myheight}}%
 \usebox{\mybox}}

\newcommand\BackgroundPic{
    \put(-2,-3){
    \includegraphics[keepaspectratio,scale=0.3]{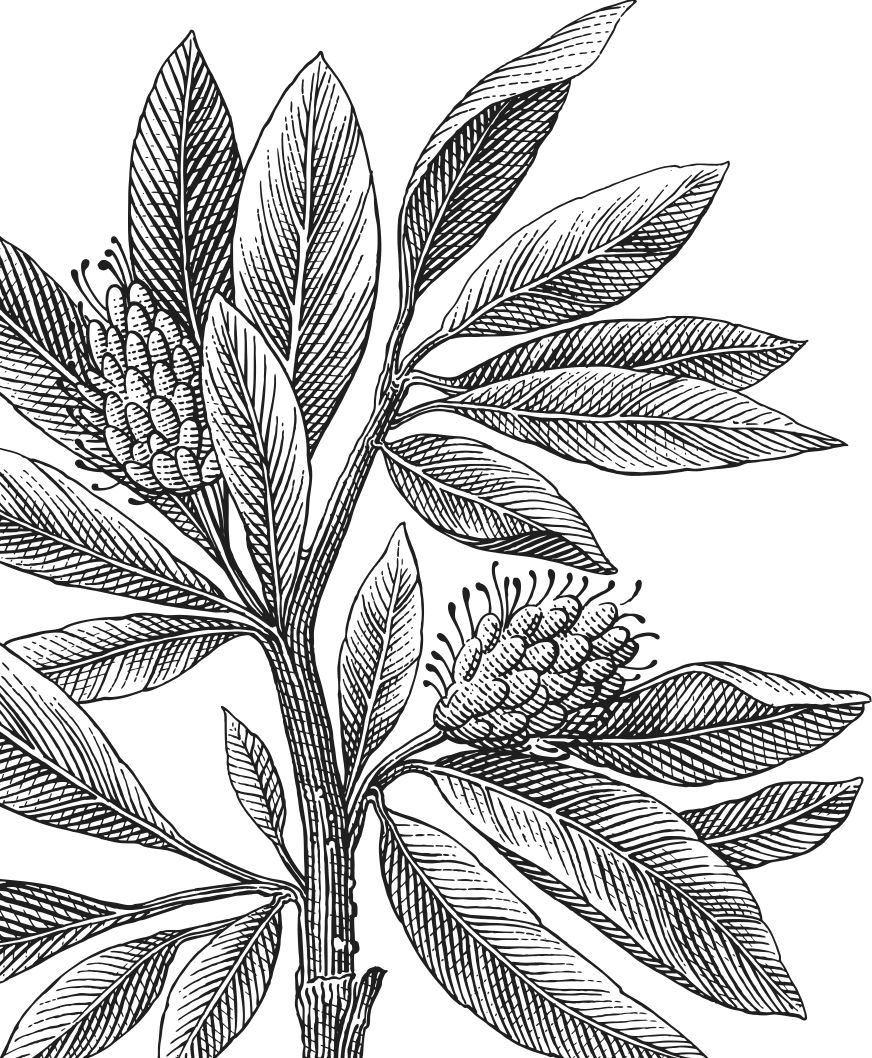} 
    }
}
\newcommand\BackgroundPicLogo{
    \put(30,740){
    \includegraphics[keepaspectratio,scale=0.10]{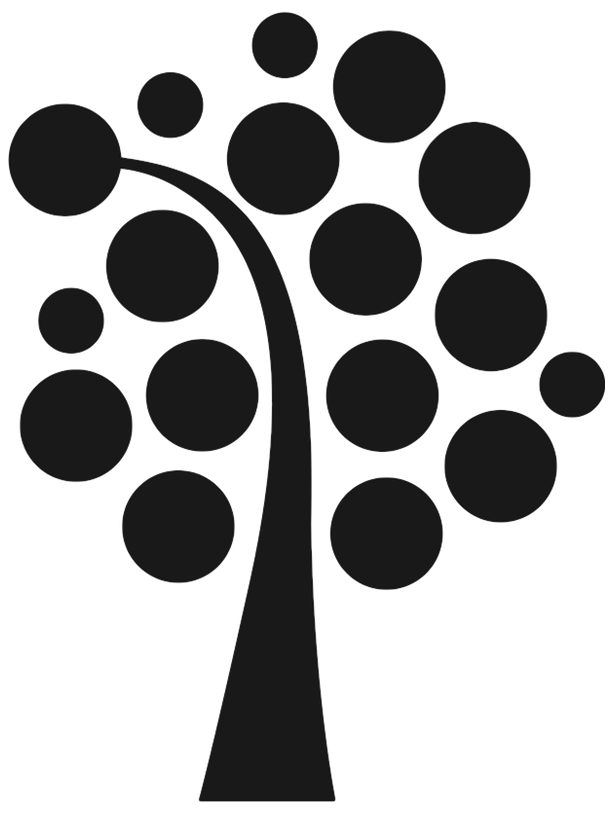} 
    }
}

\title{	
\vspace{-8cm}
\begin{sidebar}
    \vspace{10cm}
    \normalfont \normalsize
    \Huge Bachelor Degree Project \\
    \vspace{-1.3cm}
\end{sidebar}
\vspace{3cm}
\begin{flushleft}
    \huge A Visual Analytics Tool for Discovering Trajectory Patterns Using a Movement Taxonomy
\end{flushleft}
\null
\vfill
\begin{textblock}{6}(10,12)
\begin{flushright}
\begin{minipage}{\textwidth}
\begin{flushleft} \large
\emph{Author(s):}Ivan A. Hanono Cozzetti , \\
\qquad\qquad$\>\>$ Ahmad Abdou \\
\emph{Supervisor(s):} Amilcar Soares\\ 
\emph{Semester:} VT/HT 2025\\ %
\emph{Course:} 2DV50E \\ %
\emph{Subject:} Computer Science \\ 
\end{flushleft}
\end{minipage}
\end{flushright}
\end{textblock}
}

\date{} 

\begin{document}
\pagenumbering{gobble}
\newgeometry{left=5cm}
\AddToShipoutPicture*{\BackgroundPic}
\AddToShipoutPicture*{\BackgroundPicLogo}
\maketitle
\restoregeometry
\clearpage
\selectlanguage{english}
\begin{abstract}
\noindent The analysis of spatio-temporal data presents significant challenges due to the complexity and heterogeneity of movement patterns. This project proposes a data analytics tool that combines data visualization and statistical computation to facilitate spatio-temporal data analysis through a multi-level approach. The tool categorizes moving objects into distinct taxonomies using Machine Learning models, adding meaningful structure to the analysis.
Two case studies demonstrate the methodology's effectiveness. The first analyzed Arctic fox trajectories, successfully identifying and labeling foxes with Geometric or Kinematic-based behaviors, further categorized into Curvature and Acceleration groups. Statistical indicators revealed that foxes with Acceleration-based behavior showed constant, steady acceleration, while those with Curvature-based behavior exhibited acceleration peaks and sudden deceleration. The second case study examined tropical cyclone data, labeling trajectories with Speed, Curvature, and hybrid Geometric-based behaviors through unique statistical variables. Analysis of hybrid Geometric behavior (Curvature and Indentation combined) identified specific angles with the highest impact on hurricane shape and geometry.
The proposed method and tool demonstrate that spatio-temporal data, despite inherent complexity, can be analyzed and explained in detail, providing a theoretical and practical blueprint applicable to multiple domains.  \newline \newline

\end{abstract}

\newpage

\pagenumbering{gobble}
\tableofcontents 
\newpage
\pagenumbering{arabic}

%
%

\section{Introduction}

The increasing availability of GPS, telecommunication, and telemetry technologies has led to widespread movement data collection.
Such data can be derived from various sources, including humans, vehicles, ships, aircraft, animals, and even natural phenomena like hurricanes or the spread of diseases \cite{MvHiDBSCAN,UAV_patterns,disaster-management,ST_analysis_mobile_network_availability_natural_disasters,covid_patterns,song2024enhancing,spadon2024multi}. 
As a result, movement data has become a key component in several domains, including transportation, ecology, public health, urban planning, and climate studies.

A central task when working with this type of data is the discovery of patterns in how objects move across space and time. These patterns may reflect underlying behaviors, societal trends, or environmental dynamics. However, this process is challenging due to the lack of labels and the high-dimensional nature of spatio-temporal datasets.
To tackle these issues, researchers increasingly rely on visual analytics, combining automated data analysis with interactive visualizations to better understand complex datasets \cite{keim2008visual,junior2017analytic,soares2019vista}. This approach can support Exploratory Data Analysis (EDA) by helping analysts uncover patterns without predefined classifications.

This thesis study presents a visual analytics tool for spatio-temporal data exploration and the discovery of trajectory patterns using a movement taxonomy. The software tool will be developed based on a comprehensive methodological pipeline presented in section \ref{Methodology}, with the aim of addressing low and high-level details, bridging and visualizing them interactively.
The study will perform two case studies, and conduct a series of interactive experiments to analyze real-world datasets. The experiments involve two distinct real-world datasets to mitigate potential biases associated with single-source and single data-type movement, with the aim of explaining and understanding spatio-temporal data effectively. The datasets are dissimilar in volume and nature, as one dataset contains raw movement data from Arctic foxes in Canada, while the second dataset is larger and contains raw data from tropical cyclones.

A central part of this approach is using a movement taxonomy; a structured classification based on statistical and semantic characteristics of movement. 
Such a taxonomy supports interpreting and comparing movement behaviors, enhancing the analytical value of both visual and statistical methods.

\subsection{Background}

This section presents the background theory for the proposed thesis. Section 1.1.1 briefly defines trajectories and the meaning behind them. Section 1.1.2 discusses the importance of visual analytics, particularly in the field of movement data. Section 1.1.3 addresses movement taxonomies and further enhances our understanding of trajectory data, and section 1.1.4 highlights the current obstacles and challenges in the analysis of movement data.
\subsubsection{Trajectory Data}


A \textit{trajectory point} \cite{etemad2021sws}, $l^o_i$, is the location of object $o$ at time $i$, and is defined as,
\begin{equation}
l^o_i=\langle x^o_i, y^o_i \rangle
\label{not:1}
\end{equation}
\noindent where $x^o_i$  is the longitude of the location which varies from 0$^{\circ}$ to $\pm 180^{\circ}$, while $y^o_i$ is the latitude which varies from 0$^{\circ}$ to $\pm 90^{\circ}$.

A \emph{raw trajectory}, or simply \emph{trajectory} \cite{etemad2021sws}, is a time-ordered sequence of trajectory points of some moving object $o$,

\begin{equation}
\tau^o=\langle l^o_0,l^o_{1},..,l^o_n\rangle
\label{not:2}
\end{equation}

A \emph{point feature} \cite{soares2015grasp} is any numeric information that can be extracted from a raw trajectory and associated with a trajectory point (e.g., speed or bearing).
A \emph{trajectory feature} \cite{junior2018semi} is any numeric information computed from the trajectory sample and associated with a segment (e.g., average or maximal speed).

The point features are generally not immediately meaningful. 
However, appropriate pre-processing, statistical analysis, and visualization (i.e., transforming point features into trajectory features) can reveal significant patterns relevant to domains such as geography, urban studies, ecology, military logistics, or healthcare \cite{UAV_patterns,covid_patterns}.
In this thesis, we focus on making sense and analyzing trajectories with regards to trajectory features.

\subsubsection{Visual Analytics}

Visual analytics is a discipline that combines analytical reasoning with interactive visual representations \cite{keim2008visual}. 
It aims to support users in exploring data and identifying meaningful patterns, especially when dealing with large or unstructured datasets. 
In the context of movement data, visual analytics helps analysts interact with spatio-temporal data, observe patterns, compare trajectories, and generate hypotheses, even when labels are missing or limited \cite{junior2017analytic,soares2019vista,abreu2021trajectory,abreu2021local}.

This is particularly helpful for Exploratory Data Analysis (EDA), a process in which data is analyzed through visual and statistical means to reveal trends, outliers, and structures \cite{automating_EDA_with_ML,EDA_techniques_book}. 
Visual analytics tools enhance EDA by allowing users to manipulate the data visually, filter relevant dimensions, and observe linked visual effects that reveal structure or patterns otherwise hidden in raw data.

\subsubsection{Movement Taxonomies}

To make sense of the vast and complex nature of trajectory data, researchers often define taxonomies of movement behaviors. These taxonomies classify trajectories based on extracted features such as speed, direction, turning angles, or stop durations. A well-designed taxonomy provides a conceptual and statistical structure that helps analysts detect and interpret different movement types.

When combined with visual analytics, movement taxonomies can bridge the gap between low-level statistical metrics and high-level semantic understanding. For instance, statistical measures can be used to define categories in the taxonomy, while visual tools can help users explore and refine these categories by observing real data instances interactively.

\subsubsection{Challenges in Movement Pattern Discovery}

Despite substantial progress, several open challenges remain in the analysis of movement data:

\begin{itemize}
    \item \textbf{Lack of Annotations:} Many datasets lack labeled examples, making supervised learning approaches difficult to apply.
    \item \textbf{High Dimensionality:} Movement data is often characterized by a large number of features across space and time, making it difficult to analyze directly \cite{high_dimension_bayesian_model,High_Dimensional_Covariate_Spaces}.
    \item \textbf{Integration of Views:} Statistical and visual representations are often treated separately. However, true pattern discovery benefits from an integrated approach where numerical and visual information reinforce each other \cite{EDA_techniques_book}.
    \item \textbf{Limited Interaction:} Many existing tools offer static visualizations or lack coordinated views across multiple levels of abstraction, reducing their usefulness for exploratory tasks \cite{interaction_for_data_visualization,survey_making_visualizaton_better}.
\end{itemize}

Although there are efforts to develop modern visualization tools for spatio-temporal data, much of the existing literature still lacks cohesive frameworks that address all abstraction levels, from raw data to semantic interpretation, through interactive means \cite{alam_survey_spatio_temporal_data_analytics}. Tools that enable users to navigate between raw statistics, trajectory comparisons, and taxonomy-based insights are still relatively rare.

\subsection{Problem Formulation}

Although recent advances in data analysis and visualization have contributed to the development of tools for exploring complex datasets, spatio-temporal data, particularly movement trajectories, presents unique challenges that remain inadequately addressed. 
Existing approaches in the field often focus on specific aspects of the problem, such as direct trajectory comparison \cite{TS-SSPD}, higher-level taxonomy abstractions \cite{va_for_spatiotemporal_events}, or the contrast between global and local description levels \cite{DODGE_physics_of_movement}. 
Yet, a recurring limitation is the lack of integration across these methods and perspectives. 
Many existing works, particularly earlier ones \cite{spatiotemporal_relationships_for_scientific_data}, do not reflect the advances in visual analytics techniques and interaction paradigms that have emerged in recent years.

One core challenge is bridging low-level numerical descriptors, such as statistical features derived from movement data, with high-level abstractions, such as semantic taxonomies or visually driven categorizations. 
While statistical variables are essential for characterizing trajectory patterns, they are challenging to interpret in isolation. 
Analysts may struggle to identify meaningful movement patterns when relying solely on numeric comparisons, especially in high-dimensional datasets where many features interact in complex ways \cite{Yashar}. 
Conversely, visualization alone may be insufficient to justify or quantify a pattern without statistical grounding. This creates a gap between computationally derived metrics and human-centered pattern recognition, a gap particularly relevant in the context of Exploratory Data Analysis (EDA) \cite{EDA_techniques_book}.

This interplay between abstract statistical representations and more interpretable visual structures highlights the importance of linking low-level and high-level representations cohesively. 
A combined approach, where visual analytics supports the interpretation of statistical insights, and taxonomy provides a structured semantic framework, can empower analysts to navigate high-dimensional movement data more effectively \cite{visualizations_for_high_dimensional_data}. 
However, as noted by Alam et al. \cite{alam_survey_spatio_temporal_data_analytics}, many of the current tools still fall short in providing flexible, interactive, and domain-adaptable visual analytics platforms for spatio-temporal data.

Finally, interaction plays a critical role in overcoming the limitations of both purely visual and purely statistical analysis. 
Initially considered necessary mainly for large datasets, interaction has come to be seen as a central element of cognitive support in data exploration \cite{interaction_for_data_visualization}. 
Features such as parameter tuning, dynamic filtering, linked views, and zoomable timelines allow analysts to generate and test hypotheses, compare patterns, and uncover trends \cite{survey_making_visualizaton_better}. 
Yet, despite the growing body of visualization toolkits and frameworks, integrated systems that address all three aspects, semantic abstraction, statistical insight, and interactivity, remain scarce, especially for the domain of movement data.

Therefore, this thesis aims to investigate how a visual analytics tool can be employed for exploring and extracting meaningful patterns from unlabeled high-dimensional movement data, with the design and implementation of such a tool tailored specifically for spatio-temporal data analysis.

More precisely, this thesis aims to answer the following Research Questions:
\begin{itemize}    
    \item \textbf{(RQ1)} How can high-dimensional and unlabeled movement data be explored and better understood from a visual analytics tool using a taxonomy?
    \item \textbf{(RQ2)} How can a multi-level approach assist in finding patterns and representing movement data?
\end{itemize}

\subsection{Motivation}

The analysis and interpretation of spatio-temporal data are critical in various application domains. 
In healthcare, such data enables monitoring disease outbreaks by visualizing geographical and temporal trends, thereby supporting more informed public health responses \cite{healthcare-diseases}. 
In disaster management, spatio-temporal analysis facilitates understanding dynamic events. For instance, modeling how wildfires spread spatially across terrain and evolve over time, aiding decision-making and emergency planning efforts \cite{disaster-management}.

Recent work underscores the continued need for more advanced approaches to movement data analytics. 
For example, the work of \cite{MvHiDBSCAN} demonstrates substantial performance improvements over existing variants such as DBSCAN, Spatio-Temporal DBSCAN, and Higher-Dimension DBSCAN in handling movement data. 
Innovations like this have the potential to significantly impact domains like maritime traffic management, enabling more sustainable and efficient navigation practices. 
These examples reflect the broad and far-reaching societal benefits that can result from improved spatio-temporal data analysis techniques.

Despite these advances, there remains a pressing need for visual analytics tools that can support the complex demands of analyzing high-dimensional spatio-temporal data \cite{alam_survey_spatio_temporal_data_analytics,may2020challenges}. 
The challenges outlined in the previous section, namely, bridging low and high-level representations, enabling interactive exploration, and integrating statistical insights with visual structures, are still largely unresolved in available tools. 
Moreover, the interdisciplinary nature of spatio-temporal analytics means that its development overlaps with and contributes to fields such as machine learning, data visualization, explainable AI (XAI), and broader data science practices \cite{MvHiDBSCAN, task_specific_importance_XAI}.

This thesis addresses these needs by proposing the design and implementation of an interactive visual analytics tool tailored specifically to movement data. 
Such a tool will benefit researchers and analysts by providing capabilities to better understand, represent, and explore high-dimensional movement datasets. 
It will also support explanatory tasks, such as communicating insights, interpreting AI model behavior, and formulating data-driven hypotheses, across multiple domains. 
Additionally, the thesis aims to offer practical contributions to future system development through comprehensive documentation of the architectural decisions, quality attributes, toolkits, and strategies involved. 
This will serve as a reference architecture for future work, informing the development of visual analytics systems that are both technically robust and user-centered.

\subsection{Objectives}

The main objective of this thesis is to design, implement, and evaluate an interactive visual analytics tool for spatio-temporal movement data. 
The tool aims to support analysts and researchers in identifying patterns, making comparisons, and deriving insights from high-dimensional data without losing important details. More specifically, the tool will:
\begin{itemize}
    \item \textbf{Support multi-level representations by integrating movement variables with high-level taxonomic structures}, enabling users to navigate between summary views and detailed metrics. 
    It will incorporate and operationalize the taxonomy framework proposed by Tavakoli et al. \cite{Yashar}, offering a practical implementation of their classification strategy.
    
    \item \textbf{Facilitate trajectory comparison}, enhancing the interpretability of movement data through comparative visualizations across spatio-temporal contexts and movement variables.

    \item \textbf{Bridge abstraction levels by guiding users through a cyclic exploration process}: beginning with high-level overviews (e.g., taxonomy categories), progressing through mid-level representations (e.g., subcategories, feature distributions), and reaching detailed quantitative views (e.g., statistical variables). 
    The cycle then returns to higher abstraction to support hypothesis generation and validation.

    \item \textbf{Enable interaction at each stage of the analytical process}, allowing users to modify selections and refine views. These interactions will propagate through the system to ensure the consistency and coherence of analytical decisions.    
    
\end{itemize}

This approach directly addresses the core challenges outlined in the problem formulation: connecting low and high-level representations, supporting interactive data exploration, and enabling a cohesive analytical workflow for complex movement datasets.

\subsection{Contributions of the Work}

This thesis contributes to the field of spatio-temporal data analytics by designing, implementing, and evaluating an interactive visual analytics tool that addresses key challenges in movement data exploration, particularly concerning abstraction levels, interpretability, and interaction.
The main contributions are:
\begin{itemize}
    
    \item A unified visual analytics environment that enables analysts to explore movement data through a cyclic, multi-level approach, starting from high-level taxonomic abstractions, moving through intermediate representations, and arriving at detailed statistical features. 
    This cyclic process enhances interpretability and supports iterative sense-making.
    
    \item Operationalization of taxonomy-based analysis by implementing the taxonomy framework proposed by Tavakoli et al. \cite{Yashar} in a practical, visual form. 
    The tool allows users to apply, explore, and compare taxonomic classifications as part of their analysis process.
    \item Support for comparative exploration, including functionality for comparing zones (i.e., visual areas where patterns may be extracted by grouping trajectories with similar behavior scores) and trajectories across time and space. This contributes to the broader goal of enabling a nuanced understanding of movement patterns.

    \item Integration of statistical and visual representations in a cohesive interface, bridging low-level quantitative features and high-level semantic groupings. 
    This supports both exploratory data analysis and hypothesis formation based on interpretable evidence.
    
\end{itemize}

\subsection{Scope and Limitations} \label{Scope and Limitations}

This thesis focuses on the design, implementation, and evaluation of an interactive visual analytics tool for the exploration of movement data, with particular emphasis on trajectory analysis using a taxonomy-based framework. 
The scope includes the full development of the pipeline, from methodology and design decisions to implementation and analysis, supported by case studies across multiple datasets.

The system is designed to handle movement datasets enriched with features such as speed, acceleration, turning angles, bearing, and distance, enabling multi-level exploration of trajectories through statistical summaries and high-level taxonomic labels. 
The datasets used in this study include Arctic fox movements and tropical cyclone paths.
These were selected due to their spatio-temporal richness and diversity in domain characteristics, demonstrating the tool’s versatility in different application contexts.

To ensure consistency and meaningful visual interpretation, the tool assumes the availability of the aforementioned feature variables. 
As such, a crucial part of the pipeline is data pre-processing, including normalization, standardization, and outlier handling. 
These steps are necessary to reduce the risk of misrepresentation or biased visual outputs due to data irregularities. 
Datasets that diverge significantly in structure or feature availability may result in loss of functionality or analytical resolution, and thus fall outside the immediate scope of this project.

While the tool supports a cyclic, multi-level analytical process and interaction-driven refinement of views, its performance may degrade when applied to extremely large-scale datasets. 
As such, it is more suitable for medium-sized datasets in research, prototyping, or exploratory analysis contexts, rather than for large-scale production environments without further optimization.

Another important limitation is that no user studies were conducted to evaluate usability or cognitive effectiveness. 
Instead, the evaluation is based on a series of case studies that demonstrate the tool’s capabilities and alignment with the intended analytical workflow.
These case studies serve to validate the tool’s functionality, interpretability, and capacity to support insight generation, but they do not substitute for systematic user testing with domain experts.

Finally, the system is intended to support sense-making and pattern discovery, but it does not provide automated semantic interpretation or domain-specific meaning attribution.
Interpreting the significance of visual patterns and taxonomic groupings requires domain expertise and access to contextual or semantic information not captured within the tool. 
This layer of interpretation remains the responsibility of the analyst and is beyond the scope of the current work.

\subsection{Target Audience}
This thesis project and tool primarily target data analysts and researchers in the computer science field, with a particular focus on researchers in the spatio-temporal data analysis domain.
Researchers from other fields, such as wildlife and animal biology, epidemiology, geography, transportation engineering, aerospace engineering, and vessel traffic engineering, among many other researchers with the aim of understanding, explaining, and finding patterns on movement data will benefit from the utilization of the data analytics tool itself, as well as the scientific formulation, motivation, and methodology of this thesis study.    \\

\subsection{Thesis Structure}

The thesis is structured as follows. Chapter 2 summarizes key contributions and findings from previous studies relevant to the problem at hand, emphasizing the key role of the tool and the existing gap. 
Chapter 3 describes the methods and techniques used to conduct the study. 
Chapter 4 presents and analyzes the findings of the study. 
Chapter 5 interprets the results, discussing their implications, limitations, and possible improvements. 
Chapter 6 summarizes the study’s contributions and suggests directions for future research.


\newpage

\section{Related Work} \label{Related Work}

Several articles have been published in this domain, covering different aspects of movement data. 
However, some related works that have been referenced in this thesis either adopt different approaches or utilize tools that struggle to reflect the advancements in today's devices that are used to collect movement data. 
Below, we discuss the works related and with significant overlap to ours.



\subsection{Review of Existing Research}

The field of spatio-temporal data analysis, particularly trajectory mining and visual analytics, has evolved significantly over the past two decades. 
This thesis builds on recent developments in the area, most notably the work by Tavakoli et al. \cite{Yashar}, which serves as the foundation for the proposed visual analytics tool.

Tavakoli et al. \cite{Yashar} introduced the Taxonomical Description of Unlabeled Movement Data (TUMD), an innovative method for analyzing unlabeled, high-dimensional spatio-temporal datasets through a taxonomy-driven and outlier-based approach. 
Their methodology is structured in three core stages: first, transforming raw trajectories into a vectorized dataset using a set of predefined movement variables (i.e., trajectory features), second, generating outlier scores that reveal deviations based on these variables, and third, providing descriptions that contextualize the outliers with respect to the original trajectories. 
The taxonomy is organized into two high-level groups (i.e., Geometric and Kinematic), which are further subdivided to have a more fine-detailed data analysis. 
Geometric features include Indentation and Curvature, while Kinematic features include Acceleration and Speed. 
These groups are visualized in a two-dimensional outlier space, enabling intuitive pattern detection. 
The effectiveness of the TUMD framework has been demonstrated across multiple datasets, and this thesis project extends it by developing a visual interface that leverages these taxonomical representations for exploratory analysis.

Earlier foundational work on trajectory comparison includes the study by Dodge et al. \cite{DODGE_physics_of_movement}, which analyzed movement patterns across various object types by extracting both global and local movement features. 
Their three-step process involved data pre-processing, computation of global indicators (e.g., speed, acceleration, turning angle), and local feature extraction based on sinuosity and deviation. They used dimensionality reduction through Principal Component Analysis (PCA) and classification via Support Vector Machines (SVM) to distinguish between different modes of transportation. 
Their contribution illustrates how statistical movement features can serve as reliable inputs for classification tasks, a principle shared by TUMD and applied in this thesis for unsupervised exploration.

Complementary work by Silva et al. \cite{va_for_spatiotemporal_events} presented VAST, a visual analytics system for spatio-temporal phenomena such as crimes and forest fires. By allowing users to explore multiple Levels of Detail (LoD), the system supports dynamic identification of patterns across both synthetic and real-world datasets. This concept of multi-level exploration is echoed in the current thesis, which supports movement data exploration from high-level overviews down to fine-grained feature analysis.

Other studies have highlighted the importance of integrating computational techniques with visual interfaces. Andrienko et al. \cite{andrienko_visual_analytics_framework} proposed a flexible visual analytics framework that models spatio-temporal variation and enables the detection of spatial patterns through interactive cartographic displays and statistical analysis tools. 
Their work demonstrates the value of coupling algorithmic processing with visual reasoning, a core motivation behind this thesis’s visual analytics tool.

A similar integration of semantic reasoning and data clustering was presented by Shen et al. \cite{theoretical_framework_activity_groups_profiles_police}, who developed a framework for extracting spatio-temporal group behavior profiles. 
Their five-step process, from region of interest extraction to semantic validation, produced meaningful subgroup classifications of patrol officers based on GPS data. 
Although their goal was behavioral profiling, and this thesis does not seek semantic labeling, both approaches share an interest in pattern discovery from unlabeled movement data.

In terms of trajectory similarity, Cao et al. \cite{TS-SSPD} introduced the Time-Synchronized Symmetric Segment-Path Distance (TS-SSPD) to account for temporal alignment in movement trajectories. Their improved similarity measure outperformed earlier methods like SSPD in applications involving taxi and cyclone trajectories. 
Their work underscores the importance of the temporal dimension in trajectory analysis, which is reflected in the kinematic features (e.g., speed and acceleration) used in the TUMD taxonomy and adopted in this thesis.

Finally, Gao et al. \cite{gao_human_mobility_stkde_stv_staa} developed a framework for human mobility analysis that combined statistical and visualization techniques such as Spatio-Temporal Kernel Density Estimation (ST-KDE), Spatio-Temporal Autocorrelation Analysis (STAA), and interactive 3D visualizations. 
Applied to mobile phone data, their work demonstrated how integrating multiple techniques can yield insights into movement dynamics and urban interactions. Like this thesis, their approach supports exploratory analysis by combining computational processing with interactive visualization, although their system emphasized large-scale human mobility and urban flows.

\subsection{Discussion and Comparison of Approaches}

Researchers in the field of spatio-temporal analysis continue to develop new methods, algorithms, and visual interfaces to help analysts uncover meaningful patterns in movement data. 
A key role of data analysis and visualization is to support such discoveries by allowing analysts to interact with the data, reason about the patterns they observe, and communicate their insights more effectively \cite{andrienko_visual_analytics_framework}.
The studies reviewed in the previous section have each contributed distinctive methodologies and tools to this goal, ranging from statistical modeling and semantic profiling to dimensionality reduction and visual pattern discovery. Building on this foundation, the present section offers a more focused discussion of the similarities, differences, and limitations in relation to the approach proposed in this thesis.

A prominent limitation observed across the reviewed literature is the lack of interactive, comprehensive visual analytics tools that support the layered exploration of spatio-temporal features extracted from movement data. 
While several studies introduce compelling techniques for trajectory analysis, few combine flexible visualization layers with machine learning models in a way that enables users to shift fluidly between high-level overviews and detailed statistical representations. 
This thesis project addresses that gap by proposing a tool that integrates modern web-based technologies such as D3 and DeckGL to present trajectories and movement descriptors interactively, and to support detailed comparisons across multiple levels of abstraction.

The method proposed by Tavakoli et al. \cite{Yashar} offers a strong conceptual foundation for this kind of analytics. Their TUMD framework introduces a movement taxonomy, subdivided into geometric and kinematic components, and provides a robust set of statistical variables to describe each trajectory. 
While their work successfully lays out a structured approach to describing unlabeled movement data, it leaves room for further development in terms of feature importance analysis, interactive visualization, and user-guided exploration. 
These are precisely the areas where this thesis seeks to extend the TUMD framework, transforming it from a computational method into a fully interactive visual analytics tool.

Compared to Dodge et al. \cite{DODGE_physics_of_movement}, who used dimensionality reduction techniques like PCA to summarize movement patterns, this thesis avoids such transformations to preserve the integrity of the original data. 
Although Dodge’s pipeline is valuable for identifying object types from movement, it introduces a trade-off between interpretability and computational efficiency. 
By contrast, this thesis emphasizes transparency and interpretability, offering tools for side-by-side trajectory comparison that retain and reveal important differences in their statistical and geometric characteristics.

Several studies share methodological commonalities with this thesis, including the statistical approaches and emphasis on exploratory visual analytics. 
For example, both Andrienko et al. \cite{andrienko_visual_analytics_framework} and Shen et al. \cite{theoretical_framework_activity_groups_profiles_police} employ machine learning and statistical modeling to analyze spatio-temporal data. 
However, Andrienko et al. \cite{andrienko_visual_analytics_framework} rely on techniques such as K-means clustering and statistical residuals to evaluate patterns, while Shen et al. \cite{theoretical_framework_activity_groups_profiles_police} emphasize semantic validation of detected groups. In contrast, this thesis takes a more flexible and visual route, eschewing hard clustering or semantic labeling in favor of an interface that allows users to dynamically explore movement patterns across both high-level visualizations and low-level statistical indicators.

This layered approach is also inspired by Silva et al.'s VAST system \cite{va_for_spatiotemporal_events}, which supports multiscale exploration of temporal events. While Silva’s work focuses on detecting and explaining specific spatio-temporal phenomena (e.g., crime or forest fires), the current thesis applies a similar interface logic to generic movement data. 
Instead of modeling specific events, the thesis emphasizes pattern comparison and movement descriptors adopting the TUMD taxonomy, offering a reusable tool for broader application domains.

Gao et al. \cite{gao_human_mobility_stkde_stv_staa} proposed a mathematically rigorous system for analyzing human mobility using tools like spatial autocorrelation and 3D visualization. While this thesis is conceptually aligned in its emphasis on interactive data views, it diverges in its aim to prioritize usability, interpretability, and taxonomy-guided analysis over mathematical depth. Rather than focusing on autocorrelation or weighted adjacency matrices, the thesis tool aims to visualize statistical movement descriptors, such as curvature, acceleration, and indentation, alongside map views and comparison dashboards, bridging the gap between complexity and accessibility.

Taken together, these comparisons reveal significant overlap between the present thesis and foundational works such as those by Tavakoli et al. \cite{Yashar} and Dodge et al. \cite{DODGE_physics_of_movement}, particularly in the use of statistical descriptors and movement feature extraction. 
However, the thesis distinguishes itself by emphasizing interactive visual analysis over pure computational processing. It further builds on the conceptual infrastructure laid out by Silva et al. \cite{va_for_spatiotemporal_events}, Shen et al. \cite{theoretical_framework_activity_groups_profiles_police}, and Andrienko et al. \cite{andrienko_visual_analytics_framework} by offering a tool that links multiple levels of detail in real time, from map-based overviews to fine-grained movement variable distributions.

In positioning this thesis within the broader literature, it contributes to the growing interest in taxonomical and visual methods for movement data analysis. 
By operationalizing the TUMD framework within a user-centric visual tool, it provides a novel contribution to the field, one that supports flexible exploration, comparative reasoning, and transparent pattern discovery in unlabeled trajectory datasets. 
Future researchers may benefit from both the tool and the methodological insights presented here, using them to extend work on spatio-temporal analysis and to bridge the interpretability gap between statistical data descriptors and human-centered understanding.

\newpage 

\section{Methodology} \label{Methodology}




This thesis's methodology consists of a number of comprehensive steps, with the intent to convey a clear view and motivation for each of the method decisions and steps taken towards this thesis proposal. This chapter is logically divided into the following subsections: section \ref{Research Approach} presents the concrete research approach that will be followed, section \ref{conceptual-method} discusses the key conceptual considerations of the proposed method, 
and section \ref{Analytics tool components} presents the analytic tool components for the thesis project and visualization tool workflow. 
Section \ref{Data Collection} addresses the datasets used for the analytics tool,
section \ref{Data Analysis} presents the data analysis, section \ref{Reliability and Validity} discusses reliability and validity, section \ref{Ethical Considerations} discusses ethical considerations, and lastly, section \ref{Methodology Summary} provides a summary of the method.


\subsection{Research Approach} \label{Research Approach}

\noindent The thesis will follow a case study approach to investigate, explore, and describe the taxonomy-based analytics tool. This includes making a number of interactive experiments with the software tool, highlighting findings, and presenting descriptive results.

The main objective of the case study is to perform a series of experiments with the software tool proposed to analyze real-world datasets, showcasing and reviewing the sequence of individual steps taken to analyze such data, with the final intent to explore, describe, and understand spatio-temporal data comprehensively. Both the process and final findings will be documented in detail, presenting individual interpretations for each of the components discussed in the subsection \ref{Analytics tool components}, as well as the correlative linkage between the decisions made at each analysis stage, with an overall interpretation of the results.




\subsection{Conceptual Method Considerations} \label{conceptual-method}

Some conceptual decisions need to be presented in order to comprehend the components, structure, and approach for the current thesis project. 
As previously discussed, the thesis project proposes a visual analytics tool for discovering trajectory patterns using a movement taxonomy. The analytics tool aims to facilitate and enable the understanding of movement objects with an interactive scope, which allows the comparison of trajectories through both high-level and low-level details. 

It must be noted that this thesis project is built on top of Tavakoli et al.'s \cite{Yashar} work, with the objective of expanding on the understanding of spatio-temporal data from an applied analytical approach. 
Therefore, many of the concepts used throughout the methodology are based on the studies they have conducted thus far.

\subsubsection{Taxonomy and Labeling} \label{taxonomy}
    
A taxonomy is often presented as a scheme that allows classification. For the current thesis project, a taxonomy acts as a method to label movement, used to characterize and discern in a more precise manner how a trajectory may be different from another one. The taxonomy used, originally proposed by Tavakoli et al. \cite{Yashar}, is constituted by the two following groups: \textit{Kinematic}, aggregating spatio-temporal variables associated with the motion of an object, and \textit{Geometry}, as the group variables concerning the trajectory's shape. 
Each of these taxonomy groups is subdivided into two groups: \textit{Kinematic} is subdivided into \textit{Acceleration} and \textit{Speed}, while \textit{Geometry} is subdivided into \textit{Indentation} and \textit{Curvature}. 
This structure is defined in Fig. \ref{fig:tree-taxonomies}.

Without the consideration of a taxonomy, comparing trajectories may become trivial or difficult to discern; which is the main motivation behind adopting one. 
A taxonomy that is too detailed, or with too many subdivisions and subgroups, does not necessarily facilitate the understanding of movement, and may even confuse the analyst trying to understand a new dataset, as the intention for these groups is to provide a higher-level view, yet with specific label characterizations. 
Additionally, having a larger number of groups and subgroups extends the possible number of combinations, increasing the complexity during the decision-making process for the analysts (subsection \ref{Combinations}). 
Note that during the analysis and workflow process, the taxonomy groups may also be referred to as \textit{parameters}, as these represent a condition selection during analysis. 


\subsubsection{Statistical Variables} \label{Statistical Variables}
As discussed by Tavakoli et al. \cite{Yashar}, the taxonomy groups describing movement may be represented with one individual statistical variable, such as \textit{average curvature, average indentation, average acceleration, and average speed}. This could be a viable and valid approach. Nonetheless, movement data and the individual taxonomy groups presented here have several dimensions (i.e., several features). 
High-dimensionality presents challenges and limitations, making it difficult to use only individual variables, such as an \textit{average}, to represent and explain spatio-temporal data, and therefore, a lower-level view and analysis are needed.
In addition, using a single feature may not reflect the distribution and behavior performed by a moving object in a reasonable way, making the unique characteristics of the movement to get lost or overlooked. 

Therefore, to explain movement data, a total of 72 statistical variables are used for the proposed thesis project. 
Each of these variables provides a deeper insight into the taxonomy subgroups previously mentioned. 
For instance, for the subdivision \textit{Speed}, some of the statistical variables found are \textit{speed mean}, \textit{speed minimum quantile}, \textit{speed quantile 10}, \textit{speed quantile 25}, among many others. 

One of the key intents of this analytics tool is to provide proper contextualization and present the relative relationship and impact of low-level details over high-level ones. 
To achieve this, statistical variables will be linked to the trajectory map visualizations to convey how such statistical representation of a parameter reflects on the movement object itself. \\

\subsubsection{Taxonomy Combinations}\label{Combinations}

\begin{figure}
    \centering
    \includegraphics[width=0.55\linewidth]{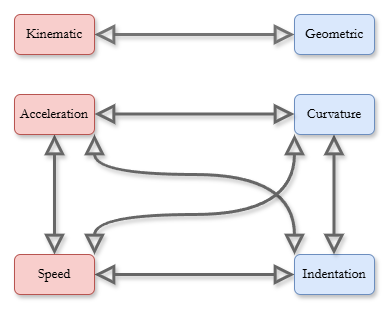}
    \caption{Taxonomy Combinations}
    \label{fig:tax_combination_conceptual}
\end{figure}

From the groups and subgroups defined in the Taxonomy subsection \ref{taxonomy}, a total of 7 valid combinations are presented below and in Fig. \ref{fig:tax_combination_conceptual}:

\begin{enumerate}
    \item Geometric and Kinematic
    \item Acceleration and Speed
    \item Curvature and Indentation
    \item Curvature and Speed
    \item Indentation and Speed
    \item Acceleration and Curvature
    \item Acceleration and Indentation
\end{enumerate}

\noindent The selection of any of these 7 combinations enables further investigation under the scope of two unique behaviors. 
As an example, the combination number 3, \textit{Acceleration and Speed}, will further show and denote those trajectories that are mainly characterized by their \textit{Acceleration and Speed} behavior (i.e., \textit{Kinematic behavior}). 
Further investigation through visual and analytical steps may then show the most significant indicator that characterizes the movement of such trajectory, for instance \textit{Acceleration Quantile 90} (i.e., a statistical variable, which is an even lower-level detail and more specific, to which it can be attributed as the main characterization of the trajectory under review). 
These combinations play a significant role in deeper levels of analysis, as these will be the two dimensions used for outlier detection, to derive patterns, and identify those trajectories that present anomalies. 

Combinations between top taxonomy groups (i.e., Kinematic and Geometric) and subdivisions of the same groups (i.e., speed, acceleration, curvature, and indentation) are not within the scope of this thesis project, and therefore, are not considered valid options (e.g., Kinematic and Speed is not a valid combination selection).

\subsubsection{Outlier Detection and Zones} \label{Outlier detection and Zones}

Tavakoli et al. \cite{Yashar} presented an approach to represent outlier scores in 4 zones. 
Outlier detection is a core consideration of the proposed method in this thesis project, as it enables a way to measure and identify the trajectories showing anomalies, and consequently, exposes the trajectories that follow a common behavior, as well as enabling pattern finding. 
Such a score is utilized to quantify the degree of interestingness of a movement pattern, as it manifests unprecedented and possibly substantial behaviors that seem exceptional from the rest of the data.

For this project, Distance-Based Outlier Score (DBOS) is used \cite{Yashar, knorr1997_unified_approach_for_mining_outliers}, which produces a score between 0 and 1. The DBOS algorithm produces the scores based on a neighbor count within a radius. The radius has a fixed value by calculating, for all data instances, the average distance between them (i.e., the distribution of the points). In other words, the spacing between points is measured to produce such scores. Effectively, a score closer to 1 represents an outlier, which has higher interestingness and presents a movement-based pattern, while a score closer to 0 represents no outlier, and hence, a less interesting behavior \cite{Yashar}.

The outlier scores are generated for the taxonomy combination selected by the analyst, which means that 2 scores are produced. 
The taxonomy combination is represented in a plane as a decision boundary, where one of the parameters selected lies on the $x$ axis, and the second one on the $y$ axis. 
Both axes in the plane are scaled from 0 to 1, representing the outlier score scale \cite{Yashar}. 
Assume that the current taxonomy selection is \textit{Geometric-Kinematic}. The \textit{Kinematic} parameter will be represented in the $x$ axis, while the \textit{Geometric} parameter will lie on the $y$ axis, each one from 0 to 1, as shown in Fig. \ref{fig: Scatter-plot}.

To facilitate the understanding and impact of the outlier scores, the decision boundary presented by Tavakoli et al. \cite{Yashar} is followed, which consists of 4 zones:
\begin{itemize}
    \item \textbf{Zone 0:} The trajectory points within zone 0 represent a score below 0.5 for both $x$ and $y$ taxonomy parameters. This means that neither of the parameters has a significant manifestation of the taxonomical elements chosen.
    \item \textbf{Zone 1:} The trajectory points within zone 1 present a score above 0.5 for the $y$ parameters, while the score for the $x$ parameter must satisfy the following linear inequality: \(x < (y - 0.5)\). 
    This means that the behavior for the trajectories lying in that zone is mainly due to $y$ behavior only (i.e., the element in the taxonomy selected as $y$).
    \item \textbf{Zone 2:} The trajectory points within zone 2 present a score above 0.5 for the $x$ parameters, while the score for the $y$ parameter must satisfy the following linear inequality: \(y < (x - 0.5)\). This means that the behavior for the trajectories lying in that zone is mainly due to the element selected as $x$ only.
    \item \textbf{Zone 3:} The trajectory points within zone 3 may have one of the following 3 cases:
    \begin{itemize}
        \item $x$ and $y$ \textbf{both} scored above 0.5
        \item $y$ scored above 0.5 and \(x > (y - 0.5)\)
        \item $x$ scored above 0.5 and \(y > (x - 0.5)\)
    \end{itemize}
   
   This means that both of the parameters have an impact on the trajectories within that zone range, having a hybrid behavior from both taxonomy parameters in the combination selected.
\end{itemize}

\begin{figure}[H]
    \centering
    \includegraphics[width=0.7\linewidth]{ 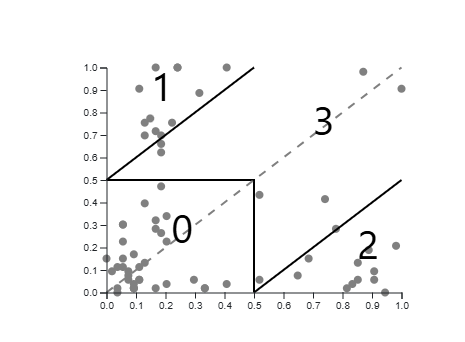}
    \caption{The decision boundary zones with scatter point trajectories.}
    \label{fig: Scatter-plot-1}
\end{figure}

Presenting the outlier scores through this decision boundary facilitates the understanding and interpretation of all trajectories under analysis. 
When populated with several trajectory points, a scatter plot is produced, as shown in Fig. \ref{fig: Scatter-plot-1}. 
The main logical input to zone matching is also presented in Algorithm \ref{alg:zone-and-logic-matching}.

\begin{algorithm}
    \caption{Zones and Logic Matching}\label{alg:Zones and Logic Matching}
    \begin{algorithmic}
        \Require $x,y = [0,1]$
        \State $0 \gets (x < 0.5$ \textit{and} $y < 0.5)$
        \State $1 \gets (y > 0.5$ \textit{and} $x < (y - 0.5))$
        \State $2 \gets (x > 0.5$ \textit{and} $y < (x - 0.5))$
        \State $3 \gets $ \textbf{not}$((x < 0.5$ \textit{and} $y < 0.5)$ \textit{or} $(x < 0.5$ \textit{and} $y > 0.5$ \textit{and} $x < (y - 0.5))$ \textit{or} $(x > 0.5$ \textit{and} $y < (x - 0.5)))$
    \end{algorithmic}
     \label{alg:zone-and-logic-matching}
\end{algorithm}



\subsubsection{Zone Comparisons, Statistical Variables and Feature Importance} \label{Zone Comparisons, Statistical Variables and Feature Importance}

In subsection \ref{Statistical Variables}, statistical variables were discussed. 
However, the relationship and source of these values must be further detailed. 
During subsection \ref{Outlier detection and Zones}, an explanation of the zones based on outlier detection and taxonomy manifestation was presented. 
In addition to the implementation of zones, a \textit{zone comparison} will be included, which would allow the comparison between all points from selected \textit{zone A} against the points in \textit{zone B}. 
This process will enable a two-group comparison for an entire set of individual trajectories sharing the same behavior, against all other trajectories in a second group with a different behavior, based on the defined zones. 
Such a comparison is done by implementing a \textit{feature importance} algorithm, which will analyze the trajectories from the zones selected, effectively deriving the highest contributing features for a trajectory to have been categorized as a member of such a zone. 
At this point, the statistical variables presented in subsection \ref{Statistical Variables} are introduced, as the analytics tool aims to present those statistical variables via feature importance in descending order, from the most significant variables to the lowest. 
To further motivate the characterization of such variables in contrast and connection to the taxonomical description, the statistical variables will be presented in two different columns; one presenting those variables for the first taxonomy parameter selected, and a column showing the variables for the second parameter of the combination, as presented in Figure \ref{fig:feature-importance}, for a combination example of \textit{Kinematic} and \textit{Geometric} parameters.

The incorporation of zone comparisons effectively facilitates the integration of lower-level detail analysis, such as statistical variables. These variables provide a deeper insight into what contributed the most to a specific trajectory to have been categorized as a member of a specific zone, which further conveys that a trajectory's behavior was linked to a taxonomy combination. The zone comparison and feature importance process not only allows analysts to determine which taxonomy (e.g., \textit{Kinematic}) or specific subdivision of such taxonomy (e.g., \textit{Speed}) is a key parameter, but more importantly, finds the most significant variables (e.g., \textit{Speed Quantile 90}), effectively heightening the analyst's understanding of the dataset and trajectories under review. Note that the concrete algorithm and implementation details are discussed later on.

\subsubsection{Trajectory Comparisons and Map Visualizations}

The individual side-by-side comparison of two trajectories is of significant interest for the proposed visual analytics tool. The selection of these trajectories will be incorporated at the decision boundary scatter plot axes, allowing the analyst to select and choose the trajectories to be further reviewed.

The selection of two trajectories will trigger the map visualizations. Two side-by-side map visualizations will be produced, one for the first selected trajectory and the corresponding visualization for the second trajectory selected by the analyst.

Map visualizations will provide two viewing options: a 3D and a 2D view. Both views aim to represent the trajectories over time, with additional details for key feature values for the respective trajectories, namely \textit{speed, acceleration, angle, distance, and bearing}.
The 3D view will introduce Tominski et al.'s wall representation for a detailed spatio-temporal view \cite{the-great-wall-of-space-time}, presenting all key movement features within the wall view, as shown in Fig. \ref{fig: the-wall}. Each of the layers in the wall represents a feature over time (e.g., speed, acceleration, bearing). The 2D visualization will show the trajectories' paths over time with a heatmap representing a single feature instead of all of them together, to simplify analysis over features individually, with arrows indicating the trajectories' direction, which proved to be a visually helpful addition, presented by Soares et al. \cite{junior2017analytic}.

Additional interactions will be included, such as the selection of a given statistical variable, from the feature importance visualizations mentioned in subsection \ref{Zone Comparisons, Statistical Variables and Feature Importance}, which will effectively update the trajectory map visualizations to highlight only the specific variable chosen, over time. The 2D and 3D trajectory visualizations will then be simplified: instead of showing the entire trajectory, only a segment of it will be shown. The shorter trajectory will be produced by picking the trajectories previously chosen by the analyst (i.e., trajectory IDs) and the statistical variable from the feature importance table selected (e.g., speed kurtosis). The tool will find, in the trajectories selected, the generalized group from the statistical variable (e.g., speed), calculate the variable selected (e.g., kurtosis), and find the closest trajectory point from the variable calculated (e.g., the speed instance that has the closes value to the speed kurtosis). The trajectories in the map will then be generated by centering on the closest value found, and showing a number of trajectory points before and some after it (e.g., 5 points before and 4 after, effectively showing a short trajectory with only 10 points).

This is of key relevance, since two of the main objectives of this thesis project and the concrete implementation of the analytics tool are to facilitate understanding and to bridge low-level details (i.e., statistical variables) with high-level views (i.e., map visualizations).

\subsubsection{Iterative and Incremental Analytical Approach}
As a final conceptual consideration, the individual approach taken towards spatio-temporal data analysis with the proposed thesis project and tool must be discussed.

\begin{figure}[htbp]
    \centering
    \includegraphics[width=1\linewidth]{ 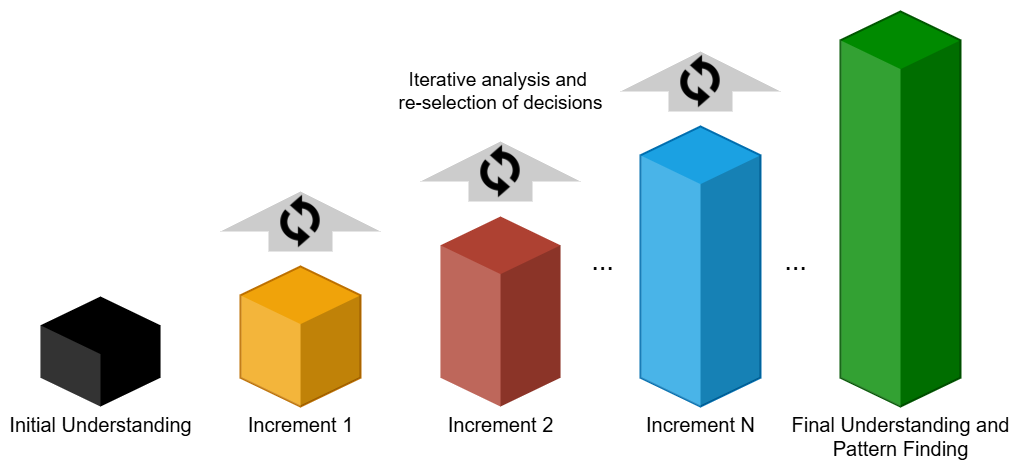}
    \caption{Iterative and Incremental data analysis.}
    \label{fig:iterative-incremental}
\end{figure}

The analytical process for such complex data must not be addressed as a \textit{one-shot analysis}, but rather as an aggregation of analytical steps that \textit{incrementally} build the understanding over the spatio-temporal dataset under review by \textit{iteratively} re-assessing previous decisions, making selection changes, and moving towards a deeper interpretation. Such a process may require repeating and re-analyzing previous steps, to refine and redirect the analyst's view until reaching a clear understanding and recognizing potential patterns in a dataset, as visualized in Fig. \ref{fig:iterative-incremental}. This conceptual approach does not necessarily disregard the possibility of understanding or finding patterns in a single iteration, however, it promotes the refining of the analyst's decisions throughout their utilization of the tool and knowledge presented in this thesis project.

\newpage

\subsection{Analytics tool components} \label{Analytics tool components}
This thesis project implements the visual analytics tool components described below:
\begin{itemize}
    \item A \textbf{\textit{taxonomy tree}} that allows a combination selection between movement variables and their corresponding subdivisions. Namely, the selection for \textit{Kinematic}, with subdivisions \textit{Speed} and \textit{Acceleration}, and \textit{Geometric}, with \textit{Curvature} and \textit{Indentation}, Fig. \ref{fig:tree-taxonomies}, Fig. \ref{fig:combination-selection}.
    
    \item A \textbf{\textit{frequency heatmap}} presenting all taxonomy and subdivision combinations allowed against all decision boundary zones, presenting a count for the trajectories of each instance, and highlighting the combination selected against the respective \textit{outlier-score based zones}, Fig. \ref{fig: Heatmap-selection}.

    \item A \textbf{\textit{scatter plot with decision boundaries}} based on the respective outlier scores, 0 to 1, for the specific combination selected in the taxonomy tree and then on the frequency heatmap, Fig. \ref{fig: Scatter-plot}. At this visualization, a selection of the two trajectory points for comparison is allowed, and zone comparison can be done. Upon the selection of 2 scatter points, the selected trajectories get displayed on a map, \textit{side-by-side}, for visual contextualization. After selecting 2 zones to be compared, the feature importance bar chart is produced.

    \item A \textbf{\textit{feature importance bar chart}} generated from the comparison over the zones selected in the scatter plot, which showcases the top statistical variables that characterize the points on the selected zones, in descending order, represented by the selected combination as shown in Fig. \ref{fig:feature-importance}.

    \item A \textit{\textbf{2D and 3D trajectory visualization map}} that shows the two selected trajectories for side-by-side comparison, reflecting the feature importance variables through a heatmap (2D) and a \textit{Great Wall of Space-time} (3D), Fig. \ref{fig: the-wall}, \cite{the-great-wall-of-space-time}.
    
\end{itemize}

\begin{figure}[htbp]
    \centering
    \includegraphics[width=1.15\linewidth]{ 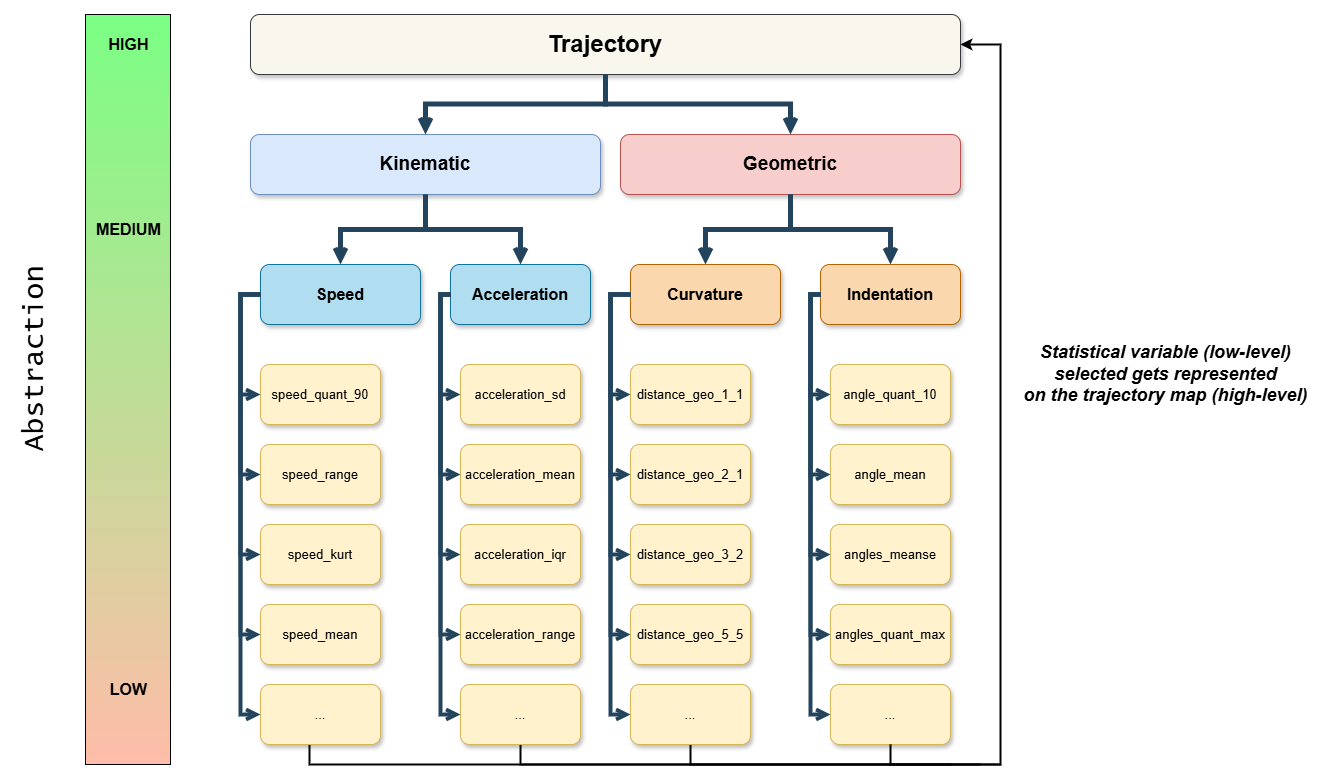}
    \caption{Detail levels.}
    \label{fig:levels}
\end{figure}

The visual components provided in this thesis project aim to address the need for interactive visualization, with a highly cohesive decision-making process, allowing comparison and understanding between the high-level and low-level details, as shown in Fig. \ref{fig:levels}.


\begin{figure}
    \centering
    \includegraphics[width=0.95\linewidth]{ 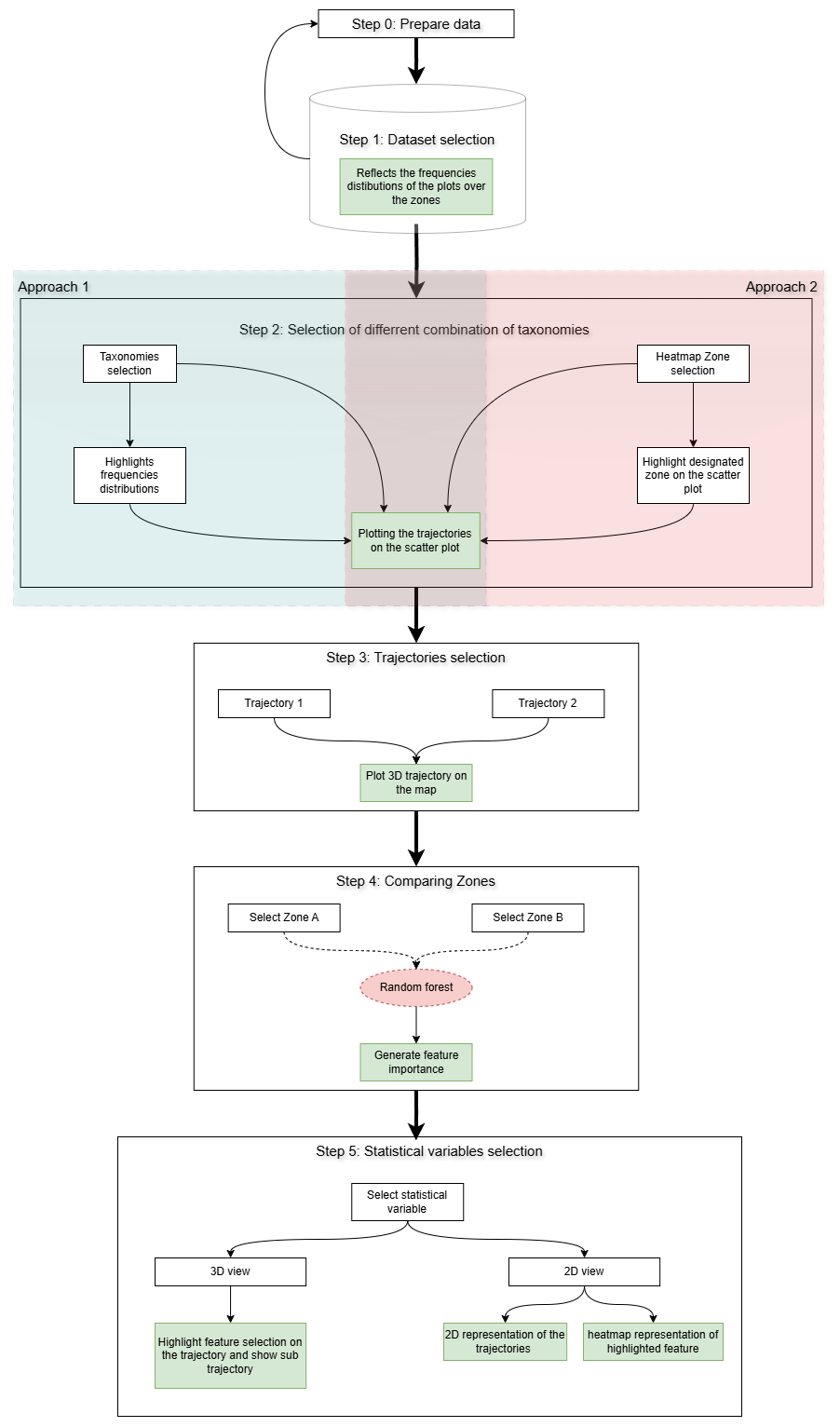}
    \caption{The tool's workflow.}
    \label{fig:tool-workflow}
\end{figure}

The visual analytics tool workflow is schematically represented in Fig. \ref{fig:tool-workflow}, addressing the visualizations adopted, besides providing an explanation about different approaches and paths that may be taken during analysis, in addition to demonstrating the interaction process of the tool and how analysts may address results.

\newpage

\subsubsection{\textit {Step 0: Data preparation}} 

This step is crucial to achieving representative and accurate results from the analysis performed. The analyst has to adapt the data to make it suitable for the system standards and requirements. Considering the fact that the tool is primarily designed for visual analysis over spatio-temporal data with a taxonomical approach, analysts must ensure data quality by starting with pre-processing, including normalizing, standardizing, and scaling.
This essential practice will minimize the impact of the outliers that may add biases to the visualization and will enhance the overall accuracy of the analysis output.

\subsubsection{\textit {Step 1: Dataset selection}} 
This is the first step from within the analytics tools, where the analyst selects the pre-processed dataset to be investigated. The selection is made using the first top left button, and once a dataset is selected, the label name for that dataset will be displayed. For the selection process to be correctly performed, certain features must exist in the dataset, including speed, acceleration, angle, distance, and bearing. 

The selected dataset will be reflected on the heatmap by showing the distribution over all taxonomy and subdivision combinations through the 4 zones discussed in \ref{Outlier detection and Zones}, from 0 to 3, as shown in Fig. \ref{fig: Heatmap-selection}.

\subsubsection{\textit {Step 2: Taxonomy and subdivision taxonomy combination selection}} 
\begin{figure}[htbp]
    \centering
    \begin{minipage}[b]{0.49\textwidth}
        \centering
        \includegraphics[width=\linewidth]{ 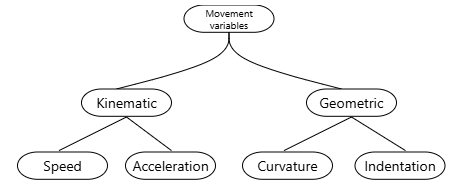}
        \caption{Tree taxonomies.}
        \label{fig:tree-taxonomies}
    \end{minipage}
    \hfill
    \begin{minipage}[b]{0.49\textwidth}
        \centering
        \includegraphics[width=\linewidth]{ 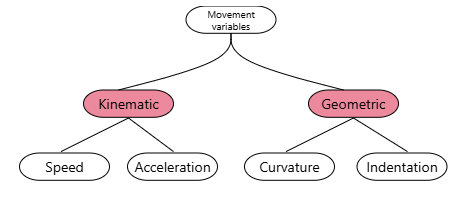}
        \caption{Combination selection.}
        \label{fig:combination-selection}
    \end{minipage}
    \hfill
\end{figure}

This combination selection can be approached in 2 different ways. The first approach is to follow the hierarchical steps starting from the taxonomy tree selection. In such a case, the analyst selects two different parameters from the taxonomy-based tree representation. A total of 7 different combinations are allowed, which may be selected from the higher-level taxonomy domain (i.e., \textit{Kinematic} and \textit{Geometric}), and/or from the subdivision of the taxonomies (i.e., the two children nodes for each of the high-level domains), namely \textit{speed}, \textit{acceleration}, \textit{curvature}, and \textit{Indentation}, as shown in Fig. \ref{fig:tree-taxonomies}. The analyst may select a combination of two parameters for further visualization and investigation, as shown in Fig. \ref{fig:combination-selection}. It is important to note that selecting a parent parameter and a subdivision parameter from the same parent taxonomy is not allowed (e.g., selecting \textit{Kinematic} as parameter 1 and \textit{Acceleration} as parameter 2, which is a subdivision of \textit{Kinematic}, is not allowed).

\begin{figure}[htbp]
    \centering
    \begin{minipage}[b]{0.45\textwidth}
        \centering
        \includegraphics[width=\linewidth]{ 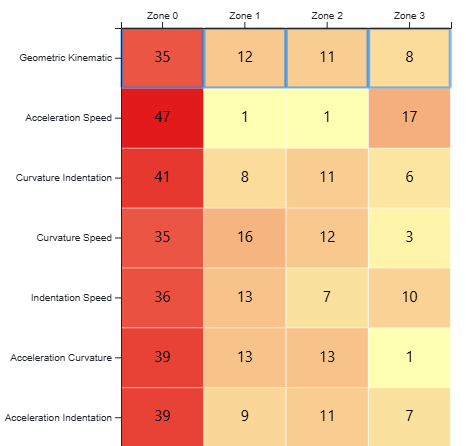}
        \caption{The frequency heatmap displaying the trajectory distribution over different zones for all parameter combinations.}
        \label{fig: Heatmap-selection}
    \end{minipage}
    \hfill
    \begin{minipage}[b]{0.5\textwidth}
        \centering
        \includegraphics[width=\linewidth]{ 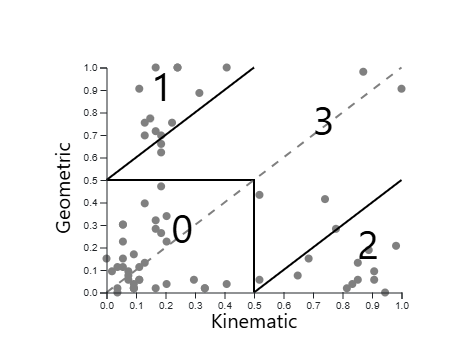}
        \caption{Scatter plot displaying trajectory distribution over 4 zones based on the score of the outlier detection algorithm results.}
        \label{fig: Scatter-plot}
    \end{minipage}
    \hfill
\end{figure}

\noindent Selecting the desired parameters will highlight the combination row in the frequency heatmap, as shown in Fig. \ref{fig: Heatmap-selection}, and the trajectory outlier scores will be generated and populated in the scatter plot, based on a scores produced from the outlier detection algorithm, representing the trajectory distribution values over the four different zones, Fig. \ref{fig: Scatter-plot}.

The second approach is to skip the hierarchical taxonomy tree selection and navigate to the heatmap directly, to then select a cell from one of the different taxonomies combinations, which will produce the same result as in the first approach.

\subsubsection{\textit {Step 3: Trajectories selection}} 
\begin{figure}[htbp]
    \centering
    \includegraphics[width=0.5\textwidth]{ 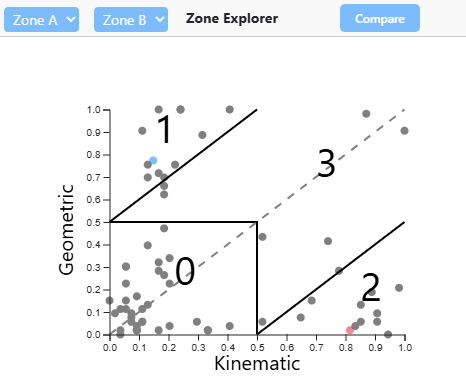}
    \caption{Selection of trajectory 1 (blue), located in \textit{zone 1} with ID 401, and trajectory 2 selection (red), located in \textit{zone 2} with ID 352.}
    \label{fig: Selecting-trajectories}
\end{figure}

\begin{figure}[htbp]
    \centering
    \includegraphics[width=0.66\textwidth]{ 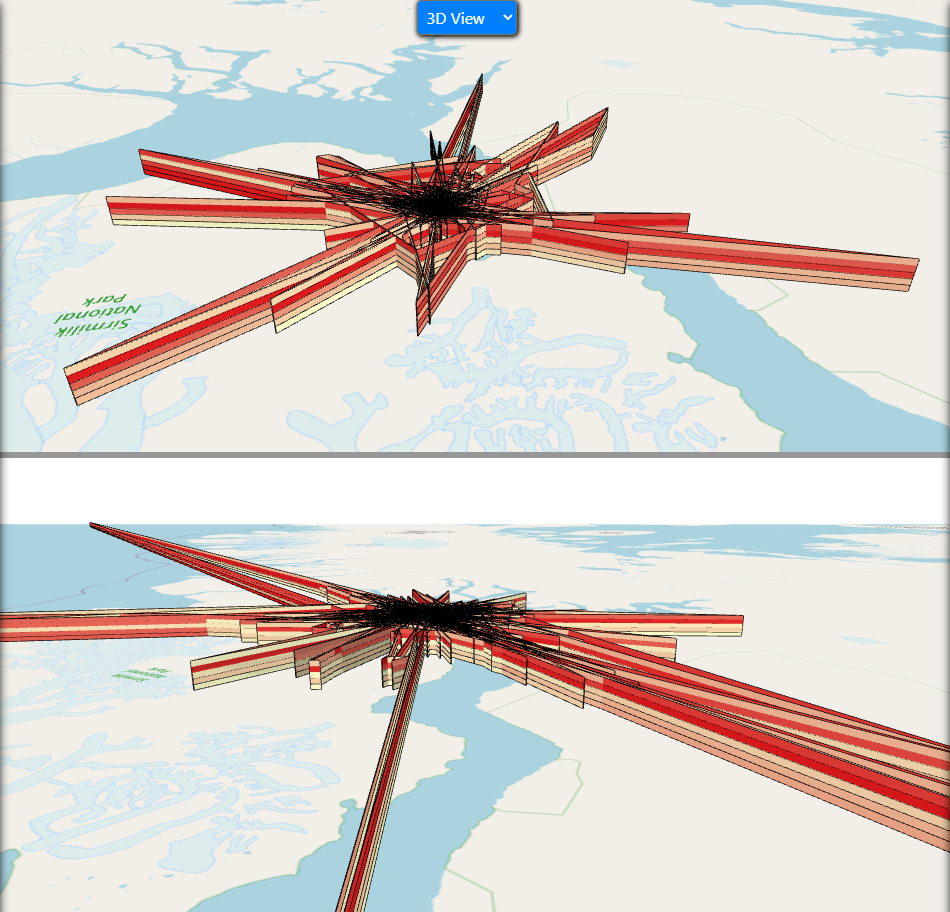}
    \caption{A 3D representation of the selected trajectories on separate maps. Each wall layer represents a feature (i.e., speed, acceleration, angle, distance, and bearing).}
    \label{fig: the-wall}
\end{figure}

At the third step, the analyst may select two different trajectories to be compared from the scatter plot, Fig. \ref{fig: Selecting-trajectories}. The trajectories selected will then be displayed on 2 separate map visualizations as 3D walls colored using a sequential color scale, with red color indicating high values and yellow color indicating low values. The wall is constructed from multiple stacked layers, where each layer represents a different feature variable. From top to bottom: speed, acceleration, angle, distance, and bearing, as illustrated in Fig. \ref{fig: the-wall}. 
\newpage

\subsubsection{\textit {Step 4: Comparing zones}} 
\begin{figure}[H]
    \centering
    \begin{minipage}[b]{0.55\textwidth}
        \centering
        \includegraphics[width=\linewidth]{ 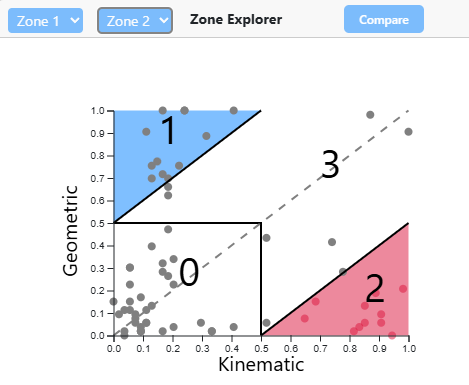}
        \caption{\textit{one-vs-one} comparison between zone 1 and zone 2.}
        \label{fig:zones-comparison}
    \end{minipage}
    \hfill
    \begin{minipage}[b]{0.37\textwidth}
        \centering
        \includegraphics[width=\linewidth]{ 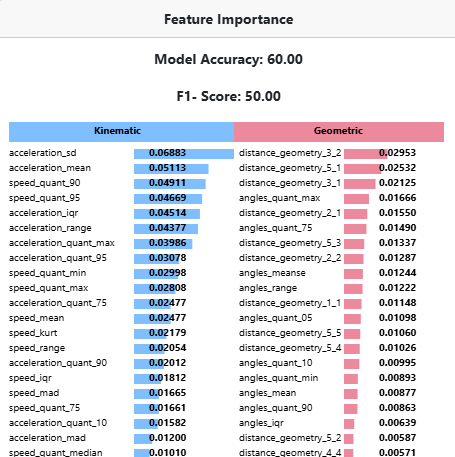}
        \caption{Feature importance of each selected taxonomy based on the trajectories located within the highlighted zones.}
        \label{fig:feature-importance}
    \end{minipage}
    \hfill
\end{figure}

During the 4th step, the analyst may then select 2 zones to be compared from the top left of the decision boundary scatter plot axes. The zones selected will be highlighted, as shown in Fig. \ref{fig:zones-comparison}, presenting a \textit{one-vs-one} comparison between zone A and zone B. The execution of the comparison triggers a random forest model, which handles the execution process by generating feature importance results, presented in descending order, indicating the features with the highest impact on the trajectory's behavior and shape at the top.
F1-score and model accuracy are both displayed, as shown in Fig. \ref{fig:feature-importance}.

\subsubsection{\textit {Step 5: Statistical variable selection}} 
\begin{figure}[H]
    \centering
    \includegraphics[width=1\linewidth]{ 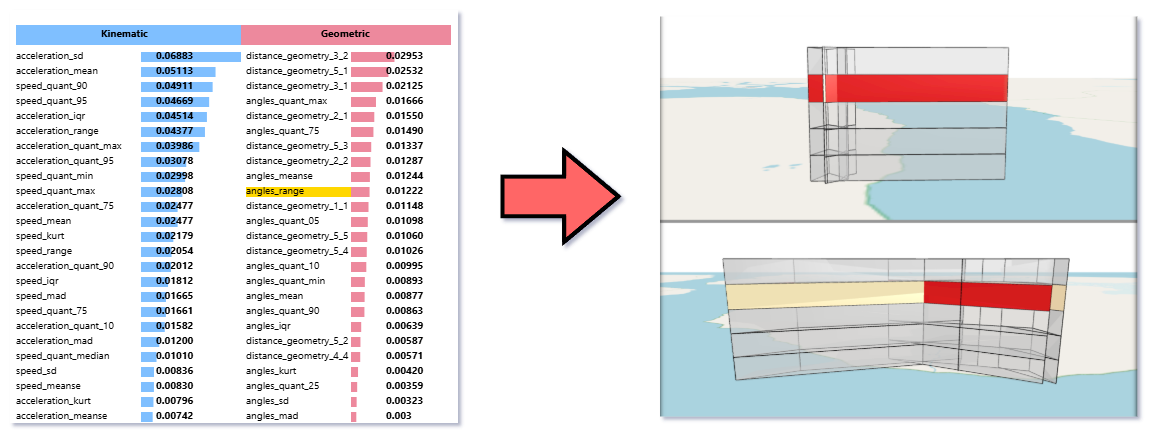}
    \caption{The reflection from selecting angles-range statistical variable on the trajectories, the yellow color on the map represents low angle, and the red color represents high angle.}
    \label{fig:selecting-statistical-variable-show-subtrajectory}
\end{figure}

The bar chart exhibited in Fig. \ref{fig:feature-importance} presents statistical variables generated from the random forest model. Selecting a single variable from the feature importance bar chart will be reflected in the representation of the trajectories over the maps, as shown in Fig. \ref{fig:selecting-statistical-variable-show-subtrajectory}.
This step allows the analyst to investigate the trajectories in-depth by avoiding projecting the entire trajectory at once, and instead, only visualizing a selected segment of it, in addition to highlighting the selected feature on the 3D wall, emphasizing the visual representation of the statistical variable selected onto the trajectory view, and effectively linking low-level variables with high-level views.

\subsubsection{\textit {Step 6: Iteration}} 

At this point, the analyst may decide which re-selections could take place, refining and redirecting the analytical process to increase the understanding of the selected dataset in step 1 as much as possible, finding patterns, and effectively describing them.

\subsection{Data Collection} \label{Data Collection}
In this work, a total of 2 real-world datasets will be used to avoid limiting the interpretation effects of the proposed system from a single source \cite{guidelines_for_conducting_and_reporting_case_study}. The sources follow some of Tavakoli et al.'s datasets presented in their study \cite{Yashar}. The datasets belong to dissimilar categories with different volumes.

The first dataset is collected in Canada through satellites, tracking 66 Arctic foxes over six years. It encompasses the movement of these foxes on Bylot Island during the sea ice time \cite{(fox_dataset)_by_lai_joël_bêty_berteaux}. This dataset is of relevant interest due to the potential for analysis over the role of animal movement in modeling the ecosystem at multiple levels.


The second dataset is gathered by \textit{The International Best Track Archive for Climate Stewardship} (IBTrACS), which contains the movement trajectories of 11229 tropical cyclones.
Data samples were given at 3-hour intervals, spanning the time from the 1840s to the present, 2025 \cite{knapp_kruk_levinson_diamond_neumann_2010, centers_centers_2019}.

Utilizing the data in its raw form is not efficient and may risk losing critical details, due to the presence of outliers, resulting in side effects during analysis and adding biases. Therefore, pre-processing for each dataset is required.

\subsubsection{Pre-processing} \label{Pre-processing}
The pre-processing step will be done by employing various methods, libraries, and algorithms, including Pandas, Min-Max Scaler from Scikit-Learn, and Density-Based Outlier Score (DBOS), among others.

The implementation of these tools supports this work by manipulating the datasets, making them feasible to leverage a more accurate and high-quality analysis. Next, the roles of the mentioned tools during the pre-processing phase are highlighted.

Pandas employed different techniques, including reading, selecting, dropping, concatenating, and saving the data. The incorporation of Pandas allows the implementation of several machine learning methods, such as outlier detection with the implementation of DBOS, the labeling of zones through score-to-zone matchers, the hyperparameter tuning, and the selection for random forest, as well as filtering and \textit{dataframe} handling with vectorized and optimized solutions.  

The data was scaled through 2 different algorithms: \textit{Min-Max Scaler} and \textit{Quantile Transformer}.
Min-Max Scaler is used to normalize features by scaling them to a specific range between 0 and 1, while Quantile Transformer is used to transform decision scores with a uniform distribution.

DBOS is used for scoring outliers based on how each data point is isolated from its neighbors in the feature space.
Data transformation was involved through extracting data points from GeoJSON files, calculating movement features, and saving the result in CSV files for further analysis and visualization.




\subsection{Data Analysis} \label{Data Analysis}
For data analysis, the proposed thesis project employs machine learning with data mining as the main analysis method. 
During the methodology, a number of pre-processing and machine learning algorithms have been briefly presented. In this section, further details are given concerning implementation decisions and technical overviews.

\begin{itemize}
    \item DBOS: Distance-Based Outlier Score is key as a scoring technique for the proposed project. Several outlier detection and scoring techniques may be used for such a purpose, and for the current proposal, DBOS is utilized due to its simple implementation yet effective results. 
    Other techniques can be used, however, they introduce additional complexity to the outlier score calculation without necessarily providing more accurate or efficient results.
    
    \item Min-Max Scaler: When performing outlier detection and producing outlier scores, applying \textit{Min-Max Scaler} is critical to produce a common range for the results achieved. Without it, values produced from the DBOS algorithm would be difficult to compare, making the scatter points plotted in the decision boundary axes too distant, and as a consequence, affecting the analysis process due to the lack of relativity between the points in the visualization. Therefore, \textit{Min-Max Scaler} is introduced, keeping the value range between 0 and 1.
    
    \item Random Forest: Random forest is a widely used machine learning algorithm that implements several decision trees in order to find a solution by averaging all trees' results. In the current proposal, random forest serves as the algorithm used for feature importance when deriving the statistical variables with the highest impact on the decision of each trajectory being included within a given zone, through the specific taxonomy combination previously selected. In other words, it provides a finer-grained statistical indicator over a given trajectory. For the current proposal, the algorithm is applied to the \textbf{outlier score results} for the \textbf{combination} selected and the \textbf{zones} selected. The dataset is split into 80\% training and 20\% testing, with a random state value of 42. The results from the feature importance algorithm are presented with a bar chart table, distributed in two columns, one for each of the combination parameters selected, as shown in Fig. \ref{fig:feature-importance}.

     \item Hyperparameter tuning and selection: For the random forest algorithm, a number of changes and tests have been made in order to increase the model's accuracy. The initial tests were performed with random forest default parameters, and F1-scores resulted in an average of 50\%, with some individual cases resulting in scores below 20\%. Before the first adjustments, grid search was performed, fitting 5 folds with 24 candidates each, with a total of 120 fits. Once grid search was completed, the parameters found were applied to real datasets to verify accuracy; however, the more complex datasets still showed low accuracy. Manual tuning was then performed, finally achieving high accuracy throughout several datasets, with the following parameters:
    \begin{itemize}
        \item n\_estimators=200
        \item max\_depth=10
        \item class\_weight='balanced'
        \item random\_state=42
    \end{itemize}
    Additionally, the test and train splitting included \textit{stratify=y} to avoid imbalances during the train and test distribution. With the split and hyperparameters selected above, the accuracy of the models increase to an average of 80\% for F1-scores, with real-world datasets that will be utilized in this thesis project. It must be noted that such models dealing with spatio-temporal datasets are often less predictable, and higher accuracy may not always be possible, as when dealing with other datasets or supervised models.

\end{itemize}



\subsection{Reliability and Validity} \label{Reliability and Validity}
The credibility and findings of this thesis project consider both reliability and validity to ensure reproducibility, reduce technical bias, and ensure supporting conclusions from the data and method proposed.\\

\subsubsection{Reliability}
To ensure reliability, this thesis study will use 2 considerably different datasets, each representing moving objects in diverse environments. The first dataset concerns the analysis of Arctic foxes in nature, 
while the second dataset concerns tropical cyclones. Moving animals, 
and natural disasters represent diverse enough characteristics to ensure that the core representation of the results during analysis is formulated solely from movement, and not from external characterizations or features unrelated to spatio-temporal analysis.

Furthermore, the analytics tool by design will not allow or compute any other features, statistical variables, or taxonomies outside of movement parameters during processing or execution, making the system resilient to technical biases.

\subsubsection{Validity}
In order to ensure  the validity of this thesis project, the following challenges are addressed:
\begin{itemize}
    \item Feature importance accuracy: achieving high accuracy in machine learning models, such as the random forest algorithms used for feature importance, can be challenging when dealing with spatio-temporal datasets, especially larger datasets with high-dimensionality. To mitigate this issue, hyperparameter selection will be done when training the random forest model, with the aim of increasing accuracy. 
    \item Data pre-processing: the process of preparing and adapting the datasets crucial for data analysis, to both achieve accurate results, while also maintaining the data as unchanged as possible, to avoid losing representative information from the original data. To pre-process data, this project follows a combination of techniques as presented in subsection \ref{Pre-processing}, including scaling, individual feature translation, and manual dropping of rows when needed, among other methods. 
\end{itemize}



\subsection{Ethical Considerations} \label{Ethical Considerations}
The proposed thesis project follows a case study approach, without the involvement of external human participants or any other personal data that should be ethically considered.



\subsection{Summary} \label{Methodology Summary}
As a summary, the following significant highlights from this section are presented: 
\begin{itemize}
    \item The thesis project methodology outlines a well-defined set of components to produce an effective visual analytics tool to discover patterns on spatio-temporal datasets with a taxonomy.
    \item The taxonomy utilized presents two groups, \textit{Kinematic} and \textit{Geometric}, further subdivided into \textit{Speed}, \textit{Acceleration}, \textit{Indentation}, and \textit{Curvature}.
    \item The main components are a taxonomy tree, frequency heatmap, a scatter plot with decision boundaries, a feature importance bar chart, 2D and 3D trajectory visualizations over maps.
    \item The analysis workflow outlines the cohesion and logical linkage between high-level and low-level details in a cycle approach, where low-level details get reflected onto high-level visualizations. 
    \item The main analysis method is machine learning and data mining, with the incorporation of DBOS for outlier detection and random forest for feature importance analysis.
    \item The method used for evaluating the thesis's tool will be case studies, using different datasets.
     \item The datasets used are from 2 different moving objects types, aiming at achieving reliable results.
    \item Potential challenges related to pre-processing and model accuracy were discussed, which will be addressed by following a number of steps and techniques to increase the validity as much as possible.

\end{itemize}

\noindent In the following chapter, the results for the case studies will be presented for the datasets discussed in this section.


\newpage

\section{Results and Analysis}
\label{ResultsAnalysis}

In this chapter, two case studies will be presented, with further review and analysis of the datasets discussed and the methodology outlined in the previous chapter. 
The first case study \ref{Case Study 1: Arctic Foxes} will analyze and present findings on the Arctic Foxes dataset, while the second case study \ref{Case Study 2: Tropical Cyclones} will present an analysis over the Tropical Cyclones dataset.


\subsection{Case Study 1: Arctic Foxes} \label{Case Study 1: Arctic Foxes}
\begin{figure}[H]
        \centering
        \includegraphics[width=0.3\linewidth]{ 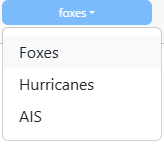}
        \caption{Arctic Foxes dataset selection.}
        \label{fig:foxes-dataset-selection}
    \end{figure}

    During the first step, the desired dataset is selected from the top left button on the analytics tool. For the first case study, the Arctic foxes are selected, labeled as "Foxes", as shown in Fig. \ref{fig:foxes-dataset-selection}.
    
\begin{figure}[H]
        \centering
        \includegraphics[width=0.7\linewidth]{ 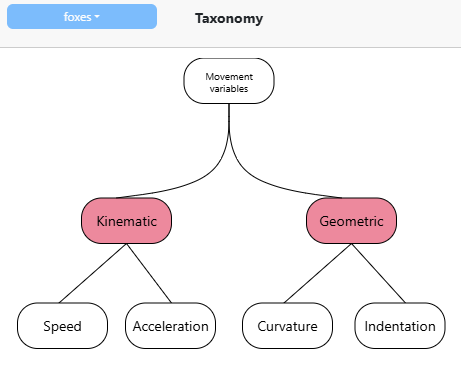}
        \caption{Taxonomy selection of Geometric and Kinematic group combination.}
        \label{fig:foxes-taxonomy-selection}
    \end{figure}

    \noindent Upon selection, the dataset gets loaded onto the analytics tools. For this very first analysis step, there is a greater interest in the overview of higher-level taxonomy groups. Namely, the selection for this first iteration is \textit{Kinematic} and \textit{Geometric}, as shown in Fig. \ref{fig:foxes-taxonomy-selection}

    \noindent The main motivation for this initial selection is to review first the two main super-groups to find underlying details that may support and redirect further analytical steps into subgroups, once relevant indicators have been found.
    
\begin{figure}[H]
        \centering
        \includegraphics[width=0.7\linewidth]{ 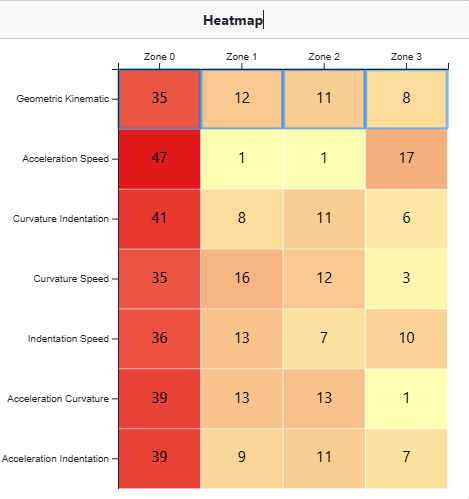}
        \caption{Foxes frequency heatmap.}
        \label{fig:foxes-heatmap}
    \end{figure}

    \noindent As the foxes dataset is selected, the frequency heatmap gets calculated and populated, highlighting the combination selected, in this case, the first row. The frequency represents the number of trajectories for the entire dataset, for each of the possible combinations against all 4 zones.
    For the Foxes dataset, Fig. \ref{fig:foxes-heatmap} shows the frequencies over the combination \textit{Geometric Kinematic} as follows:
    \begin{itemize}
        \item Zone 0: 35 trajectories 
        \item Zone 1: 12 trajectories 
        \item Zone 2: 11 trajectories 
        \item Zone 3: 8 trajectories 
    \end{itemize}

    \noindent The 35 points within \textit{zone 0} represent the majority of the foxes' trajectories (53\%), which are categorized within such zone due to the outlier score results. Particularly, these trajectories are represented by having an outlier score below 0.5 for both the combination parameters selected; \textit{Geometric} and \textit{Kinematic} scores are below 0.5, within a scale from 0 to 1, which represents that neither behavior characterizes the trajectory individually or combined.
    
    Zones 1 and 2, show 12 and 11 trajectories, respectively (for a total of 35\%). 
    The scores for the selected combination in zone 1 are characterized by having a score above 0.5 for the first parameter (Geometric), but a low representative score for the second parameter (Kinematic). The trajectories within zone 2 effectively represent the opposite, having a score above 0.5 for Kinematic behavior. 
    In other terms, all 12 points within zone 1 are highly characterized by Geometric behavior, but without significant Kinematic behavior, and all 11 trajectories in zone 2 have low Geometric behavior, but high Kinematic behavior. The final 8 trajectories in zone 3 represent hybrid results: both Kinematic and Geometric are representative, characterized by having a combined behavior (12\%).
    
\begin{figure}[H]
        \centering
        \includegraphics[width=0.7\linewidth]{ 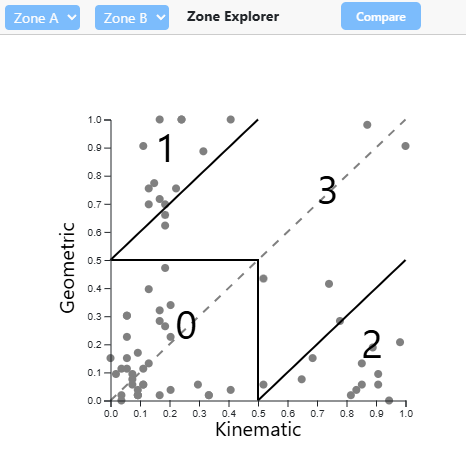}
        \caption{Arctic Foxes scatter plot with decision boundaries: (zone 0) 35 trajectories, (zone 1) 12 trajectories, (zone 2) 11 trajectories, (zone 3) 8 trajectories.}
        \label{fig:foxes-scatter-1}
    \end{figure}

    From the selected taxonomy combination, the outlier scores for each trajectory are then plotted as scatter points in the decision boundary plane, as shown in Fig. \ref{fig:foxes-scatter-1}. 
    The distribution of the scores presented in the previous step (i.e., on the heatmap) is now shown on a plane, where the y-axis is represented by the Geometric parameter, while the Kinematic parameter is shown on the x-axis.  
    The points are spread out in the decision boundary plot from the score results of the DBOS algorithm.

\begin{figure}[H]
        \centering
        \begin{minipage}[b]{0.325\textwidth}
            \centering
            \includegraphics[width=\linewidth]{ 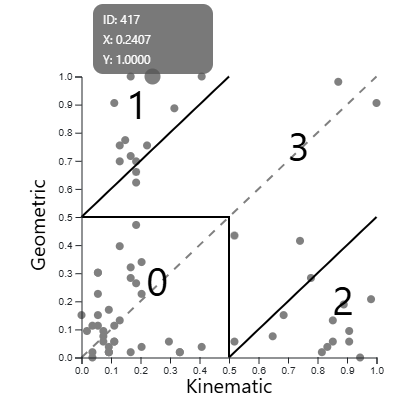}
            \caption{Trajectory \\with Geometric behavior.}
            \label{fig:foxes-scatter-hovering-1}
        \end{minipage}
        \hfill
        \begin{minipage}[b]{0.325\textwidth}
            \centering
            \includegraphics[width=\linewidth]{ 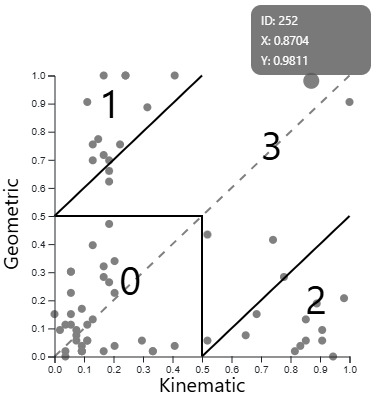}
            \caption{Trajectory \\with hybrid behavior.}
            \label{fig:foxes-scatter-hovering-2}
        \end{minipage}
        \hfill
        \begin{minipage}[b]{0.325\textwidth}
            \centering
            \includegraphics[width=\linewidth]{ 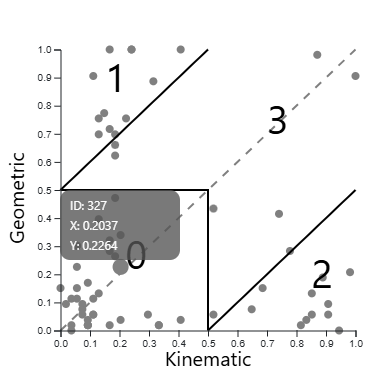}
            \caption{Trajectory without significant behavior.}
            \label{fig:foxes-scatter-hovering-3}
        \end{minipage}
        \hfill
    \end{figure}

    \noindent From the decision boundary visualization and interaction, the distribution and behavior become simplified. For instance, identifying the 3 trajectories with the highest Geometric behavior can easily be derived by simply hovering over these points, as shown in Fig. \ref{fig:foxes-scatter-hovering-1}. Similarly, it is possible to find the 2 trajectories with the highest hybrid behavior, as shown in Fig. \ref{fig:foxes-scatter-hovering-2}, and the trajectories with low representation from the combination selected, Fig. \ref{fig:foxes-scatter-hovering-3}. The IDs and scores are shown, and further analysis of the individual trajectories can be performed.

\begin{figure}[H]
        \centering
        \includegraphics[width=0.7\linewidth]{ 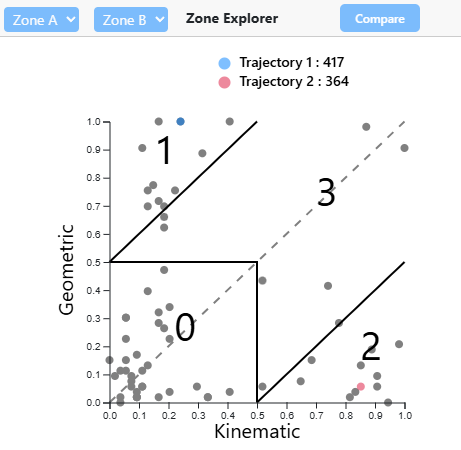}
        \caption{Arctic Foxes scatter plot with decision boundaries: first trajectory selected in zone 1 (blue) and second trajectory in zone 2 (red).}
        \label{fig:foxes-scatter-2}
    \end{figure}

    \noindent To further investigate individual trajectories from both Geometric and Kinematic parameters, two trajectory points are selected: one from zone 1 and another from zone 2. The two points selected were the following (also shown in Fig. \ref{fig:foxes-scatter-2}):
    \begin{itemize}
        \item Fox ID 417, zone 1, x=0.2407, y=1.0 (blue).
        \item Fox ID 364, zone 2, x=0.8519, y=0.0566 (red).
    \end{itemize}

     \noindent The selection is done by directly pressing on the scatter point, pressing \textit{Compare}, and then selecting the second point.

    \noindent The selected points in the scatter plot will be used in further steps to analyze and visualize the trajectories in the 2D and 3D maps, when presenting the statistical variables in context to the high-level visualizations.
        
\begin{figure}[H]
        \centering
        \includegraphics[width=0.7\linewidth]{ 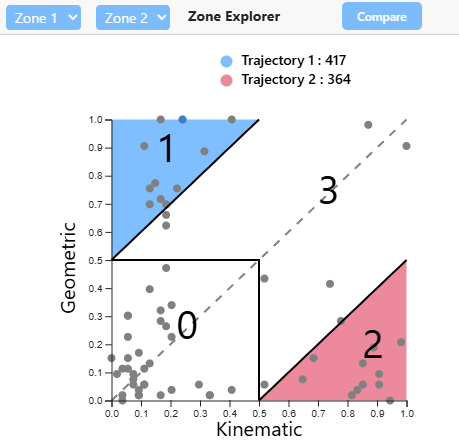}
        \caption{Arctic Foxes scatter plot with decision boundaries: zone 1 vs zone 2.}
        \label{fig:foxes-scatter-3}
    \end{figure}

    During the following stage, zone comparison is done. To further compare Kinematic and Geometric behavior, a \textit{one-vs-one} comparison is selected between zone 1 (blue) and zone 2 (red). Upon selection, both zones get highlighted accordingly, as shown in Fig. \ref{fig:foxes-scatter-3}.
    
    \noindent The zone comparison triggers the random forest algorithm, which compares all points from the selected zones and the combination of features selected from the taxonomy. This determines and denotes the statistical variables that had the most significant weight on the decision of these points being categorized within the previously shown zones. 
    The model gets trained from points selected within the combination. In this case, since the higher-level parameters were selected, namely Geometric and Kinematic, all 72 features are considered, as the combination encapsulates all statistical variables. The results of the zone comparisons are shown in the following step.

    From the comparison between zone 1 and zone 2 with the random forest algorithm, the following feature importance table is produced, with an F1-score of 80\%, Fig. \ref{fig:foxes-feature-importance}:
    
    \begin{figure}[H]
        \centering
        \includegraphics[width=0.7\linewidth]{ 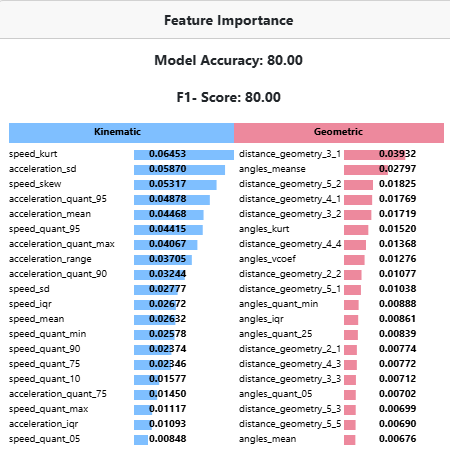}
        \caption{Zone 1 vs zone 2 feature importance results, separated in Kinematic and Geometric columns, descending order.}
        \label{fig:foxes-feature-importance}
    \end{figure}

   \noindent Through further inspection of the feature importance column, the Kinematic column describes the following 5 statistical variables with the highest impact: 
   \begin{itemize}
        \item Speed Kurtosis (speed\_kurt)
        \item Acceleration Standard Deviation (acceleration\_sd)
        \item Speed Skewness (speed\_skew)
        \item Acceleration Quantile 95 (acceleration\_quant\_95)
        \item Acceleration Mean (acceleration\_mean)
    \end{itemize}
    
    \noindent From the Geometric column, the top 5 statistical variables are: 
    \begin{itemize}
        \item Distance Geometry Signature 3\_1 (distance\_geometry\_3\_1) 
        \item Angles Mean Standard Error (angles\_meanse)
        \item Distance Geometry Signature 5\_2 (distance\_geometry\_5\_2)
        \item Distance Geometry Signature 4\_1 (distance\_geometry\_4\_1)
        \item Distance Geometry Signature 3\_2 (distance\_geometry\_3\_2)
    \end{itemize}
        
    \noindent The finding of such indicators reveals the unique statistical variables that may be attributed to the highest importance for all points within zones 1 and 2, to be included in either of them. Since the combination selected is Kinematic and Geometric for the parameter selection, zone 1 represents mainly Geometric behavior, and zone 2 mostly means Kinematic behavior. The key statistical variables were related to acceleration, speed, and distance geometry.
    Given this information, analyzing these statistical variables on the trajectory map will provide further insights, which will be useful to reiterate and analyze subgroup features related to them.
    
\begin{figure}[H]
        \centering
        \includegraphics[width=0.8\linewidth]{ 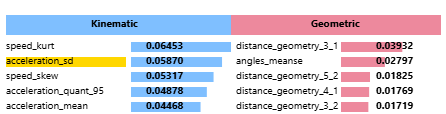}
        \caption{Selecting \textit{acceleration\_sd} (acceleration standard deviation) from the feature importance bar chart to produce sampled map visualizations.}
        \label{fig:foxes-feature-highlighted}
    \end{figure}

    From the feature importance bar chart, the Kinematic column shows that both acceleration and speed play a significant role in the overall Kinematic behavior. From the very top 5 statistical variables, speed sums up to a score of \textbf{0.1177} (the sum of Speed Kurtosis and Skewness), while acceleration sums to a score of \textbf{0.15216} (the sum of Acceleration Standard Deviation, Quantile 95, and Mean). During this analysis, \textbf{acceleration standard deviation} shows to be the most significant out of the 3 acceleration statistical variables, therefore getting selected for further analysis, and being highlighted on the feature importance bar chart, as shown in Fig. \ref{fig:foxes-feature-highlighted}.

    \noindent At this step, it is possible to further analyze the feature importance variable selected in the previous step on the 3D and 2D map views, specifically comparing the 2 trajectories chosen in the scatter plot, Fig. \ref{fig:foxes-acceleration-3D-map} and Fig. \ref{fig:foxes-acceleration-2D-map}:

    \begin{figure}[H]
        \centering
        \begin{minipage}[b]{0.49\textwidth}
            \centering
            \includegraphics[width=\linewidth]{ 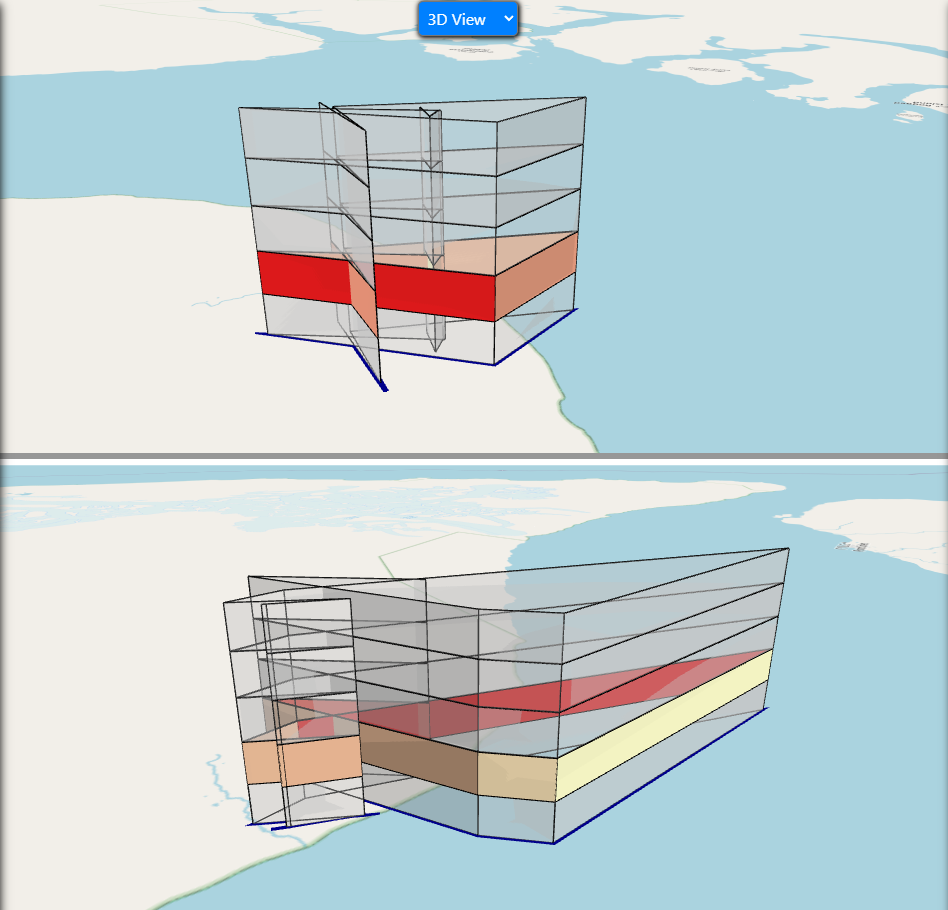}
            \caption{Trajectory of fox ID 417 (top) vs fox ID 364 (bottom), highlighting acceleration over time in 3D.}
            \label{fig:foxes-acceleration-3D-map}
        \end{minipage}
        \hfill
        \begin{minipage}[b]{0.49\textwidth}
            \centering
            \includegraphics[width=\linewidth]{ 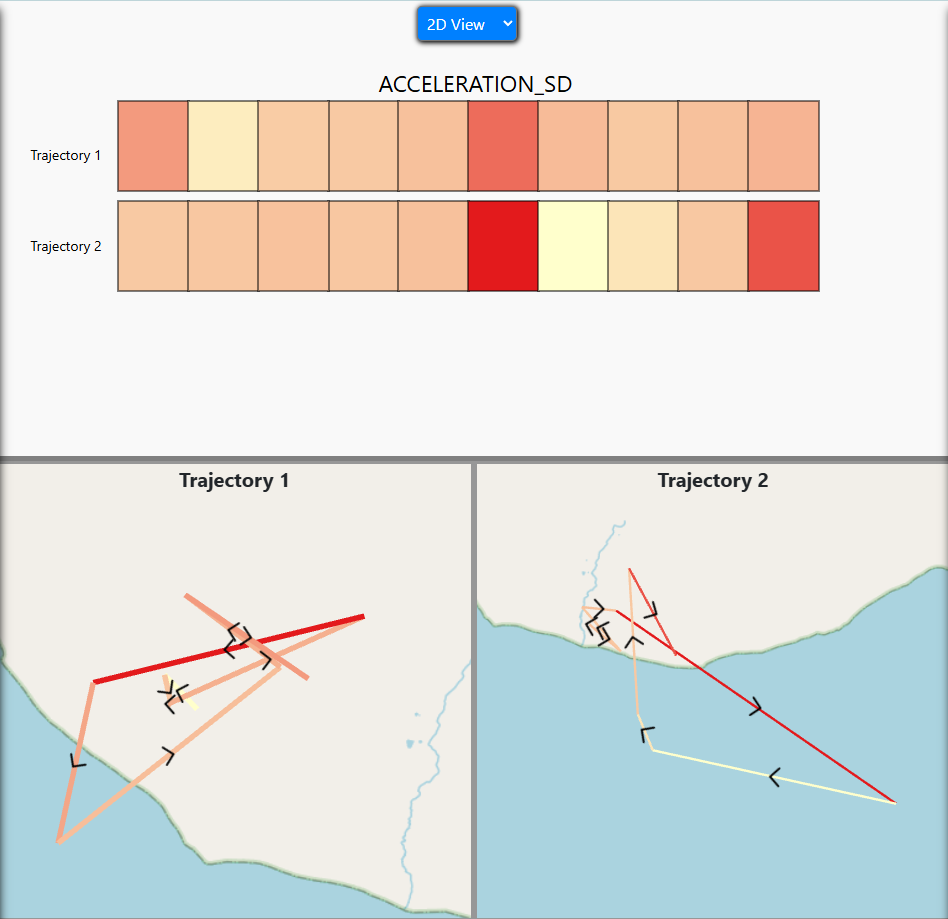}
            \caption{Trajectory of fox ID 417 (\textit{Trajectory 1}) vs fox ID 364 (\textit{Trajectory 2}), highlighting acceleration over time in 2D.}
            \label{fig:foxes-acceleration-2D-map}
        \end{minipage}
        \hfill
    \end{figure}

    \noindent The selection of the statistical variables \textit{acceleration standard deviation} entails also the significance of acceleration importance, as an overall feature. The 3D map view specifically highlights "acceleration" in order to further analyze such a feature, shown as the 2nd layer in the 3D wall highlighted, Fig. \ref{fig:foxes-acceleration-3D-map}. This view represents the trajectory path for the 2 foxes selected side-by-side. Since \textit{acceleration standard deviation} was selected, the standard deviation from the entire acceleration feature in both foxes previously selected (i.e., fox ID 417 and fox ID 364) is calculated. The dataset entry matching the closest value to the standard deviation gets fetched, as well as the 5 previous entries and 4 after, effectively displaying 2D and 3D views simplified, with only 10 trajectory points in total (i.e., the \textit{sample} view).
    
\begin{figure}[H]
        \centering
        \begin{minipage}[b]{0.49\textwidth}
            \centering
            \includegraphics[width=\linewidth]{ 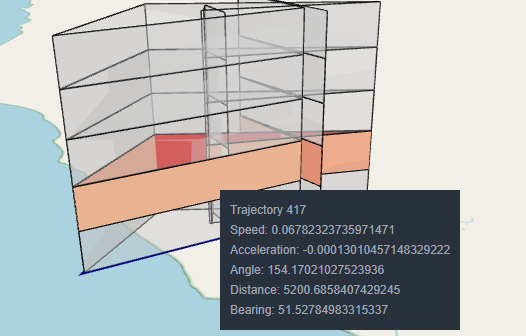}
            \caption{Trajectory of fox ID 417 highlighting acceleration over time.}
            \label{fig:foxes-acceleration-3D-hovering-1}
        \end{minipage}
        \hfill
        \begin{minipage}[b]{0.49\textwidth}
            \centering
            \includegraphics[width=\linewidth]{ 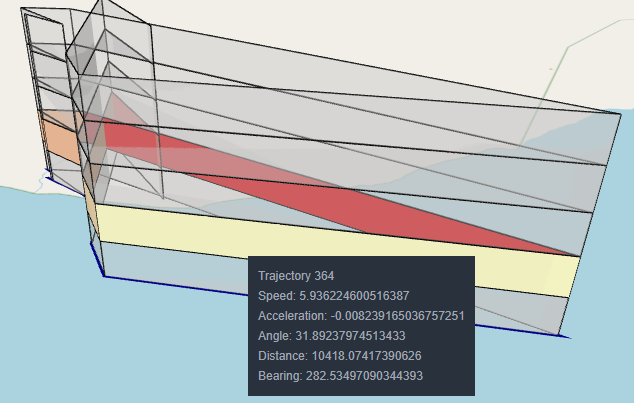}
            \caption{Trajectory of fox ID 364 highlighting acceleration over time.}
            \label{fig:foxes-acceleration-3D-hovering-2}
        \end{minipage}
        \hfill
    \end{figure}

    The 3D view also provides feature values for all 4 other features relevant to the foxes' trajectories. Hovering over a specific area of the wall displays speed, acceleration, angle, distance, and bearing at that specific space and time, for the selected foxes (IDs 417 and 364, respectively), as shown in Fig. \ref{fig:foxes-acceleration-3D-hovering-1}, and Fig. \ref{fig:foxes-acceleration-3D-hovering-2}.
    
\begin{figure}[H]
        \centering
        \includegraphics[width=0.9\linewidth]{ 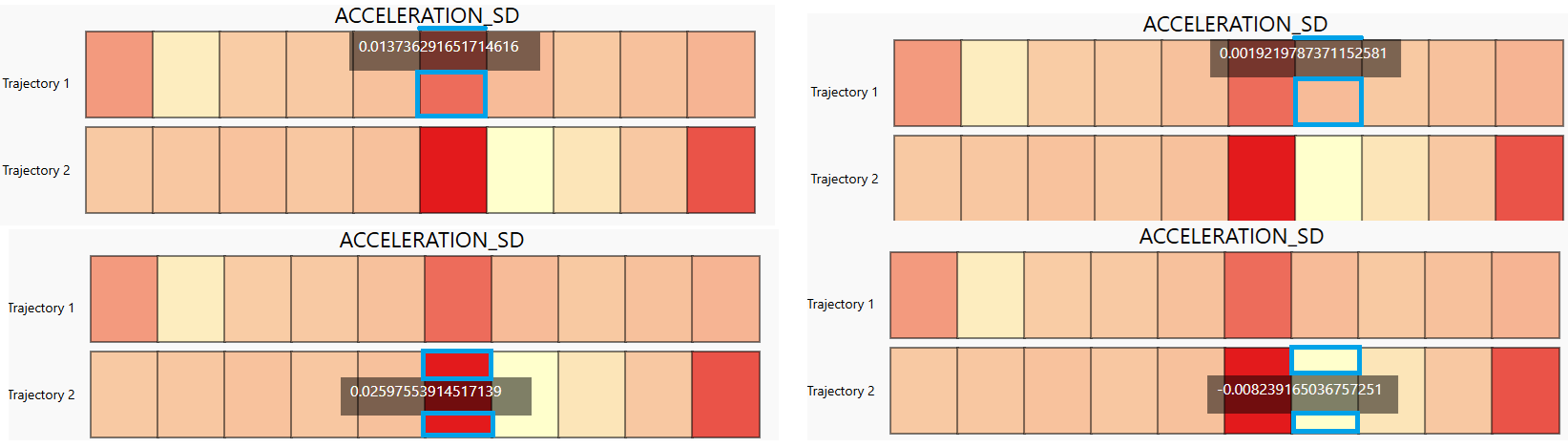}
        \caption{Fox ID 417 (\textit{Trajectory 1}) vs Fox ID 364 (\textit{Trajectory 2}).}
        \label{fig:foxes-2D-heatmap-map}
    \end{figure}

    \noindent Once the view is changed from 3D to 2D from the top button of the trajectories, the 2D view is displayed, as in Fig. \ref{fig:foxes-2D-heatmap-map}. This simplifies the view significantly, focusing on the trajectory and the single feature currently under analysis.

    \noindent Here, a heatmap from each of the trajectories is shown side-by-side. Each of the "colored rectangles" represents a unique latitude, longitude, and time for the acceleration feature, enabling the understanding of acceleration through space and time for the 2 selected foxes.

    By reviewing the analytical steps taken thus far, including combination selection, frequency, heatmap, decision boundary, feature importance, 2D and 3D map views, it can be discerned that acceleration is significant for the Kinematic behavior over the fox trajectory with ID 364 (also presented as "Trajectory 2" in the 2D view). When reviewing acceleration on the 2D heatmap, the behavior showed to be variant for fox 364: the first 5 entries showed a constant acceleration, the 6th entry present a peak of acceleration (\textbf{0,0259}, Fig. \ref{fig:foxes-2D-heatmap-map}, bottom left shown in strong red color), followed by great deceleration (\textbf{-0.0082}, Fig. \ref{fig:foxes-2D-heatmap-map}, bottom right shown with light yellow color), with the 3 final entries showing a re-acceleration behavior (i.e., the acceleration of the fox passed from a \textit{yellow} heatmap entry, which represents a negative value, towards \textit{strong red} by the end of the trajectory sample, representing a positive high value). These variations in behavior are derived from 10 entries in space and time, denoting a lot of changes in acceleration during a short sample. However, from map visual analysis, it can still be discerned that direction changes had an impact on such variance (i.e., there are changes in direction, which naturally could impact acceleration as well). Additionally, Fox 364 is represented by high Kinematic behavior, which encapsulates speed with as much importance as acceleration, making it difficult to yet convey any relevance for one specific subgroup, from higher taxonomy level results (i.e., the label from the outlier score indicates high Kinematic behavior, and not Acceleration specifically).
    
    Such insights regarding variation, in addition to the feature importance results for acceleration as an overall feature when comparing zone 1 (Geometric behavior) against zone 2 (Kinematic behavior), exhibit higher interest, leading to further reviewing \textbf{acceleration} during a new iteration, as the steps presented thus far showcase acceleration to be one of the key behaviors for foxes with high Kinematic characterization. Such findings, related to acceleration-based behavior and acceleration variables, further motivate an iteration focusing on acceleration.\\ 
    
\begin{figure}[H]
        \centering
        \includegraphics[width=0.95\linewidth]{ 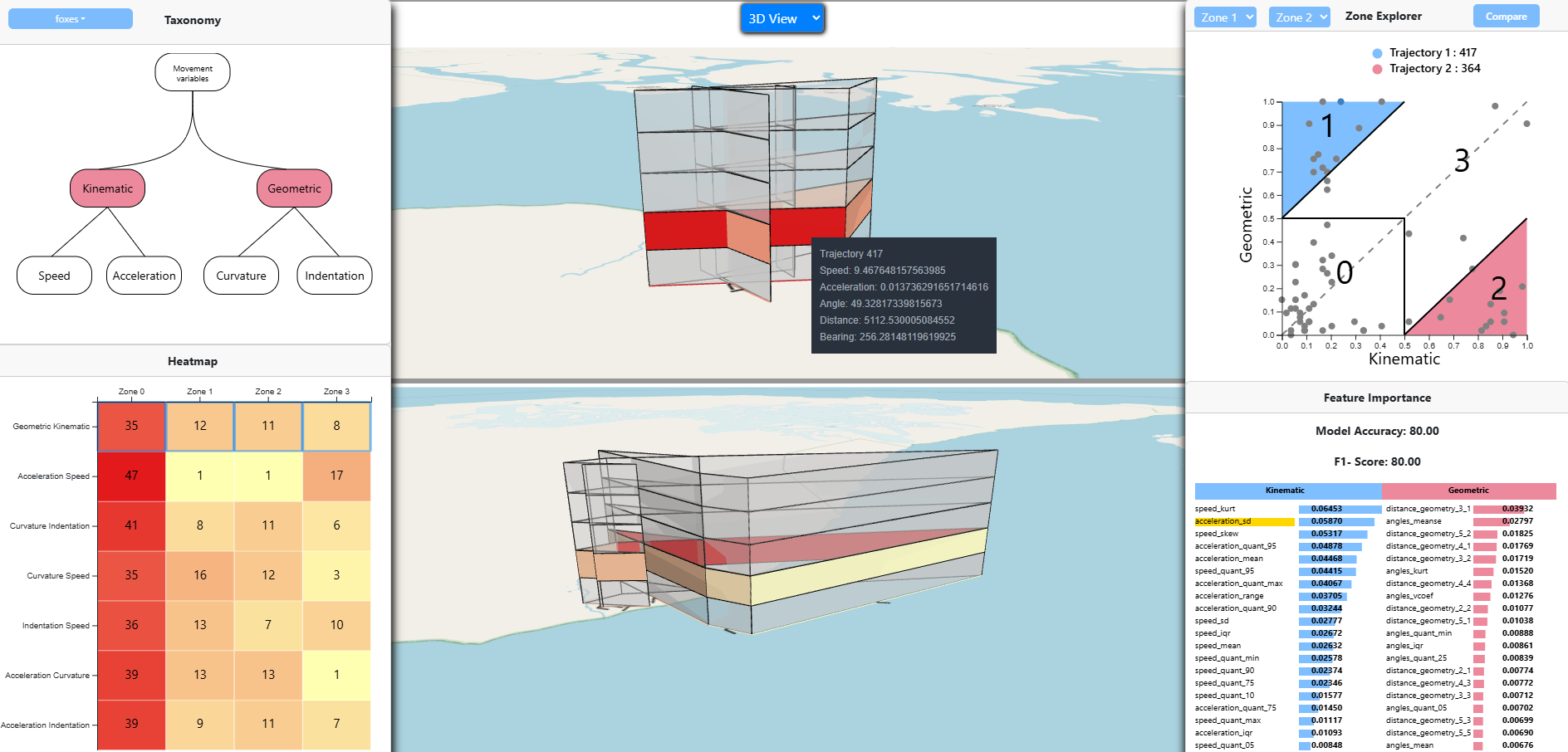}
        \caption{The full view of the tool after all analytic steps were outlined.}
        \label{fig:foxes-full-tool-3D}
    \end{figure}

    \noindent From the Geometric side, the DBOS algorithm produced the respective outlier scores, and the trajectories with mostly Geometric behavior were scattered throughout zone 2, from which Fox ID 417 was selected for individual comparison. During the review of acceleration on the 2D heatmap view (Fig. \ref{fig:foxes-2D-heatmap-map}, labeled as \textbf{Trajectory 1}, on the top of each of the 4 figures shown), it can be noticed that there is a similarity regarding acceleration, in contrast to fox 364 (selected from the Kinematic zone group). However, on a lower scale, with lower acceleration peaks, higher overall consistency, and lower deceleration values. Overall, acceleration still had relevant significance, although less than for Fox ID 364. In contrast, Fox ID 417 showed higher outlier score values for Geometric behavior. By analyzing the feature importance results, the distance geometry variables were found to be relevant for all the trajectory points in zone 2 (e.g., distance geometry 3\_1 and 5\_2, among others). However, from visual inspection of the trajectories through the 2D and 3D maps, it can be noticed that there is also high significance on the angle, as it can be discerned from trajectory 417 through its considerably "sharper turns", which indicate higher indentation as per Geometric analysis. To further review the impact of acceleration, an additional iteration is taken, which considers the combination of \textbf{Acceleration and Curvature}, however, with a deeper focus on Acceleration.
    Figure \ref{fig:foxes-full-tool-3D} presents a full tool view for some of the analysis performed thus far.

    \noindent As discussed in the previous iteration, a new combination is selected, choosing Acceleration and Curvature as shown in Figure \ref{fig:foxes-taxonomy-selection-2}, next to the frequency heatmap, highlighting the row under analysis, as displayed in Fig. \ref{fig:foxes-heatmap-2}:
    
    \begin{figure}[H]
        \centering
            \begin{minipage}[b]{0.49\textwidth}
            \centering
            \includegraphics[width=\linewidth]{ 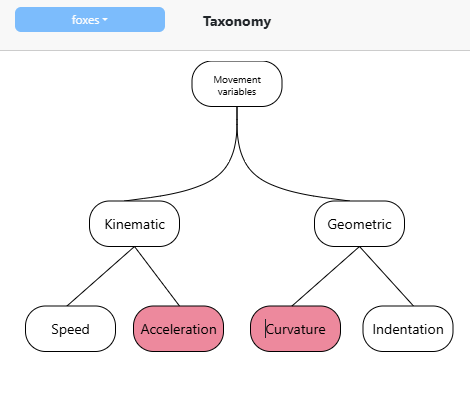}
            \caption{Arctic foxes, combination selection, Acceleration and Curvature.}
            \label{fig:foxes-taxonomy-selection-2}
        \end{minipage}
        \hfill
        \begin{minipage}[b]{0.39\textwidth}
            \centering
            \includegraphics[width=\linewidth]{ 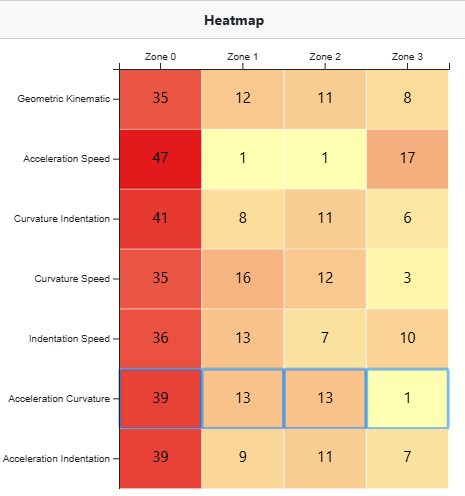}
            \caption{Combination highlight onto frequency heatmap.}
            \label{fig:foxes-heatmap-2}
        \end{minipage}
        \hfill
    \end{figure}
    
    \noindent The following values are found in the heatmap:
    \begin{itemize}
        \item Zone 0: 39 trajectories 
        \item Zone 1: 13 trajectories 
        \item Zone 2: 13 trajectories 
        \item Zone 3: 1 trajectories 
   \end{itemize}
    
    \noindent Presenting only 1 trajectory with hybrid behavior (i.e., zone 3, acceleration and curvature together) and 39 trajectories in zone 0, which show no distinctive Acceleration or Curvature-based behavior. Nonetheless, 20\% of the foxes showed distinctive Acceleration-based behavior, and 20\% showed Curvature-based behavior. 

\begin{figure}[H]
        \centering
            \begin{minipage}[b]{0.465\textwidth}
            \centering
            \includegraphics[width=\linewidth]{ 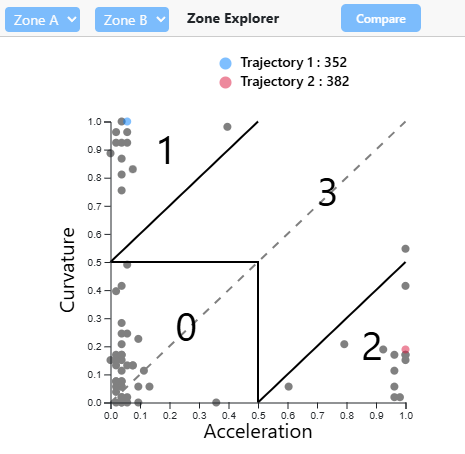}
            \caption{Arctic foxes decision boundary for Acceleration and Curvature, and trajectory selections.}
            \label{fig:foxes-acc-curv-scatter}
        \end{minipage}
        \hfill
        \begin{minipage}[b]{0.5\textwidth}
            \centering
            \includegraphics[width=\linewidth]{ 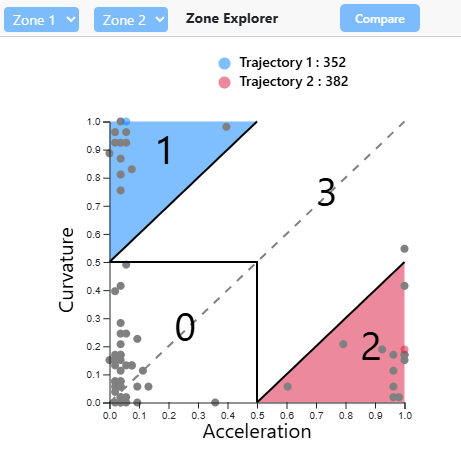}
            \caption{Arctic foxes, zone 1 (Curvature) vs zone 2 (Acceleration).}
            \label{fig:foxes-acc-curv-scatter-2}
        \end{minipage}
        \hfill
    \end{figure}

    The decision boundary is then updated, and scatter points are spread out throughout the 4 zones based on the outlier scores as presented in  Fig. \ref{fig:foxes-acc-curv-scatter}. One trajectory from each subgroup is selected: one from zone 1 with high curvature, and another trajectory from zone 2, denoted by having high acceleration. Additionally, zone 1 is selected and compared against zone 2, as shown in Fig. \ref{fig:foxes-acc-curv-scatter-2}, to generate feature importance results and analyze them in the following steps.
    
    \noindent From visual inspection, it can be noticed how the trajectories within zone 1 are considerably high in curvature (i.e., all scores being above 0.7), and very low in acceleration (i.e., below 0.1, with the exception of 1 trajectory). The same is shown in zone 2, as most trajectories with high acceleration behavior have relatively low curvature scores.
    
    In addition, the only trajectory that had hybrid behavior (i.e., curvature and acceleration together) was only slightly above 0.5 in curvature, and with a value close to 1 in acceleration.\\

    \noindent From the zones compared, the foxes selected were the following two:
    \begin{itemize}
        \item Fox ID \textbf{352}, zone 1, x=\textbf{0.0566}, y=\textbf{1.00} (blue).
        \item Fox ID \textbf{382}, zone 2, x=\textbf{1.00}, y=\textbf{0.1887} (red).
    \end{itemize}
    
    \noindent From the latest comparison between zones 1 and 2, the feature importance bar chart below is produced, with an F1-score of 100\%, Fig. \ref{fig:foxes-feature-importance-2}:

    \begin{figure}[H]
        \centering
        \includegraphics[width=0.7\linewidth]{ 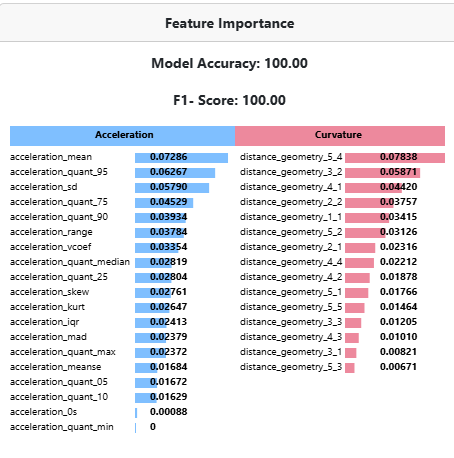}
        \caption{Feature importance zone 1 (Curvature behavior) vs zone 2 (Acceleration behavior).}
        \label{fig:foxes-feature-importance-2}
    \end{figure}

   \noindent By analyzing the feature importance columns, the following top 5 acceleration statistical variables are derived: 
   \begin{itemize}
        \item Acceleration Mean (acceleration\_mean)
        \item Acceleration Quantile 95 (acceleration\_quant\_95)
        \item Acceleration Standard Deviation (acceleration\_sd)
        \item Acceleration Quantile 75 (acceleration\_quant\_75)
        \item Acceleration Quantile 90 (acceleration\_quant\_90)

    \end{itemize}
    
    \noindent As per curvature, the top 5 statistical variables were:
    \begin{itemize}
        \item Distance Geometry Signature 5\_4 (distance\_geometry\_5\_4) 
        \item Distance Geometry Signature 3\_2 (distance\_geometry\_3\_2)
        \item Distance Geometry Signature 4\_1 (distance\_geometry\_4\_1)
        \item Distance Geometry Signature 2\_2 (distance\_geometry\_2\_2)
        \item Distance Geometry Signature 1\_1 (distance\_geometry\_1\_1)

    \end{itemize}

    \noindent As the main focus during this iteration is acceleration, further attention is given to the top 5 acceleration-related variables presented above.

\begin{figure}[H]
        \centering
        \begin{minipage}[b]{0.45\textwidth}
            \centering
            \includegraphics[width=\linewidth]{ 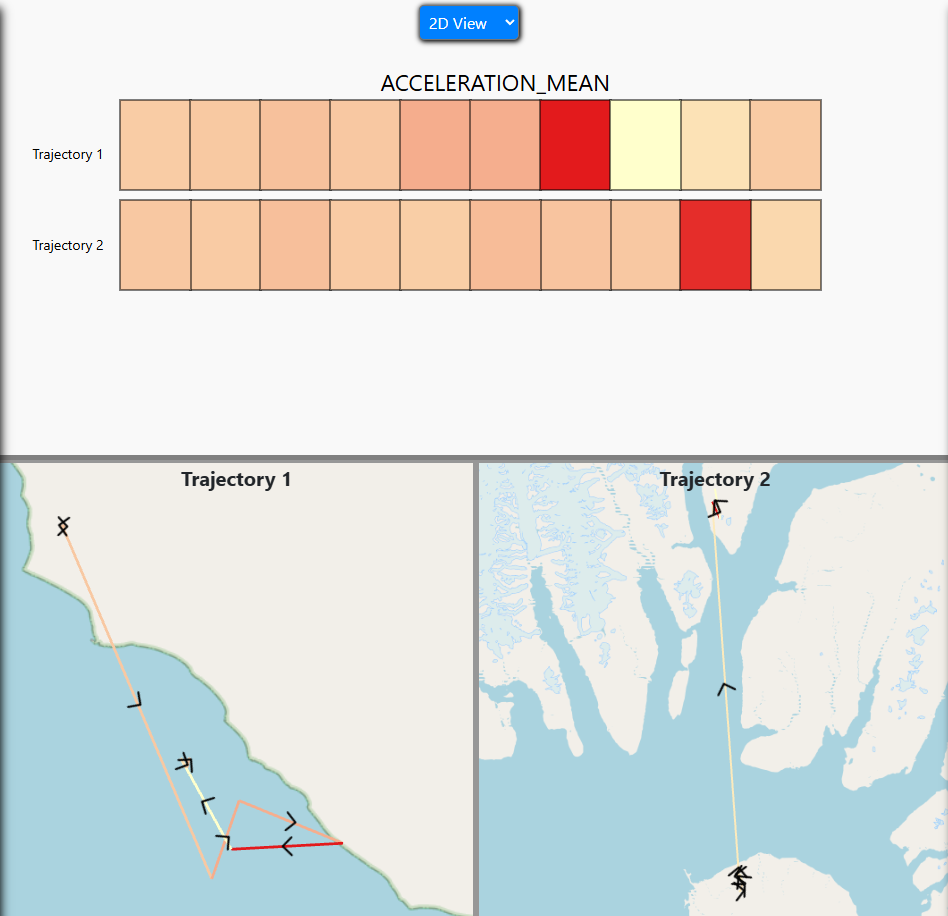}
            \caption{Mean acceleration.}
            \label{fig:foxes_curv_acc_mean_352_vs_382_2D}
        \end{minipage}
        \hfill
        \begin{minipage}[b]{0.45\textwidth}
            \centering
            \includegraphics[width=\linewidth]{ 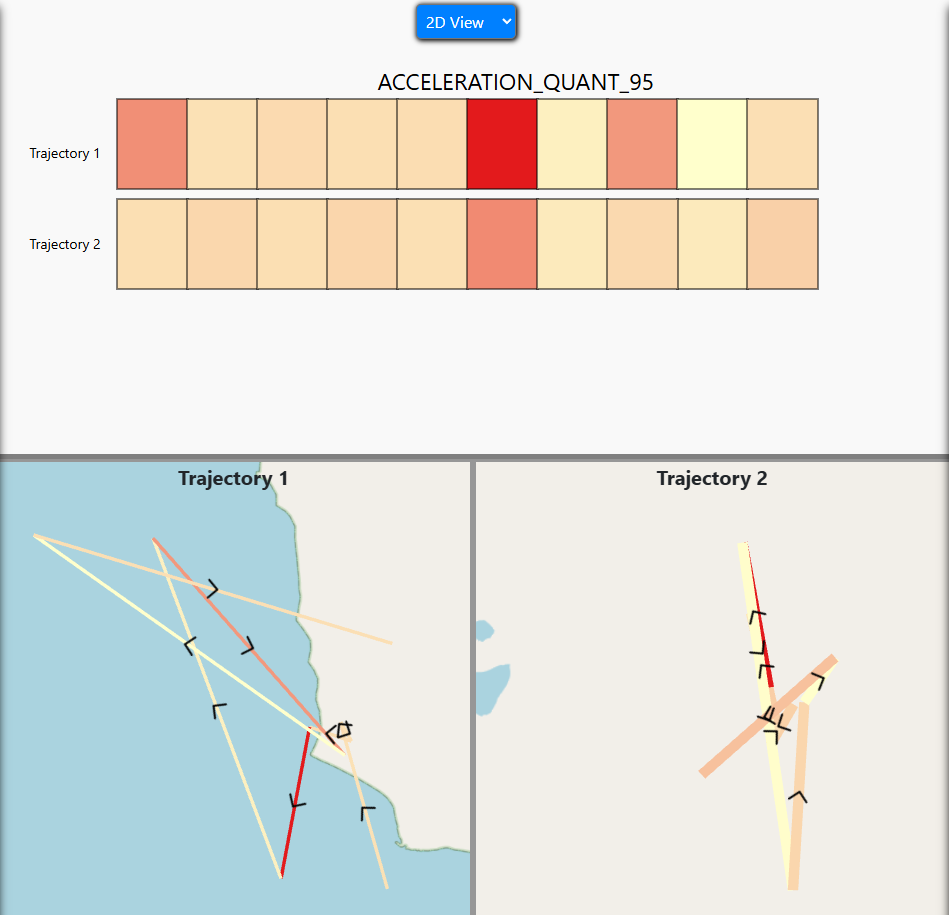}
            \caption{Acceleration quantile 95.}
            \label{fig:foxes_curv_acc_quant95_352_vs_382_2D}
        \end{minipage}
        \hfill
        \begin{minipage}[b]{0.45\textwidth}
            \centering
            \includegraphics[width=\linewidth]{ 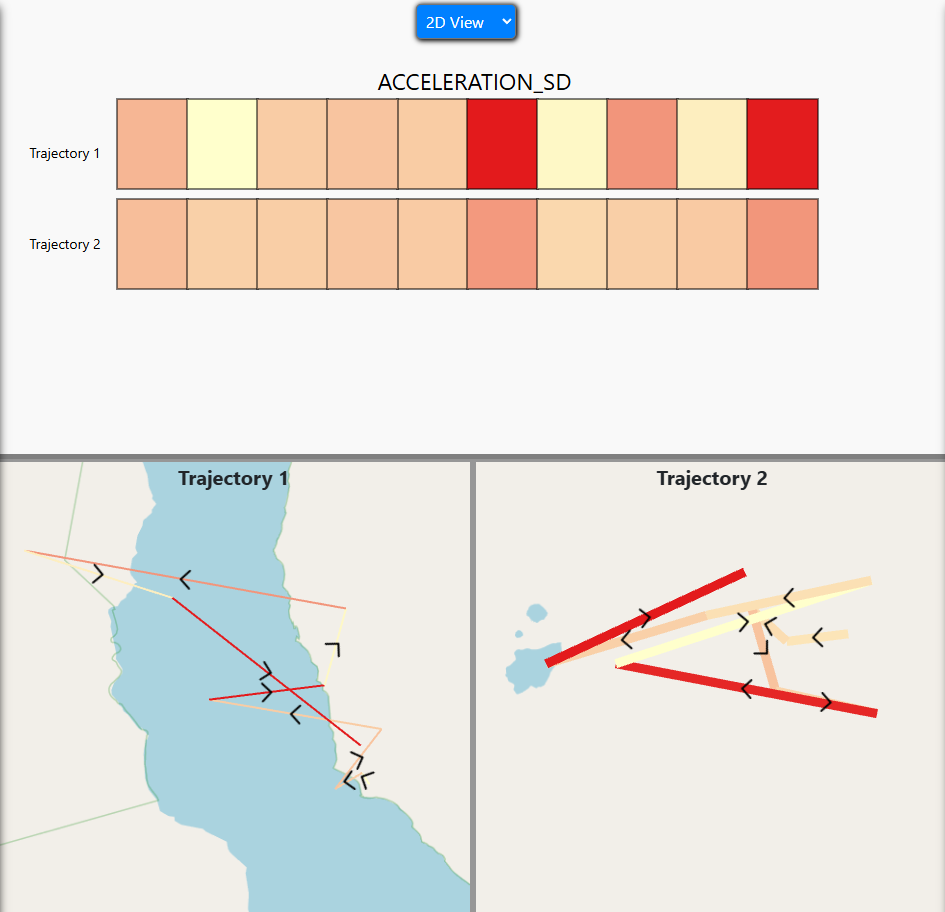}
            \caption{Acceleration standard deviation.}
            \label{fig:foxes_curv_acc_sd_352_vs_382_2D}
        \end{minipage}
        \hfill
    \end{figure}

    Next, several comparisons are made to analyze the trajectories and their behaviors, contrasting points selected from the curvature zone against those in the acceleration zone. First, the two trajectories selected before, namely fox ID 352 (curvature) and fox ID 382 (acceleration), are analyzed. The three figures below show different statistical variables for these two foxes: Fig. \ref{fig:foxes_curv_acc_mean_352_vs_382_2D} shows \textit{acceleration mean}, Fig. \ref{fig:foxes_curv_acc_quant95_352_vs_382_2D} shows \textit{acceleration quantile 95}, and the third figure, Fig. \ref{fig:foxes_curv_acc_sd_352_vs_382_2D}, represents \textit{acceleration standard deviation}. In the figures mentioned, Fox 352 appears on top, labeled as \textit{Trajectory 1}, and on the left side of the maps, while Fox 382 is shown at the bottom, with the name \textit{Trajectory 2}, presented on the right side of the maps.

    \noindent From comparing these three statistical variables samples over the same foxes, it can be noted that \textit{Trajectory 2}, fox 382, presents a more constant acceleration, that is, there is less presence of acceleration peaks or sudden decelerations, in comparison to fox 352, which was selected from the \textit{high curvature} zone. \textit{Trajectory 1}, presenting mainly curvature behavior, shows several peaks on acceleration, deceleration, and more significantly than those present in the \textit{acceleration trajectory sample} (i.e., the \textbf{high acceleration} peaks show higher results, represented in \textbf{deep red}, and stronger \textbf{decelerations, shown in light-yellow color}, which represent negative acceleration values). The figure \ref{fig:352vs382-overall} below shows a side-by-side view of these trajectories, showing the 3 statistical variables for the two foxes.

    \begin{figure}[H]
        \centering
        \includegraphics[width=0.9\linewidth]{  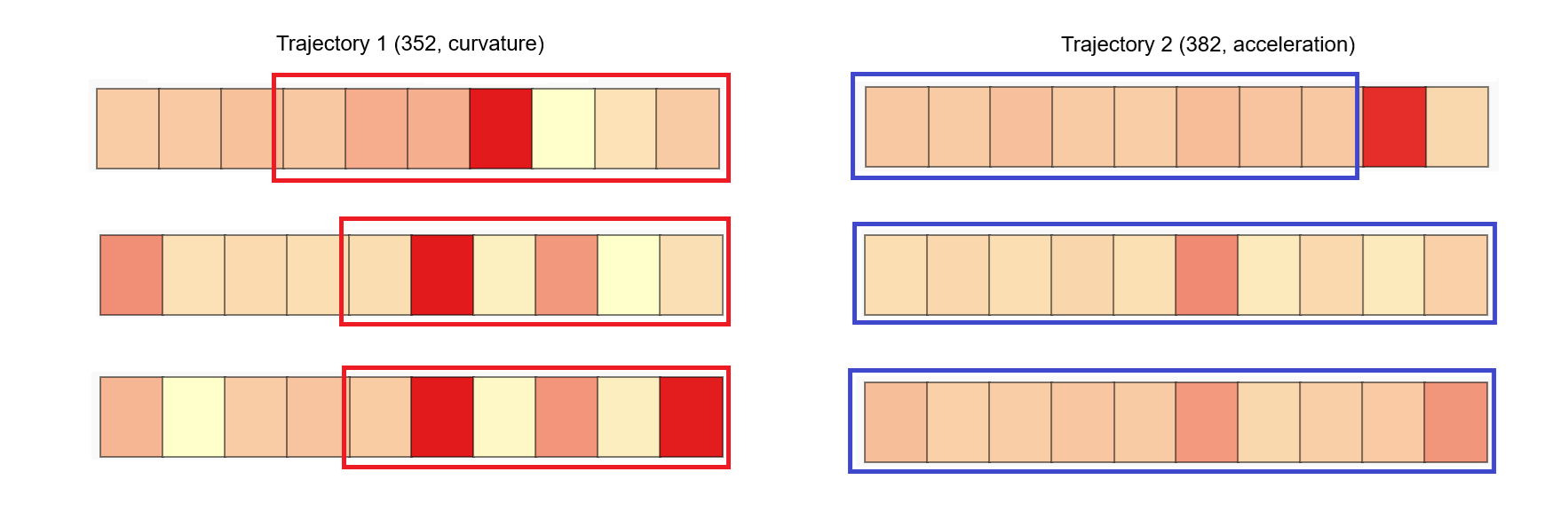}
        \caption{Trajectory 1 (ID 352, curvature zone) vs trajectory 2 (ID 382, acceleration zone), \textbf{acceleration mean, quant.95, and stand. deviation} (from top to bottom).}
        \label{fig:352vs382-overall}
    \end{figure}

    \noindent The \textbf{non-constant} behavior was derived from a fox in zone 1, which denotes \textbf{high curvature} (outlier score 1.00) and very low acceleration (score 0.0566). The \textbf{constant} behavior shows to be particular to the fox that was selected from zone 2, which represents mainly \textbf{acceleration behavior} (score 1.00) and almost no curvature (score 0.1886). This is a significant finding; however, to further determine if the behavior is particular to the single fox under review or not, further analysis is needed by comparing additional foxes in the same areas. \\

\begin{figure}[H]
        \centering
        \begin{minipage}[b]{0.45\textwidth}
            \centering
            \includegraphics[width=\linewidth]{  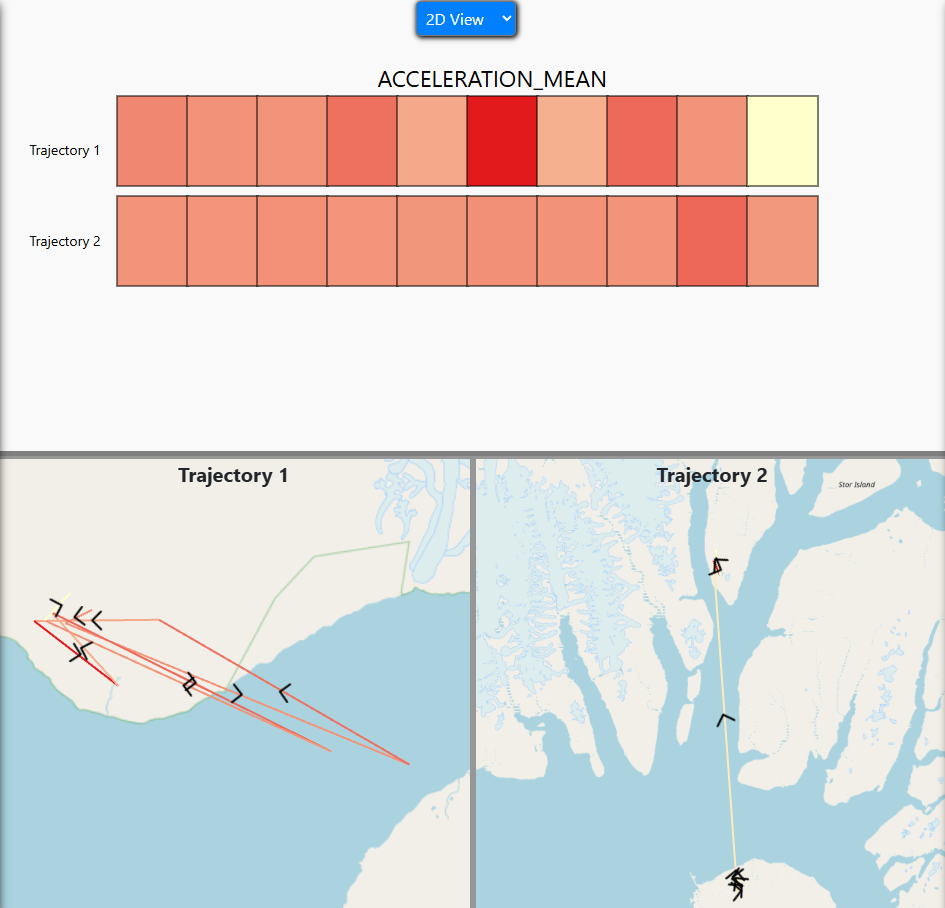}
            \caption{Mean acceleration.}
            \label{fig:foxes_curv_acc_mean_340_vs_382_2D}
        \end{minipage}
        \hfill
        \begin{minipage}[b]{0.45\textwidth}
            \centering
            \includegraphics[width=\linewidth]{  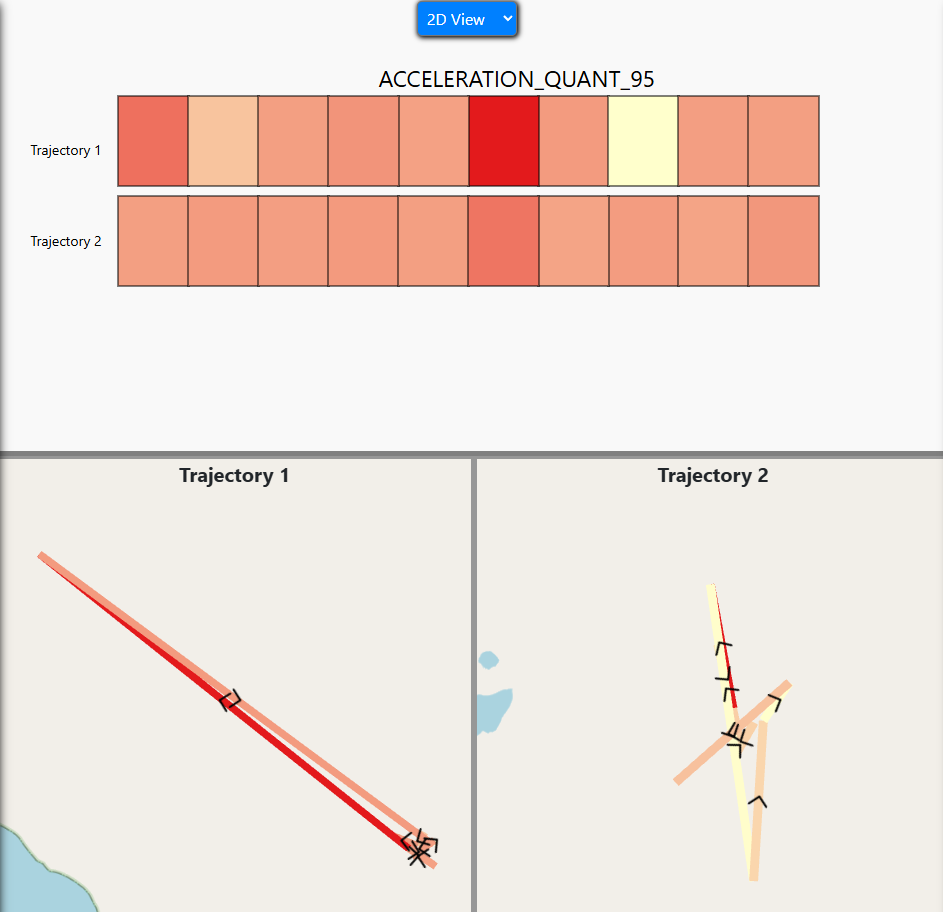}
             \caption{Acceleration quantile 95.}
            \label{fig:foxes_curv_acc_quant95_340_vs_382_2D}
        \end{minipage}
        \hfill
        \begin{minipage}[b]{0.45\textwidth}
            \centering
            \includegraphics[width=\linewidth]{  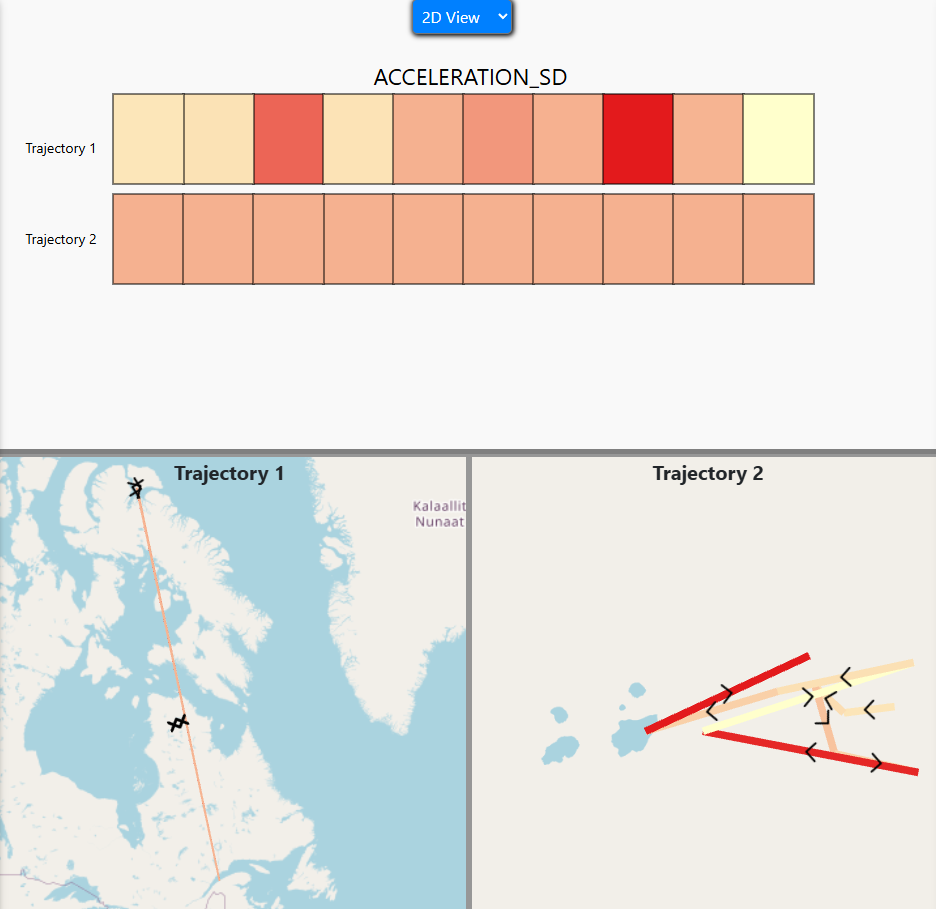}
            \caption{Acceleration standard deviation.}
            \label{fig:foxes_curv_acc_sd_340_vs_382_2D}
        \end{minipage}
        \hfill
    \end{figure}

    \noindent To further analyze the recent findings discussed, the selection of a new fox from zone 1 is done. Namely, fox ID 340 is selected from zone 1 (i.e., the \textit{curvature} zone), while fox ID 382 is \textbf{re-selected} from zone 2 (i.e., the \textit{acceleration} zone). For consistency, the same fox as in the previous acceleration case was selected, and the same statistical variables are selected and visually reviewed: First mean acceleration Fig. \ref{fig:foxes_curv_acc_mean_340_vs_382_2D}, then acceleration quantile 95 Fig. \ref{fig:foxes_curv_acc_quant95_340_vs_382_2D}, and acceleration standard deviation \ref{fig:foxes_curv_acc_sd_340_vs_382_2D}. It must be reminded that the colors shown in the 2D heatmaps are relative to the 20-point samples from both trajectories (10 points from each trajectory). Hence, the same trajectory being analyzed (i.e., 382) may have different color-tones as in previous comparisons, as the color grading is relative to the points at that current comparison.
    
    \noindent From reviewing the newly selected fox from zone 1, the same behavior is found, as shown in Fig. \ref{fig:340vs382-overall} below. The new trajectory (trajectory 1, ID 340) from zone 1, representing \textit{curvature}, shows significant acceleration and deceleration peaks, which are not present in the acceleration samples (i.e., the fox selected from the Acceleration group).

    \begin{figure}[H]
        \centering
        \includegraphics[width=0.9\linewidth]{  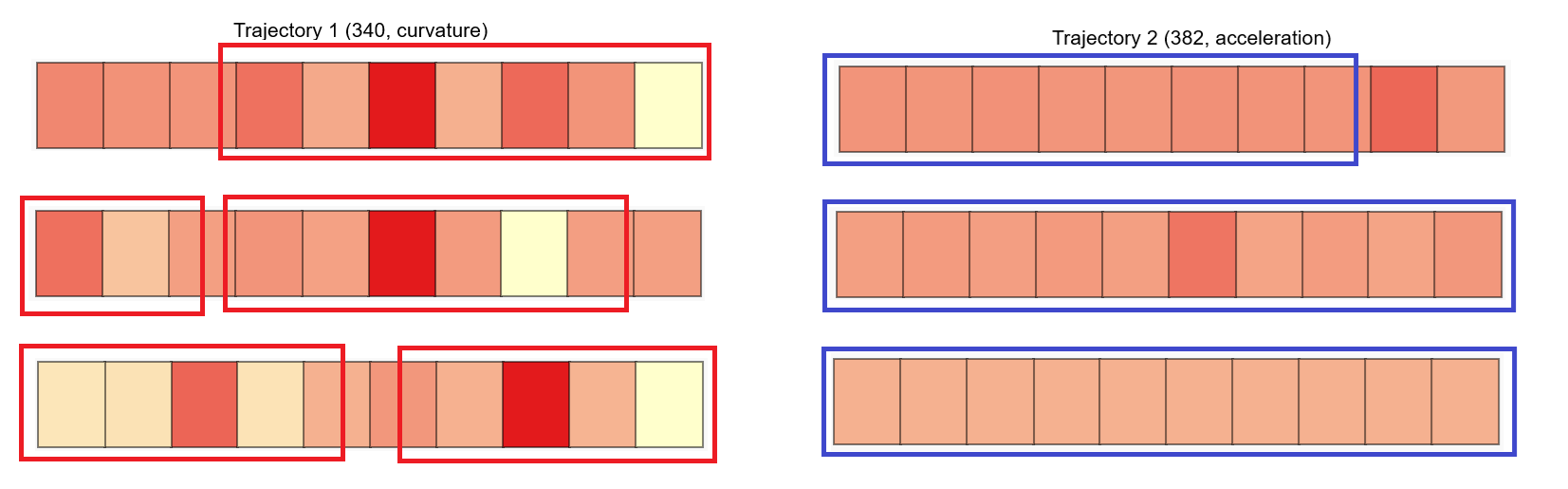}
        \caption{Trajectory 1 (ID 340, curvature zone) vs trajectory 2 (ID 382, acceleration zone), \textbf{acceleration mean, quant. 95, and stand. deviation} (from top to bottom).}
        \label{fig:340vs382-overall}
    \end{figure}

\begin{figure}[H]
        \centering
        \begin{minipage}[b]{0.45\textwidth}
            \centering
            \includegraphics[width=\linewidth]{  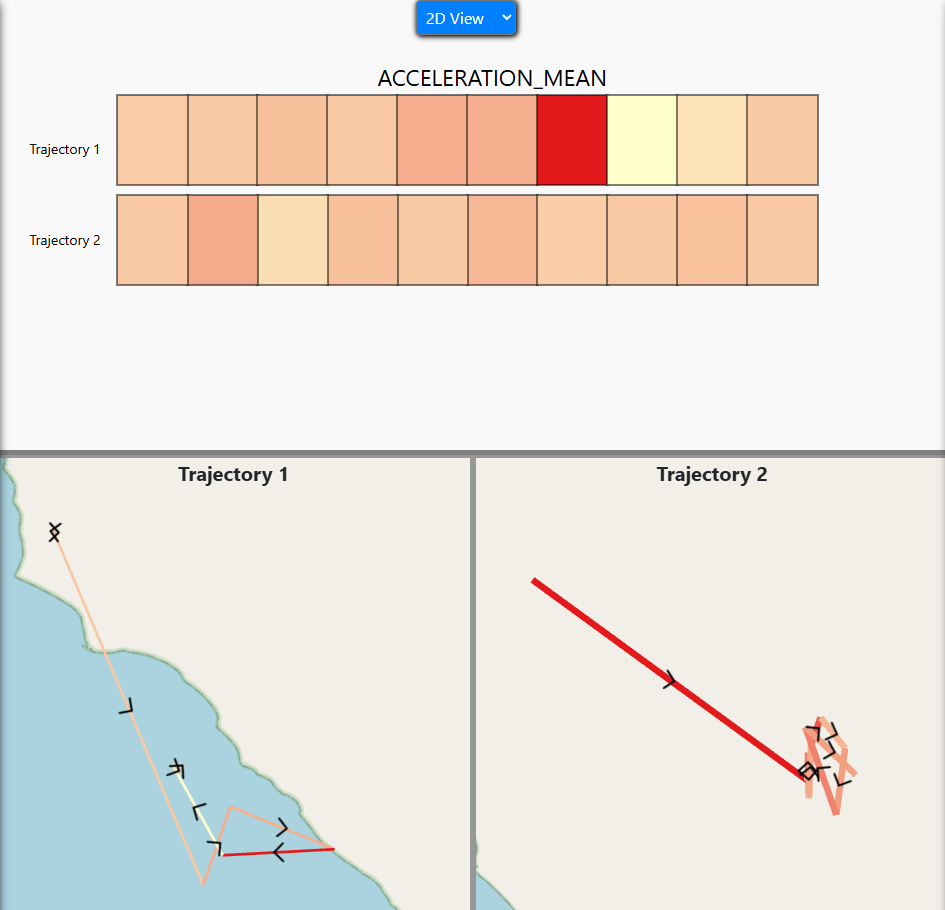}
            \caption{Mean acceleration.}
            \label{fig:foxes_curv_acc_mean_352_vs_428_2D}
        \end{minipage}
        \hfill
        \begin{minipage}[b]{0.45\textwidth}
            \centering
            \includegraphics[width=\linewidth]{  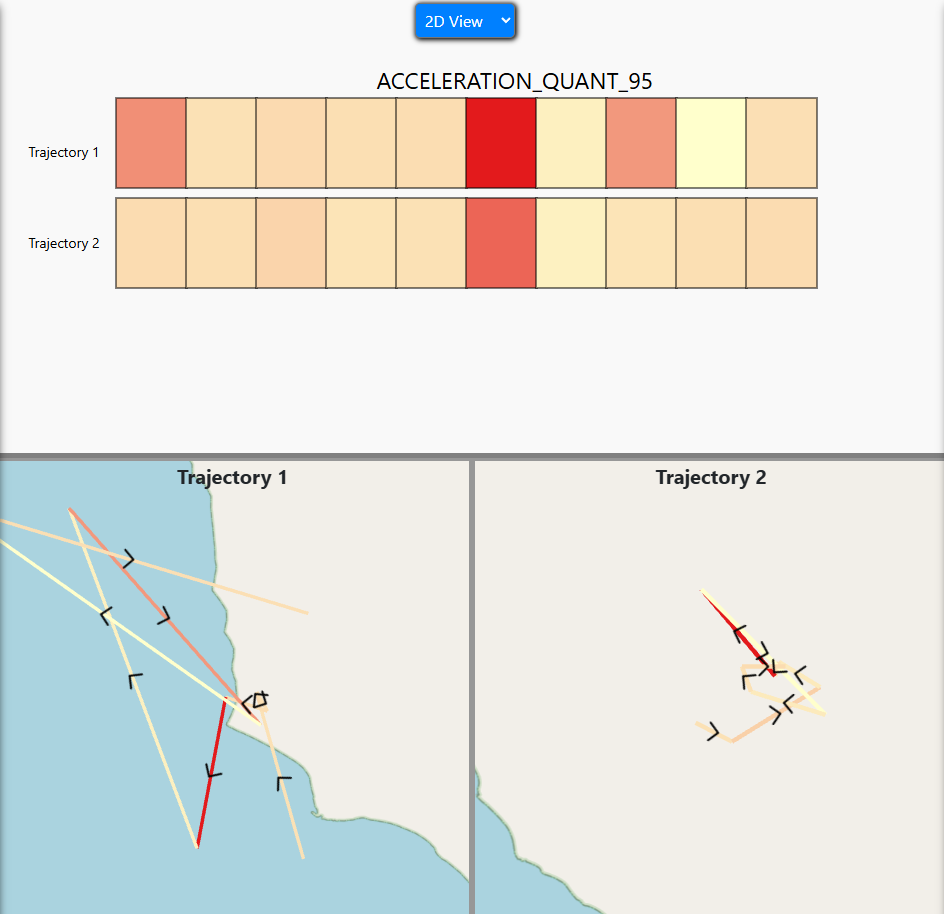}
            \caption{Acceleration quantile 95.}
        \label{fig:foxes_curv_acc_quant95_352_vs_428_2D}
        \end{minipage}
        \hfill
        \begin{minipage}[b]{0.45\textwidth}
            \centering
            \includegraphics[width=\linewidth]{  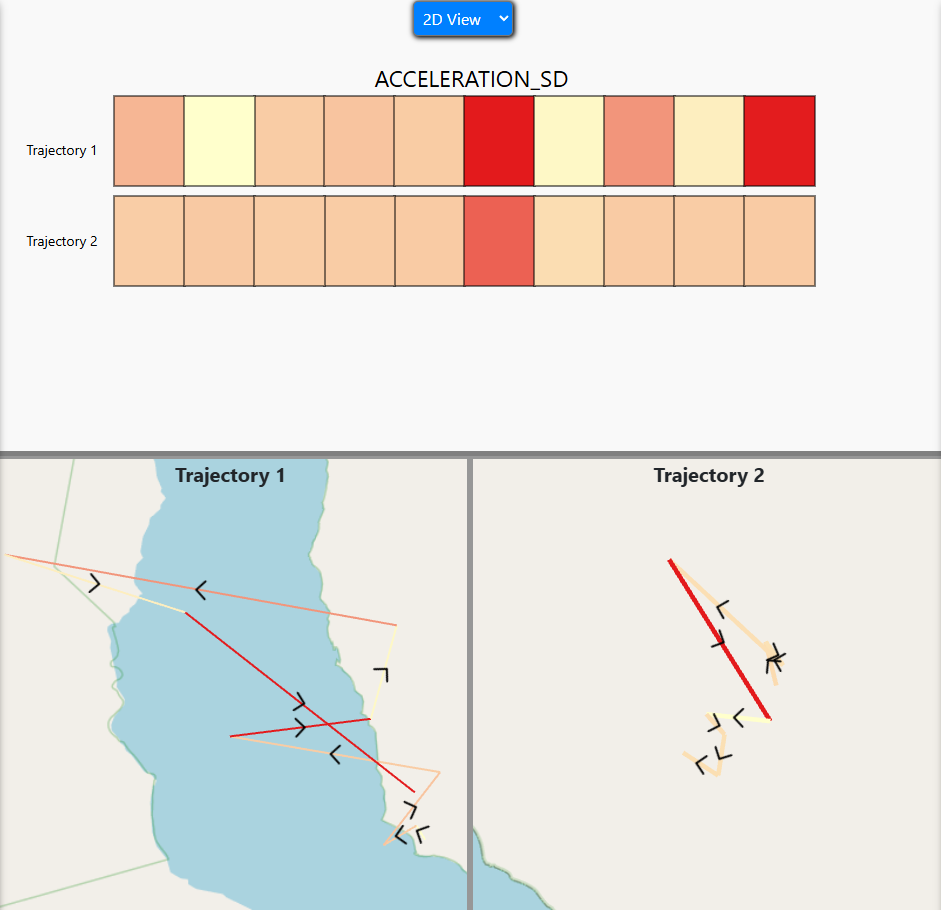}
            \caption{Acceleration Standard Deviation.}
            \label{fig:foxes_curv_acc_sd_352_vs_428_2D}
        \end{minipage}
        \hfill
    \end{figure}

    \noindent To complete the review of such constant against non-constant acceleration behaviors seen thus far, an additional comparison is done between the same fox as in the first comparison, namely fox 352 (selected from zone 1, representing \textit{curvature}) and a new fox, ID 428, from zone 2 (\textit{acceleration} behavior) are presented. The figures are presented in the same order as before; Fig. \ref{fig:foxes_curv_acc_mean_352_vs_428_2D} shows the mean acceleration, acceleration quantile 95 is shown in Fig. \ref{fig:foxes_curv_acc_quant95_352_vs_428_2D}, and acceleration standard deviation in Fig. \ref{fig:foxes_curv_acc_sd_352_vs_428_2D}.
    
    \noindent The comparison below in Fig. \ref{fig:352vs428-overall} shows the final acceleration comparison between Fox 352 and Fox 428.

   \begin{figure}[H]
        \centering
        \includegraphics[width=1.1\linewidth]{  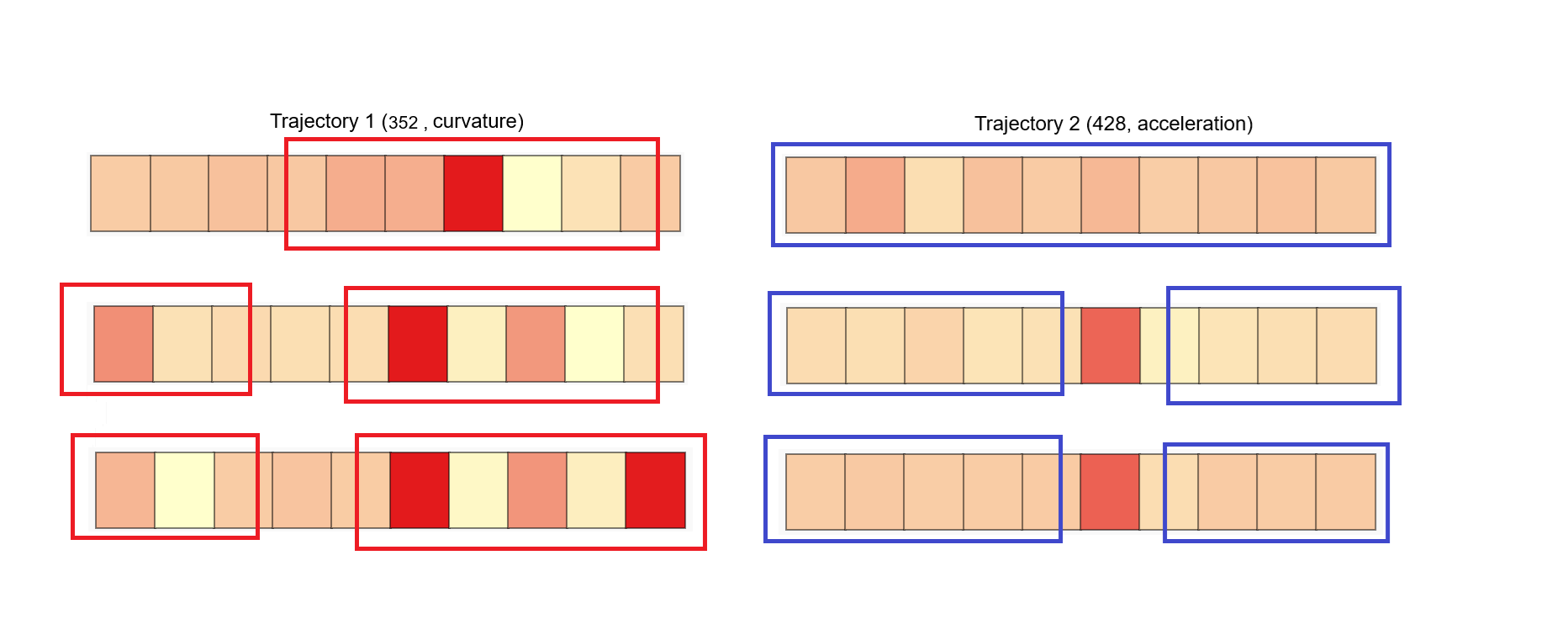}
        \caption{Trajectory 1 (ID 352, curvature zone) vs trajectory 2 (ID 428, acceleration zone), \textbf{acceleration mean, quant. 95, and stand. deviation} (from top to bottom).}
        \label{fig:352vs428-overall}
    \end{figure}

    \subsubsection{Case Study 1 Final Results}
     \noindent During the analysis of the Arctic foxes dataset, a number of steps were taken in order to expand and enable the understanding and interpretation of the data, through a taxonomical view and labeling analysis.

     There were two main analytical steps taken. First, an experiment was made comparing \textbf{Kinematic} against \textbf{Geometric} zones, as well as individual foxes on each of the zones. This initial iteration compared and presented several visual and analytical insights over individual foxes, groups of foxes within the compared zones, and a taxonomy overview towards statistical variables with linkage to 2D and 3D map visualizations. The results then presented several significant variables related to speed and acceleration, which were found to be highly significant to the categorization of foxes as \textbf{Kinematic Behavior}. Such statistical variables were \textbf{Speed Kurtosis, Acceleration Standard Deviation, Speed Skewness, Acceleration Quantile 95, and Acceleration Mean}. Other statistical variables were presented as well for the \textbf{Geometric behavior} analysis, however, further attention was given to the Kinematic taxonomy group, with the aim to internalize possible findings over \textbf{Acceleration}, a subdivision of the taxonomy utilized in this tool. The Kinematic and Geometric zones in the boundary distribution showed to account for the majority of the points with significant behavior, with a 35\% (zone 0 accounted for the majority of the trajectories' outlier scores with 53\%, however, without significance on either Geometric or Kinematic behavior).
     
     The decision for Acceleration to be further reviewed was derived from a number of reasons. First, the weight Acceleration showed on the \textbf{feature importance} results, through \textbf{38 different statistical variables} related to Kinematic behavior, with \textbf{Acceleration Standard Deviation, Acceleration Quantile 95, and Acceleration Mean} accounting for \textbf{more than 15\%} of all trajectories within the zone to be labeled as \textit{Kinematic}. In addition to feature importance results, the variability on the acceleration samples (having constant values for one group, but peaks and sudden deceleration changes in a second group), derived from the \textbf{visual inspection of the 2D and 3D maps}, increased the interest and tilted the following analytical iteration towards \textbf{Acceleration}, in contrast to Curvature.  

     The second iteration enabled a deeper overview of the \textbf{Acceleration} subgroup, compared against \textbf{Curvature}. As the main focus was centered on Acceleration, the top statistical variables derived from the feature importance algorithms were \textbf{Acceleration Mean, Acceleration Quantile 95, Acceleration Standard Deviation, Acceleration Quantile 75, and Acceleration Quantile 90}. Individual comparisons took place, where \textbf{high acceleration peaks} and \textbf{sudden deceleration} in the fox selected from the \textbf{Curvature} zone were present, in addition to non-consistent acceleration. In other words, the fox selected from the \textbf{Curvature} group showed changes in its acceleration rather often, even in the short 10-sample views presented in 2D. In contrast, the fox selected from the \textbf{Acceleration} zone showed to be constant, without sudden changes in acceleration throughout time. To further review this behavior, several more comparisons were performed with the aim of knowing if such a pattern was present from the calculation results of different statistical variables and foxes. A total of 9 additional comparisons were performed, which analyzed different foxes through the top 3 feature importance variables, namely \textbf{Acceleration Mean, Acceleration Quantile 95}, and \textbf{Acceleration Standard Deviation}, to understand if this was a behavioral pattern in their acceleration. The final findings showed indeed that those foxes within the \textbf{acceleration} group, in relation to the top 3 statistical variables selected, were often \textbf{steady and consistent in their acceleration over time} while those previously labeled \textbf{curvature-based} had \textbf{non-constant acceleration} and \textbf{sudden changes} over time.

     Overall, the case study showed that 11 Arctic foxes were categorized within zone 2, which meant they had significant Kinematic behavior, due to the outlier score results. Further analysis showed that Acceleration was one of the key indicators for the results of such scores, accounting for a highly significant part of the feature importance output. A deeper analysis of acceleration showed that Acceleration Mean, Acceleration Quantile 95, and Acceleration Standard Deviation were the most significant statistical variables, which contributed to the categorization of foxes having an acceleration-based behavior. These same statistical variables were found in the first iteration, when reviewing the top 5 variables that determined Kinematic-based behavior. Finally, from individual comparison of these statistical variables throughout different foxes, side-by-side comparisons and manual visual analysis of the trajectories over time, it was found that those foxes who were labeled as having acceleration-based behavior had a more constant and steady behavior for overall acceleration than those that were not within the acceleration group, which showed sudden and often changes in their acceleration instead. It can be conveyed that the sudden changes in acceleration may be related to a combination of angles and distance geometry changes in relation to acceleration-based behavior, however, that may require additional or separate analytical iterations focusing instead on the Geometric behavior's impact on acceleration over time.


\subsection{Case Study 2: Tropical Cyclones} \label{Case Study 2: Tropical Cyclones}

In the second case study, the tropical cyclones dataset was selected. The last 20 years of data were used, from 2004 to 2024, containing trajectories of 2119 tropical cyclones. 
When the dataset selection is done from the analytics tool, the workflow follows the same steps as in the first case study, starting by loading the data, producing the heatmap first, and showing the frequencies of the trajectories for each zone.

\begin{figure}[H]
    \centering
        \begin{minipage}[b]{0.47\textwidth}
        \centering
        \includegraphics[width=\linewidth]{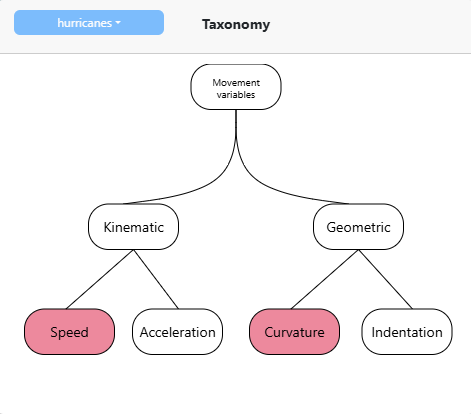}
        \caption{Taxonomy selection of Speed and Curvature group combination.}
        \label{fig:hurricane_tree_sp_curv}
    \end{minipage}
    \hfill
    \begin{minipage}[b]{0.5\textwidth}
        \centering
        \includegraphics[width=\linewidth]{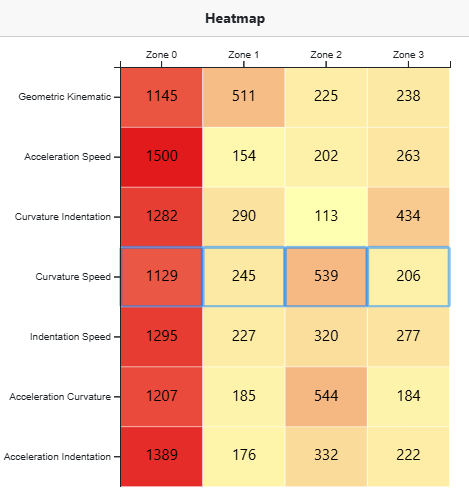}
        \caption{Heatmap showing each combination against the zones, with their respective trajectory frequencies.}
        \label{fig:hurricane_heatmap_Speed_curvature}
    \end{minipage}
    \hfill
\end{figure}

During this experiment, the taxonomy selection was different from that in the first case study, as the analysis began by studying subgroups from the first step, focused on the trajectory's speed and curvature as shown in Figure \ref{fig:hurricane_tree_sp_curv}. The frequency heatmap generated after loading the dataset is shown in Fig. \ref{fig:hurricane_heatmap_Speed_curvature}.


\noindent The highlighted row in the frequency heatmap presents the distribution of trajectories over the zones, where zone 0 had the highest number of trajectories among all, with 1129 trajectories, representing that 53.28\% of the tropical cyclone trajectories show no significant speed or curvature-based behavior.
In zone 1, 245 trajectories were distributed in this zone, with only 11.56\% of the cyclones exhibiting interesting behavior in regards to curvature.
Zone 2 had the second-highest frequency, with 539 trajectories, presenting speed-based behavior for 25.44\% of the cyclones.
Lastly, in zone 3, 206 trajectories were distributed, and 9.72\% of the cyclones showed a hybrid behavior combining speed and curvature.  

\begin{figure}[H]
    \centering
    \begin{minipage}[b]{0.49\textwidth}
        \centering
        \includegraphics[width=\linewidth]{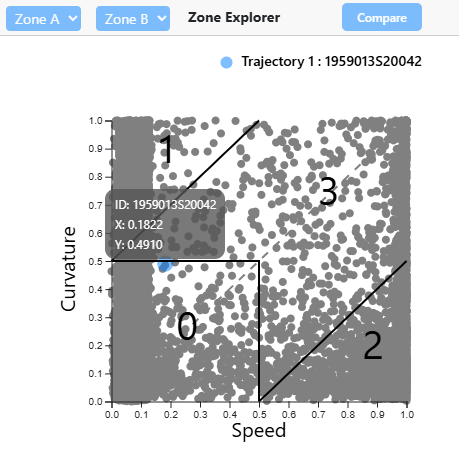}
        \caption{Selecting the first trajectory with ID 1959013S20042.}
        \label{hurricane_1959013S20042_sp_curv}
    \end{minipage}
    \hfill
    \begin{minipage}[b]{0.49\textwidth}
        \centering
        \includegraphics[width=\linewidth]{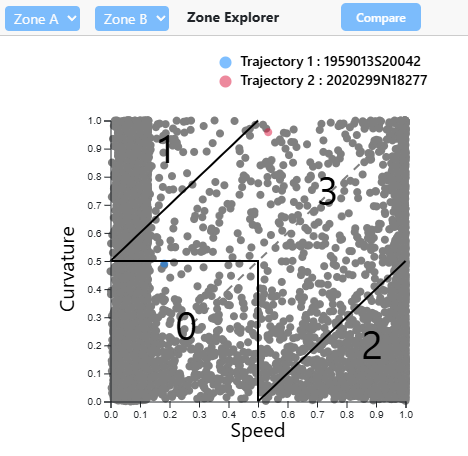}
        \caption{Selecting the second trajectory with ID 2020299N18277.}
        \label{hurricane_2020299N18277_sp_curv}
    \end{minipage}
    \hfill
\end{figure}

The frequencies presented in the heatmap are then reflected on the scatter plot, with a visual representation of the points \textit{spread out}, as per the score results from the outlier detection algorithm. At this stage, an individual comparison is made between trajectory 1 with ID 1959013S20042 (from zone 0, no significant behavior) and trajectory 2 with ID 2020299N18277 (from zone 3, hybrid behavior) as shown in Fig. \ref{hurricane_1959013S20042_sp_curv}, and Fig. \ref{hurricane_2020299N18277_sp_curv}, respectively.

\begin{figure}[H]
    \centering
    \includegraphics[width=0.7\linewidth]{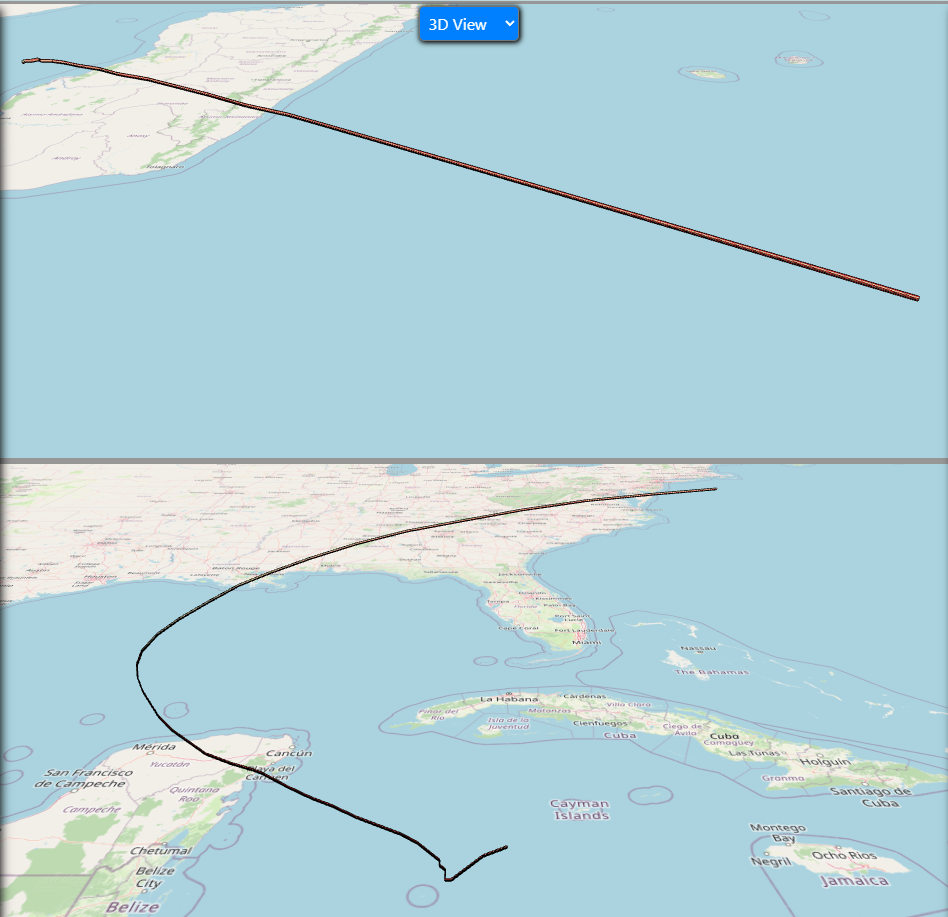}
    \caption{Showing the entire trajectory for hurricane with ID 1959013S20042 (top) vs hurricane with ID 2020299N18277 (bottom).}
    \label{hurricane_3D_eniter_trajectories}
\end{figure}

\begin{figure}[H]
    \centering
    \begin{minipage}[b]{0.49\textwidth}
        \centering
        \includegraphics[width=\linewidth]{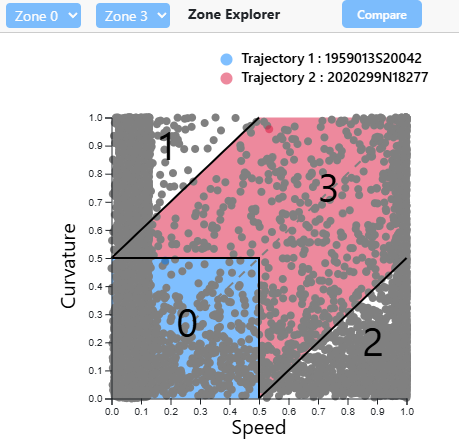}
        \caption{Comparing Zone 0\\ with Zone 3.}
        \label{hurricanes_zone_0_vs_zone_3_sp_curv}
    \end{minipage}
    \hfill
    \begin{minipage}[b]{0.49\textwidth}
        \centering
        \includegraphics[width=\linewidth]{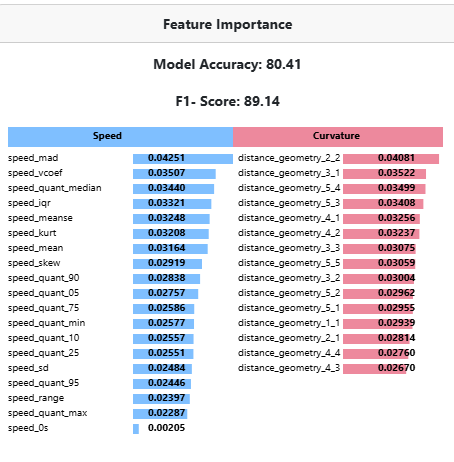}
        \caption{The generated feature importance by comparing Zone 0 with Zone 3.}
        \label{hurricanes__sp_curv_feature_importance}
    \end{minipage}
    \hfill
\end{figure}

\noindent Comparing the entire trajectory is not necessarily meaningful, as it is harder to interpret, as shown in Fig. \ref{hurricane_3D_eniter_trajectories}. Thus, it is essential to follow the same principles as in the first case study, comparing the respective zones from which each of the trajectories was selected, namely, zone 0 and zone 3, effectively running the random forest model to generate the feature importance table, as shown in Fig. \ref{hurricanes_zone_0_vs_zone_3_sp_curv}, and Fig. \ref{hurricanes__sp_curv_feature_importance}.

\noindent By examining the feature importance results, the top 5 Speed-related variables were: 

\begin{itemize}
    \item Median Absolute Deviation (speed\_mad)
    \item Variance Coefficient (speed\_vcoef)
    \item Quantile Median (speed\_quant\_median)
    \item Interquartile Range (speed\_iqr)
    \item Standard Error of the Mean (speed\_meanse)
\end{itemize}
As per curvature, the top 5 statistical variables were:
\begin{itemize}
    \item Distance Geometry Signature 2\_2 (distance\_geometry\_2\_2) 
    \item Distance Geometry Signature 3\_1 (distance\_geometry\_3\_1)
    \item Distance Geometry Signature 5\_4 (distance\_geometry\_5\_4)
    \item Distance Geometry Signature 5\_3 (distance\_geometry\_5\_3)
    \item Distance Geometry Signature 4\_1 (distance\_geometry\_4\_1)
\end{itemize}

\begin{figure}[H]
    \centering
    \begin{minipage}[b]{0.48\textwidth}
        \centering
        \includegraphics[width=\linewidth]{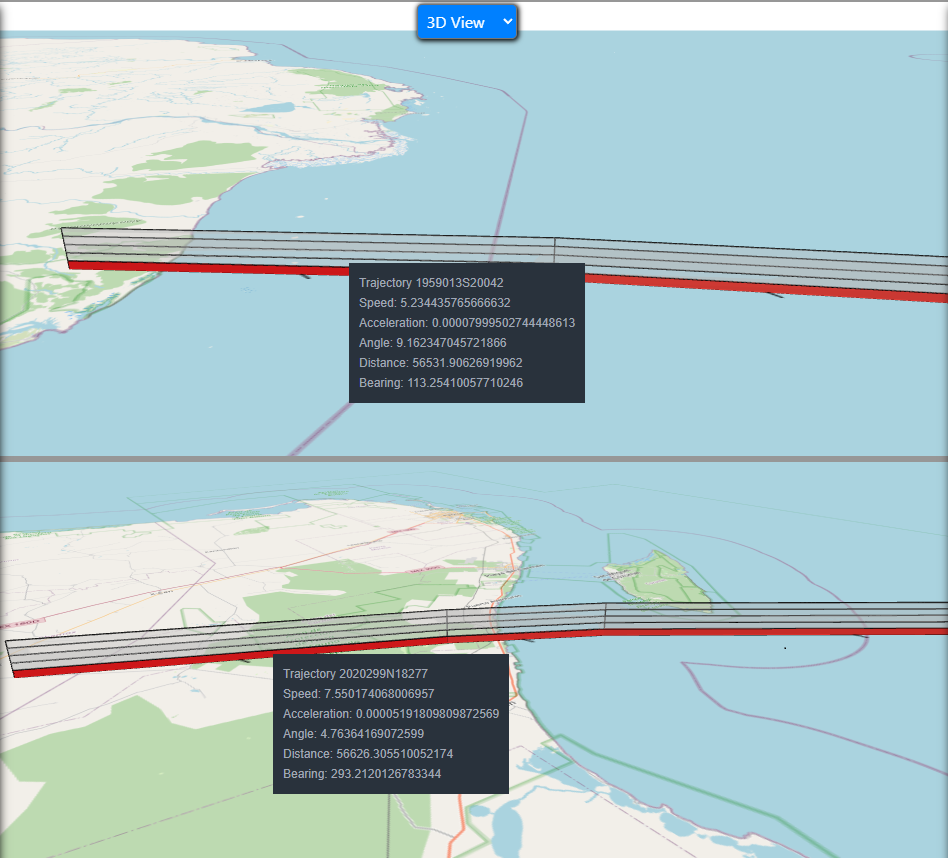}
        \caption{Showing the trajectory for the hurricanes with ID 1959013S20042 (top) vs the hurricane with ID 2020299N18277 (bottom) after selecting \textit{speed\_mad} statistical variable.}
        \label{hurricane_3D_trajectories_speed_mad}
    \end{minipage}
    \hfill
    \begin{minipage}[b]{0.49\textwidth}
        \centering
        \includegraphics[width=\linewidth]{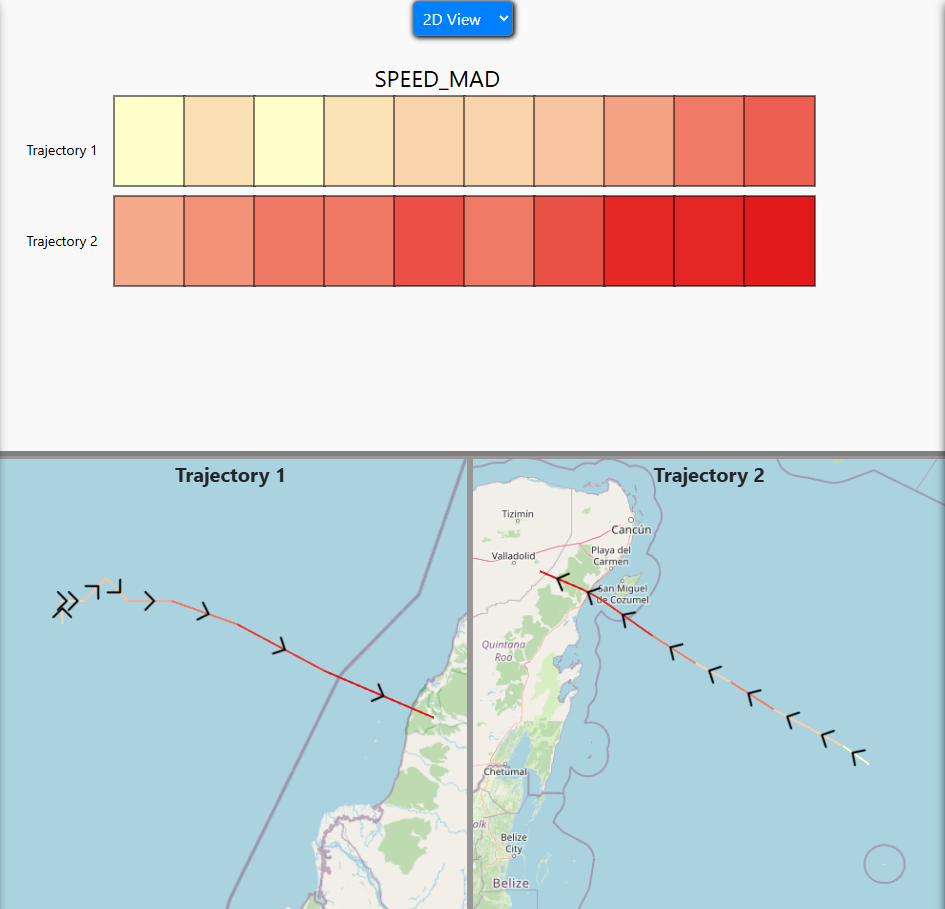}
        \caption{Trajectory of cyclone ID 1959013S20042 (left) vs cyclone ID 2020299N18277 (right), with speed heatmap over time in 2D.}
        \label{hurricanes_2D__sp_curv_speed_mad}
    \end{minipage}
    \hfill
\end{figure}

\noindent By selecting the \textit{speed median absolute deviation} variable, the trajectory gets visually represented on a map, shown in Fig. \ref{hurricane_3D_trajectories_speed_mad}, and Fig. \ref{hurricanes_2D__sp_curv_speed_mad}.

\noindent By comparing the heatmap representing Speed changes, from the \textit{speed median absolute deviation} selection, shown in Fig. \ref{hurricanes_2D__sp_curv_speed_mad}, it can be noted that trajectory 2 incrementally had higher speed than trajectory 1, also shown in the 3D representation in Fig \ref{hurricane_3D_trajectories_speed_mad}, as the fastest speed for trajectory 1 was 5.23\textit{ m/s}, while trajectory 2 reached a speed of 7.55 \textit{m/s}. Further visual inspection, however, showed that other Geometric behaviors possibly had an impact on the top speed achieved for the current sample. Fig. \ref{hurricanes_2D__sp_curv_speed_mad} also shows that trajectory 1 (i.e., the trajectory selected from zone 0, which represents \textbf{no} speed-curvature based behavior), has significant angle changes at the beginning of the sample, possibly limiting the trajectory's speed, while trajectory 2 (i.e., selected from zone 3, which represents hybrid behavior between speed and curvature) showed a significantly higher straightness during its trajectory, allowing for higher speed increments. This further promotes that the outlier score distinction for \textit{hybrid} behavior was correctly representational for speed and curvature together, as speed is significant in terms of the highest speed achieved, while curvature and the straightness of the trajectory had a relative impact on it.

\begin{figure}[H]
    \centering
    \includegraphics[width=0.9\linewidth]{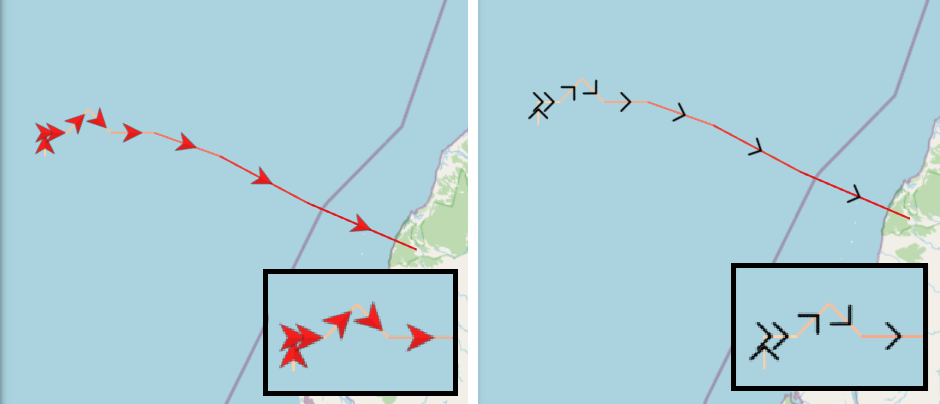}
    \caption{Trajectory 1, ID 1959013S20042, several angle changes.}
    \label{fig:iteration1-several-angle-changes}
\end{figure}

The findings thus far presented from the initial iteration proved to be rather easier to discern, which may be conceptualized from a simpler analysis than the one potentially achievable with the proposed tool. Nevertheless, it can be noted that there is a potential impact and significant relationship between indentation and curvature, as curvature does not necessarily represent trajectory 1 from zone 0 (as per the outlier score and decision boundary distribution, which was below 0.5), while the indentation effect on the trajectory over time seems to potentially affect curvature, and therefore Geometric behavior, in a significant manner, as visually represented in Fig. \ref{fig:iteration1-several-angle-changes}. 

\noindent This led to the following analytical iteration, which considers the comparison and selection of \textbf{curvature and indentation}, with the main intention to understand their relationship and the outlier detection score results in regards to both statistical variables and visual analysis, from diverse view levels.

\begin{figure}[H]
    \centering
        \begin{minipage}[b]{0.47\textwidth}
        \centering
        \includegraphics[width=\linewidth]{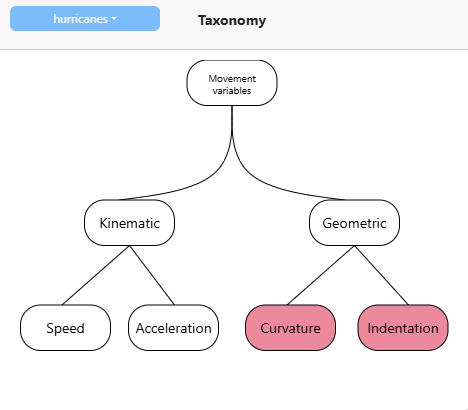}
        \caption{Taxonomy selection: Curvature and Indentation.}
        \label{fig:hurricanes-combination-selection}
    \end{minipage}
    \hfill
    \begin{minipage}[b]{0.47\textwidth}
        \centering
        \includegraphics[width=\linewidth]{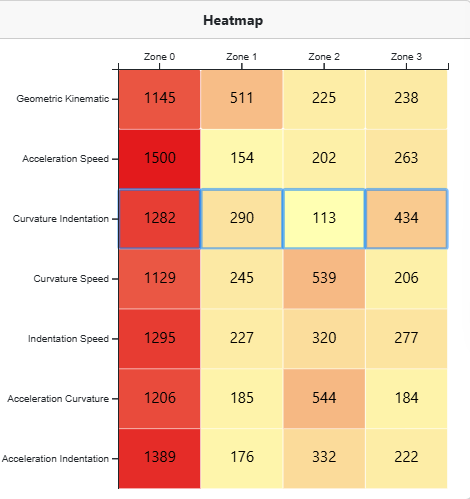}
        \caption{Heatmap showing each combination against the zones, with their respective trajectory frequencies.}
        \label{fig:hurricanes-heatmap-row-select}
    \end{minipage}
    \hfill
\end{figure}

As discussed in the first iteration, the experiment moves towards the analysis for \textbf{Curvature} and \textbf{Indentation}. The corresponding selection is done in the taxonomy tree, as shown in Fig. \ref{fig:hurricanes-combination-selection}, and the heatmap row for such selection is highlighted, as shown in Fig. \ref{fig:hurricanes-heatmap-row-select}.

\noindent The frequency heatmap presents the outlier score distribution through the 4 zones as per the 2119 tropical cyclones. In zone 0, most of the trajectories are found, with 1282 trajectories, accounting for 60.50\% of the trajectories under analysis. Zone 1 accounts for 13.69\% with 290 trajectories, while zone 2 is constituted of 113 trajectories, with a 5.33\%. Finally, zone 3 has 434 trajectories, accounting for 20.48\%.

\begin{figure}[H]
    \centering
        \begin{minipage}[b]{0.48\textwidth}
        \centering
        \includegraphics[width=\linewidth]{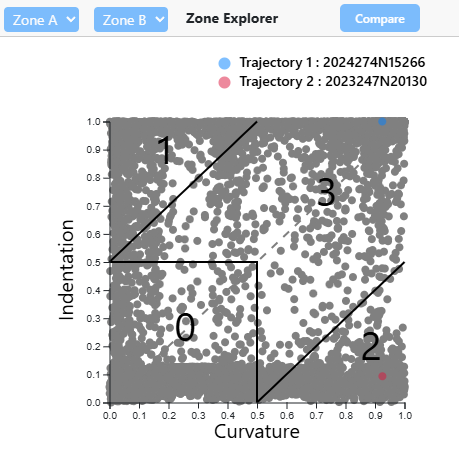}
        \caption{Trajectory selection: \textit{Trajectory 1} ID 2024274N15266 (blue) and \textit{Trajectory 2} ID 2023247N20130 (red).}
        \label{fig:hurricanes-trajectories-selected}
    \end{minipage}
    \hfill
    \begin{minipage}[b]{0.49\textwidth}
        \centering
        \includegraphics[width=\linewidth]{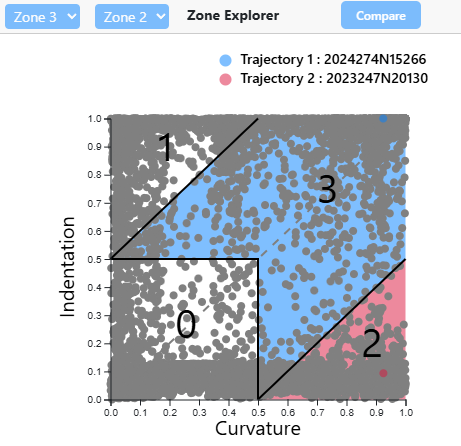}
        \caption{Zone selection: zone 2 (curvature-based behavior only) vs zone 3 (curvature and indentation, hybrid).}
        \label{fig:hurricanes-zone2-vs-zone3}
    \end{minipage}
    \hfill
\end{figure}

The boundary distribution is shown in Fig. \ref{fig:hurricanes-trajectories-selected} for the outlier scores. More specifically, it also represents the distribution of the zones in relation to the combination currently selected, outlined as follows: 

\begin{itemize}
    \item \textbf{Zone 0:}  No significant behavior from either Curvature and/or Indentation.
    \item \textbf{Zone 1:}  Significant Indentation behavior with no significant Curvature behavior.
    \item \textbf{Zone 2:}  Significant Curvature behavior with no significant Indentation behavior.
    \item \textbf{Zone 3:}  Hybrid behavior, Curvature and Indentation combined.
\end{itemize}


\noindent It must be noted that while discussing Curvature (as a measure), the main tool and indicator used is distance geometry (mathematical, using distances between points). These two terms may be used interchangeably during the following analytical steps.

As discussed thus far, the focus on this iteration is centered on the analysis of curvature and indentation. Experimenting and analyzing both individualistic and combined behaviors may further explain and motivate their distribution. During the previous iteration, curvature did not necessarily represent the movement for one of the trajectories selected, as it became apparent that the specific angles (i.e., indentation) and often angle changes had a significant effect on the overall geometry of the hurricane (which was apparent to have affected, possibly, overall speed as well). Therefore, the \textbf{first individual trajectory} selected will be from zone 3, which represents \textbf{hybrid behavior} from both curvature and indentation; namely, \textbf{hurricane ID 2024274N15266}. The overall \textbf{zone 3} will also be selected for the \textit{zone-vs-zone} comparison. As per the \textbf{second trajectory}, a hurricane from \textbf{zone 2} will be selected, namely, \textbf{hurricane ID 2023247N20130} to differentiate and understand their movement behavior, as well as \textbf{zone 2} for the zone comparison to produce the feature importance bar chart. The trajectories selected are presented in Fig. \ref{fig:hurricanes-trajectories-selected}, while the zones selected are shown in Fig. \ref{fig:hurricanes-zone2-vs-zone3}.

\begin{figure}[H]
    \centering
    \includegraphics[width=0.7\linewidth]{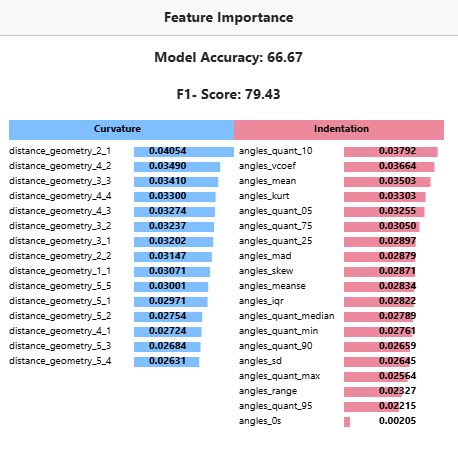}
    \caption{Feature importance for hurricanes dataset, comparing zone 3 against zone 2.}
    \label{fig:hurricane-feature-importance}
\end{figure}

\noindent From the random forest algorithm comparing all trajectory points in zone 3 against all points in zone 2, the feature importance bar chart shown in Fig. \ref{fig:hurricane-feature-importance} is produced, with a F1-score of 79.43\%.

\begin{figure}[H]
    \centering
    \includegraphics[width=0.75\linewidth]{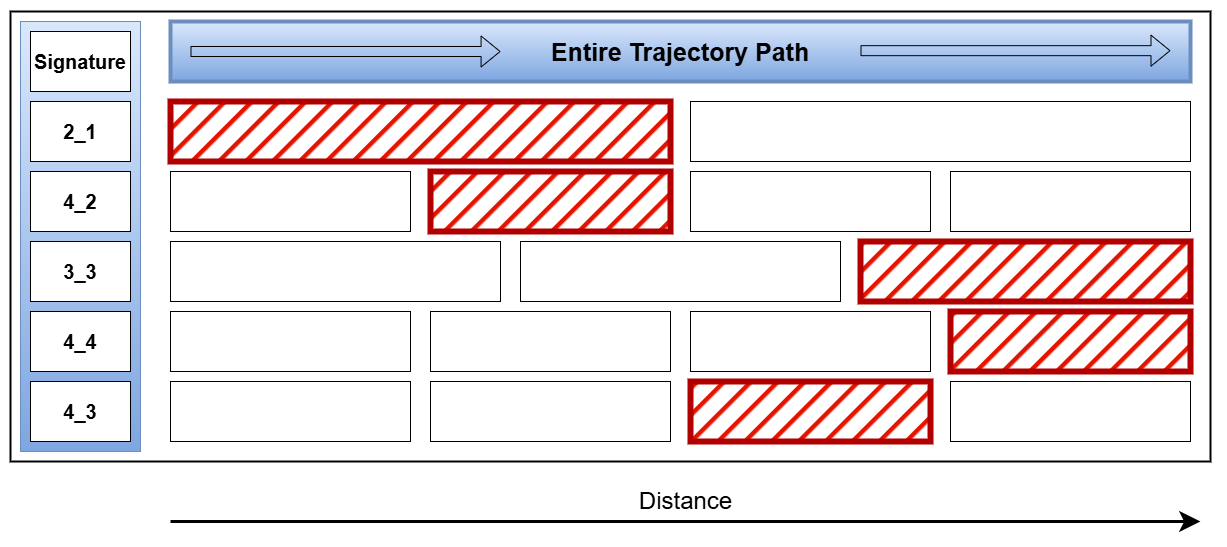}
    \caption{Top 5 distance geometry signatures, visual overview.}
    \label{fig:signatures}
\end{figure}

\noindent The feature importance bar chart presents a rather even distribution of the curvature statistical variables, as most of the distance geometry signatures have a significant weight on the curvature behavior categorization. The signatures 2\_1, 4\_2, 3\_3, 4\_4, and 4\_3 are the top 5 most significant variables, however, the 10th signature (i.e., 5\_5) has a difference of only \textbf{0.01053} with the top signature in the Curvature column (i.e., 2\_1). Fig. \ref{fig:signatures} shows the top 5 signatures from the feature importance bar chart. From such visualization, it becomes easier to discern the distribution of importance throughout all signatures; when combining all signatures, it can be noticed that the entire path is relevant to the categorization for curvature-based behavior. Therefore, additional visual analysis is needed in order to contextualize the representation of each signature.

\noindent To further investigate these trajectories, the two trajectories previously selected on the decision boundary scatter plot are analyzed. The two trajectories are referenced as follows:
\begin{itemize}
    \item Trajectory 1, ID 2024274N15266, Hybrid Behavior, Zone 3 (mainly referred to as "Trajectory 1")
    \item Trajectory 2, ID 2023247N20130, Curvature Behavior, zone 2 (mainly referred to as "Trajectory 2")
\end{itemize}

\begin{figure}[H]
    \centering
    \includegraphics[width=1\linewidth]{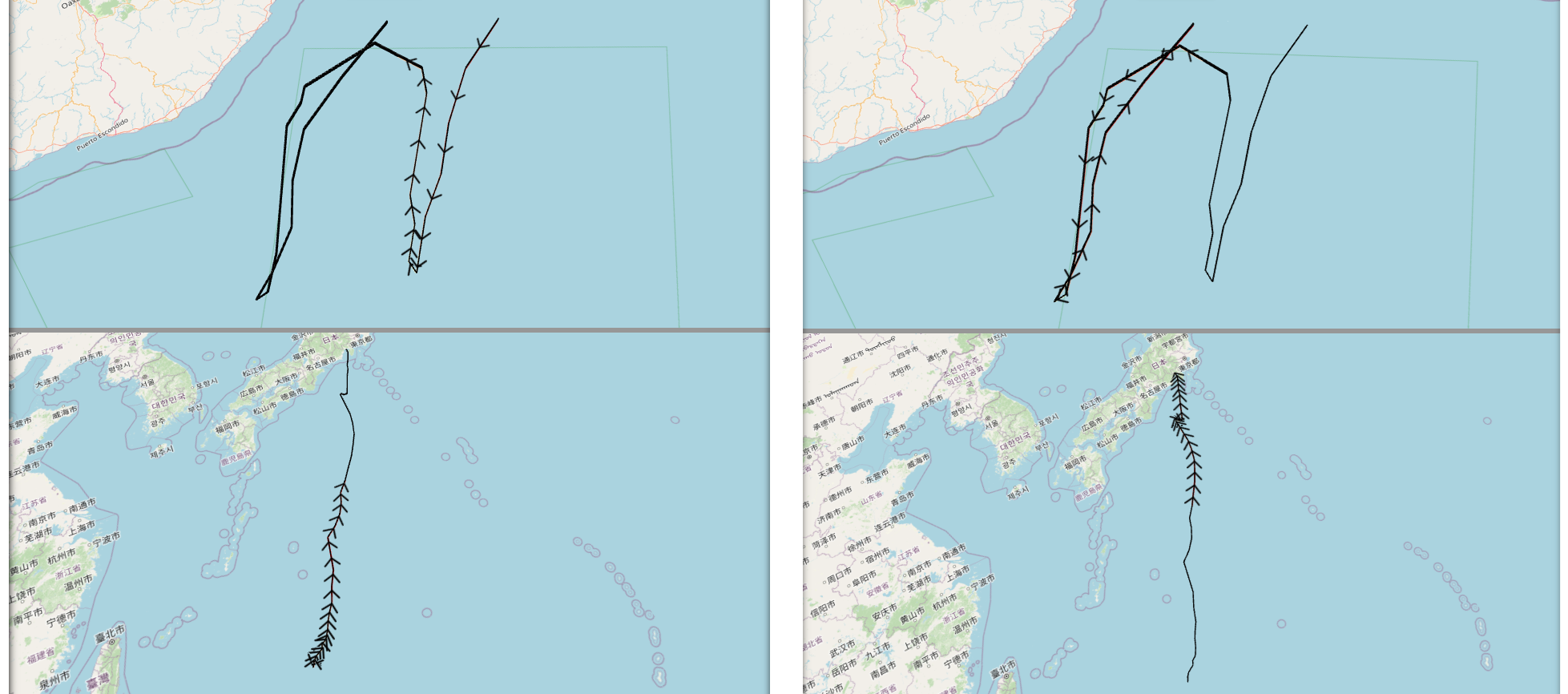}
    \caption{Trajectory 1, 2024274N15266 (top) and Trajectory 2, 2023247N20130 (bottom), presenting signatures 2\_1 (left) and 2\_2 (right).}
    \label{fig:entire_3D_2024274N15266_vs_2023247N20130}
\end{figure}

\noindent First, a visual comparison is performed over distance geometry, with respect to the entire trajectory, with arrows highlighting each of the signatures, as shown in Fig. \ref{fig:entire_3D_2024274N15266_vs_2023247N20130}.

\begin{figure}[H]
    \centering
    \includegraphics[width=0.8\linewidth]{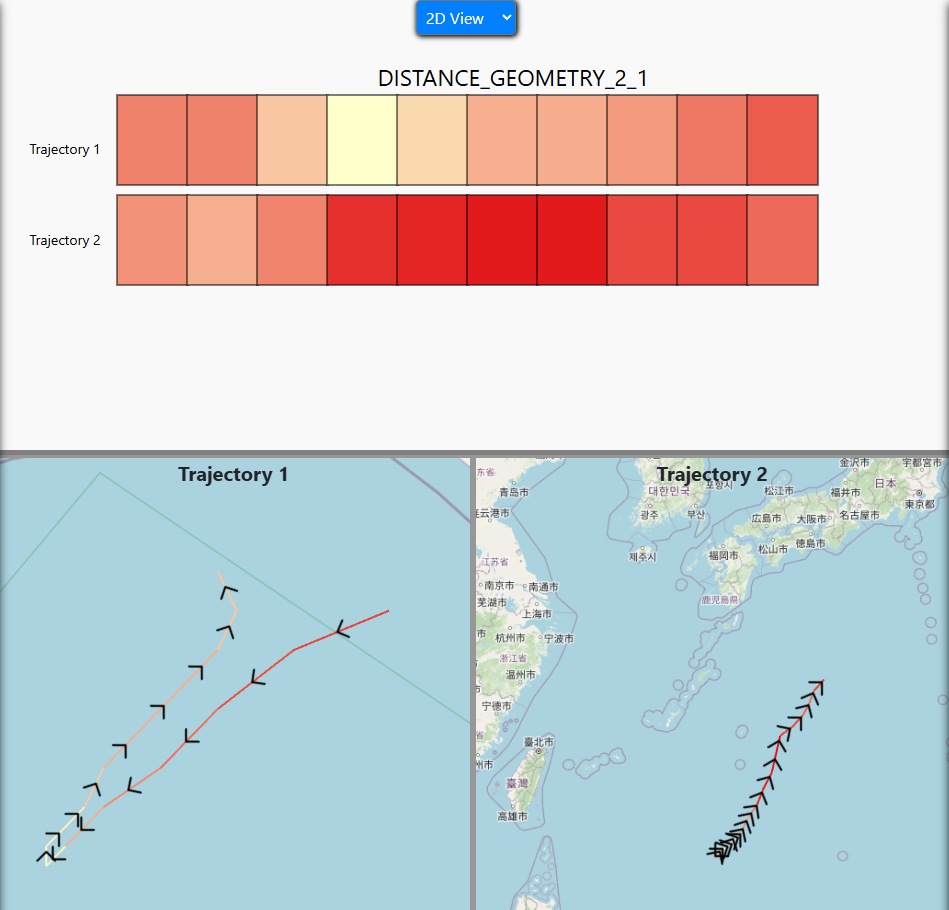}
    \caption{Trajectory 1, 2024274N15266 (heatmap top and left on the map) and Trajectory 2, 2023247N20130 (heatmap bottom and right on the map), signature 2\_1.}
    \label{fig:distance_geometry_2_1_2024274N15266_vs_2023247N20130}
\end{figure}

\noindent As previously discussed, visualizing an entire trajectory does not always provide significant insights, however, it enables further guidance and discussion for additional analytical steps. In this case, it can be seen from the entire trajectory samples that Trajectory 1 (top) has significant shape changes, while Trajectory 2 (bottom) seems to have higher straightness and less variation in angle (i.e., indentation). To emphasize the importance of the signatures over time and further review this, the focus is tilted towards the 2D view with heatmap representations, as shown in Fig. \ref{fig:distance_geometry_2_1_2024274N15266_vs_2023247N20130} for signature 2\_1 (since this signature showed to have the highest feature importance overall).

\noindent Before further review, the distance geometry measure and color heatmap representation must be briefly covered. Distance geometry is measured between 0 and 1, where zero represents a trajectory to be \textit{tortuous} (i.e., high amount of change, a non-straight path), and 1 represents a trajectory being a line \textbf{completely straight}. In the 2D map views, the heatmap on top represents exactly that; low distance geometry values are represented in light yellow (non-straight trajectory path), while deep-red color represents high distance geometry scores (a straight trajectory path).

By reviewing and analyzing the distance geometry results for signature 2\_1, it can be noticed that Trajectory 1, which is the sample point selected from the hybrid behavior zone (i.e., curvature and indentation), has significantly lower distance geometry values in comparison to Trajectory 2, which only has curvature-based behavior. For this signature, the visual overview indeed supports the previous categorization for the outlier score distribution, which recognized Trajectory 1 to have hybrid behavior (i.e., to be in zone 3), and Trajectory 2 to be highly related to curvature only (i.e., to be in zone 2). Visually, trajectory 1 showed little straightness, and trajectory 2 showed high straightness (which led to the higher distance geometry values in trajectory 2).
To further review such a relation, reviewing other signatures is needed to verify if this holds true for other signatures as well.

\begin{figure}[H]
    \centering
    \includegraphics[width=0.75\linewidth]{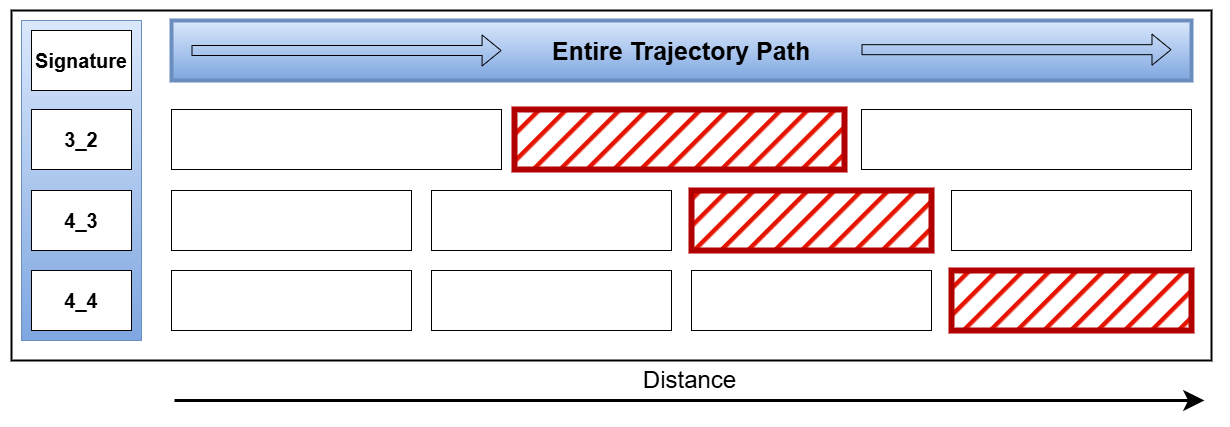}
    \caption{Next signatures under review.}
    \label{fig:signatures2}
\end{figure}

Since the previous signature was 2\_1, the first half of the trajectory was under review. Therefore, the three following 2D views and heatmap values that will be analyzed are from the signatures 3\_2, 4\_3, and 4\_4. A simplified trajectory overview significance is shown in Fig. \ref{fig:signatures2} to convey the signatures of interest.

\begin{figure}[H]
    \centering
    \begin{minipage}[b]{0.45\textwidth}
        \centering
        \includegraphics[width=\linewidth]{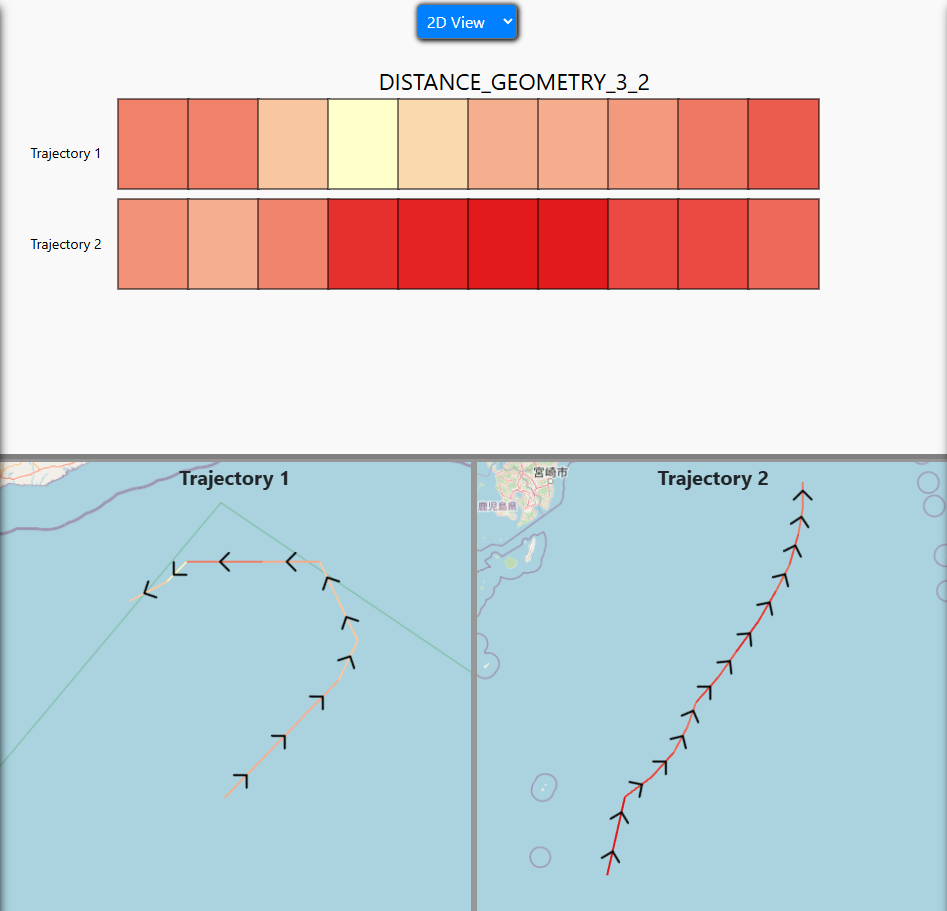}
        \caption{Signature 3\_2}
        \label{fig:hurricanes_signature_3_2}
    \end{minipage}
    \hfill
    \begin{minipage}[b]{0.45\textwidth}
        \centering
        \includegraphics[width=\linewidth]{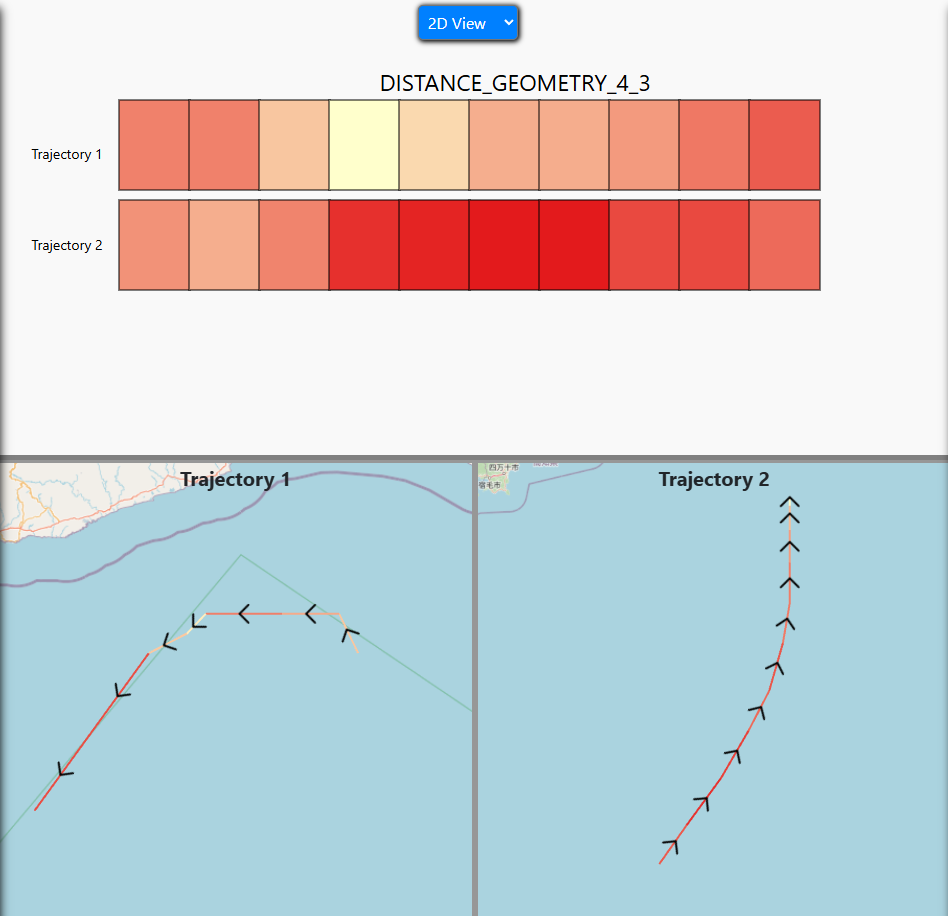}
        \caption{Signature 4\_3}
    \label{fig:hurricanes_signature_4_3}
    \end{minipage}
    \hfill
    \begin{minipage}[b]{0.45\textwidth}
        \centering
        \includegraphics[width=\linewidth]{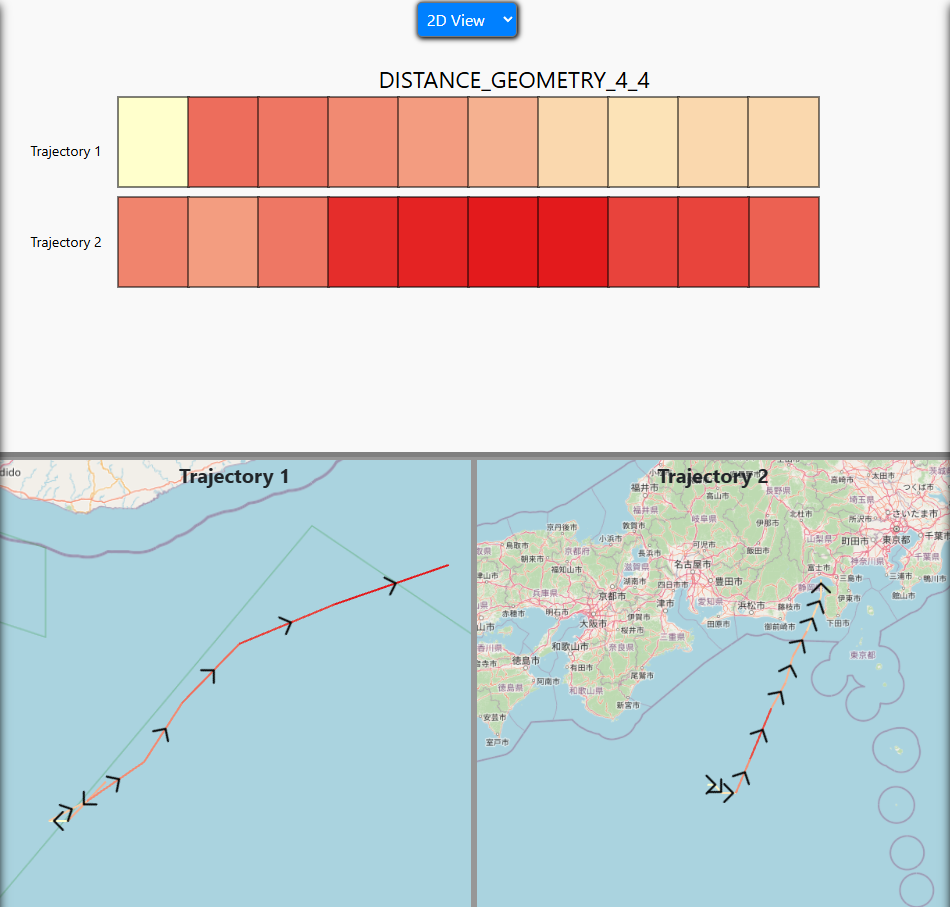}
        \caption{Signature 4\_4}
        \label{fig:hurricanes_signature_4_4}
    \end{minipage}
    \hfill
\end{figure}

\noindent The figures \ref{fig:hurricanes_signature_3_2}, \ref{fig:hurricanes_signature_4_3}, and \ref{fig:hurricanes_signature_4_4} display the correspondent views, for \textbf{3\_2}, \textbf{4\_3}, and \textbf{4\_4}, respectively. All three signatures showcase the same curvature-based behavior as previously seen in signature 2\_1; Trajectory 1 has significantly lower curvature values in contrast to Trajectory 2, which presents higher distance geometry results throughout the entire trajectory. Visually, the straightness in trajectory 2 shows to be much higher; hence, the distance geometry value is also higher, and the heatmap colors with \textit{deeper red} representation.

\noindent Thus far, Trajectory 1 showed lower overall distance geometry scores in comparison to Trajectory 2. From the visualizations on the maps, it can be noted that several direction changes occurred during the hurricane, which were less or not present at all in Trajectory 2. Since Trajectory 1 was originally categorized as \textit{hybrid} behavior, due to the outlier score results spreading the trajectory onto zone 3, it means that for the current combination, indentation-based behavior should be present. In comparison, Trajectory 2, from zone 1 (i.e., curvature-based behavior only), should show little indentation significance. To verify this, the following steps move towards the analysis of angles and angle changes, which have been only addressed from a visual overview hitherto.

\begin{figure}[H]
    \centering
    \includegraphics[width=0.7\linewidth]{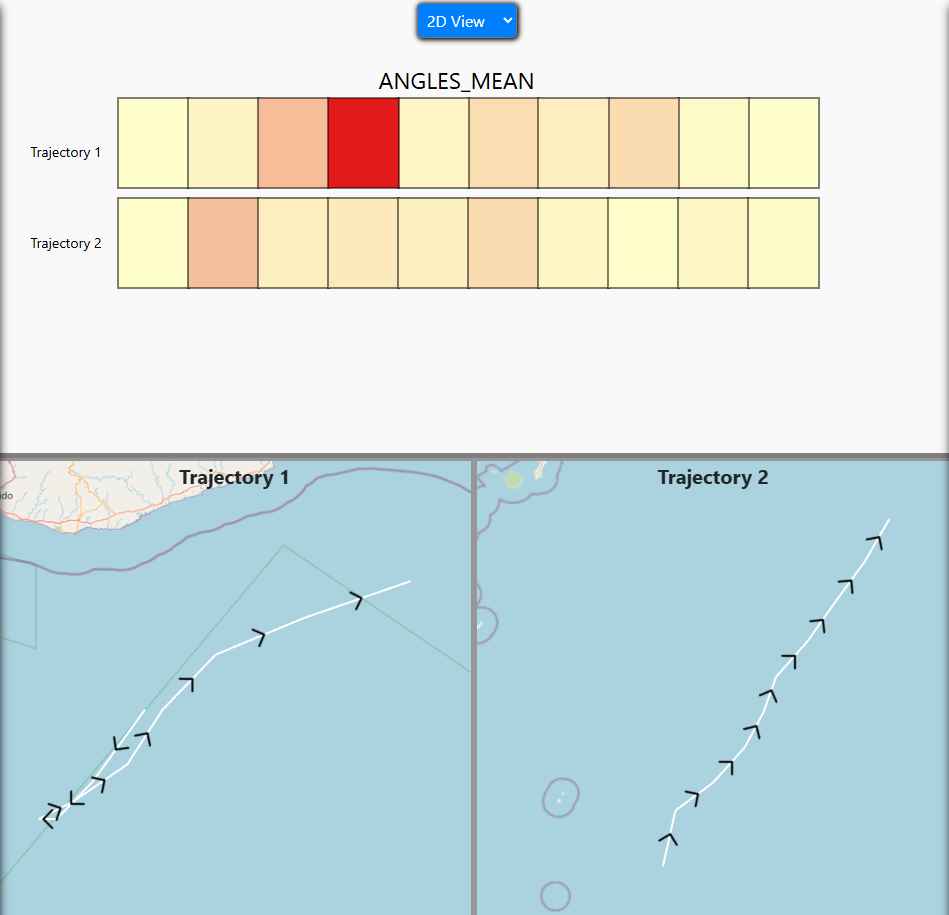}
    \caption{Trajectory 1, 2024274N15266 (heatmap top and left on the map) and Trajectory 2, 2023247N20130 (heatmap bottom and right on the map).}
    \label{fig:hurrianes-angles-mean}
\end{figure}

The first statistical variable under analysis is the mean value for the angles, which scored high in the indentation feature importance overview, and would provide a mean value representation of the angles' impact as per the two hurricane trajectories selected. Figure \ref{fig:hurrianes-angles-mean} shows the angle \textit{mean} samples for the result over each of the trajectories.

\noindent When selecting \textit{angle mean} from the feature importance bar chart, the mean value for the angles in both trajectories are calculated, and the row sample in the dataset that has the \textbf{closest value to the mean results} of each trajectory is selected, as well as a few samples before and a few after, to only show a subset of the movement, which facilitates the review of such indicators (i.e., a total 10 points only).

From the angle view, now simplified and only showing a smaller section of the trajectory, it can be found that Trajectory 1, selected from the hybrid zone (i.e., curvature and indentation), indeed presents higher angle values in comparison to Trajectory 2, which is only characterized by curvature.

\begin{figure}[H]
    \centering
        \begin{minipage}[b]{0.49\textwidth}
        \centering
        \includegraphics[width=\linewidth]{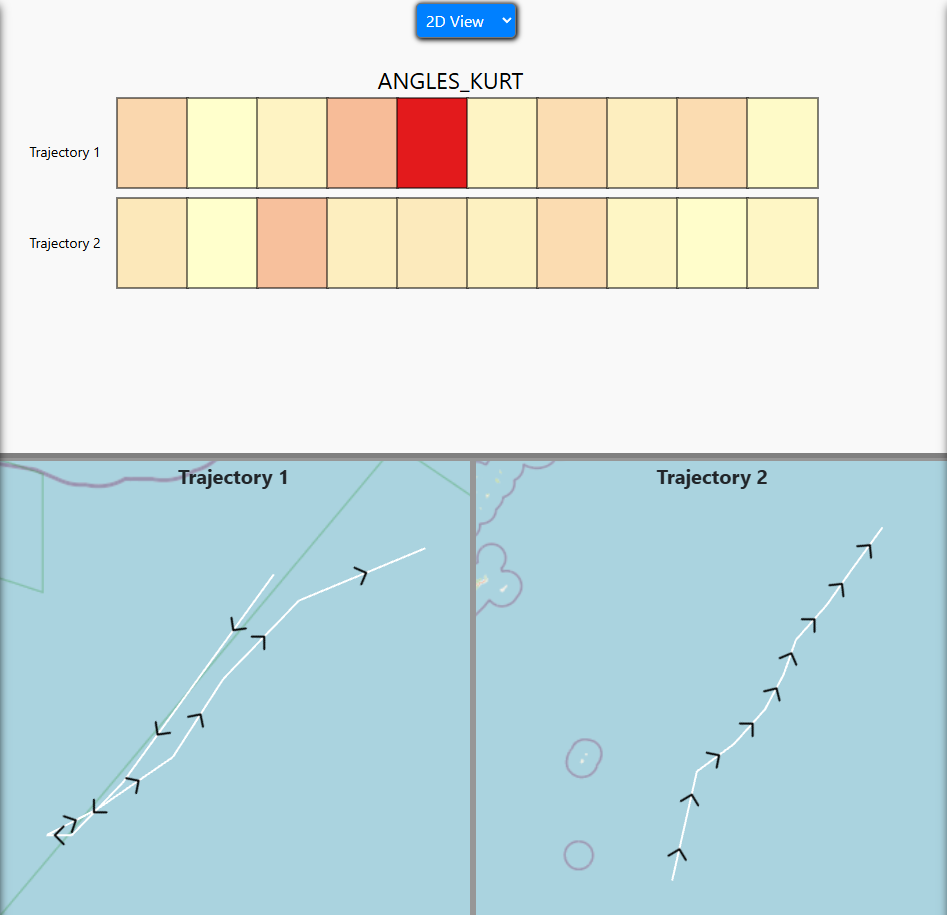}
        \caption{Angles kurtosis: Trajectory 1, 2024274N15266 (heatmap top and left on the map) and Trajectory 2, 2023247N20130 (heatmap bottom and right on the map).}
        \label{fig:hurrianes-angles-kurtosis}
    \end{minipage}
    \hfill
    \begin{minipage}[b]{0.49\textwidth}
        \centering
        \includegraphics[width=\linewidth]{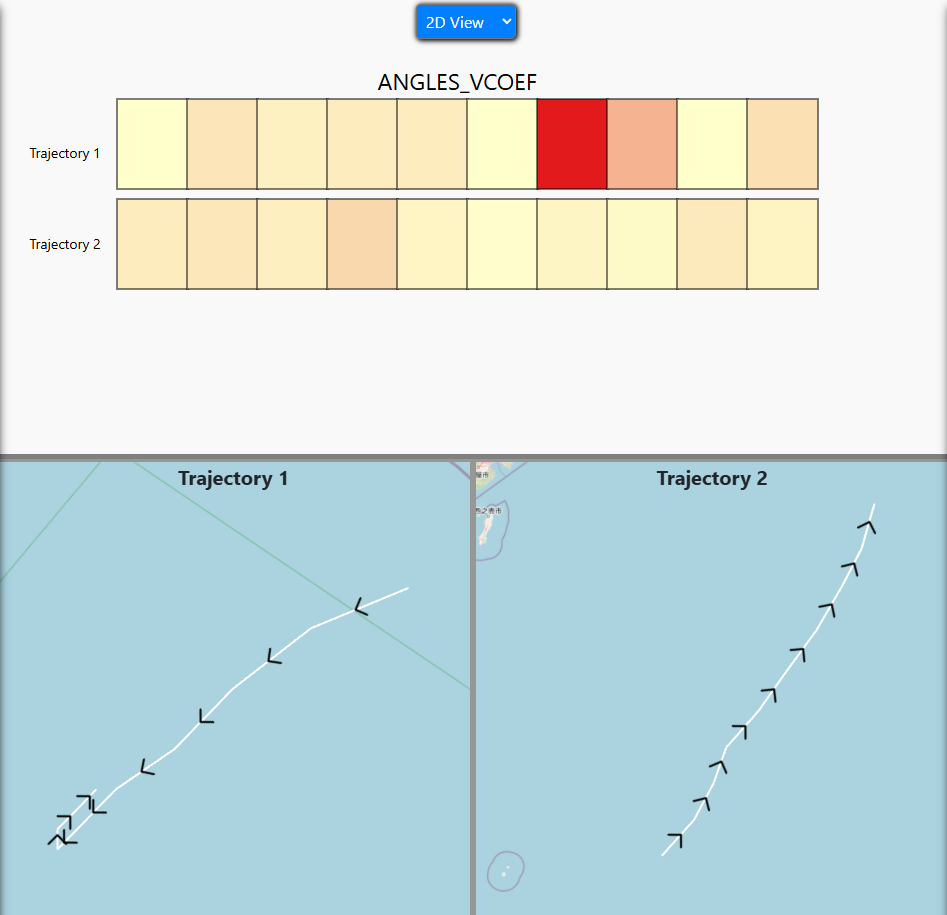}
        \caption{Angles V. Coeff.: Trajectory 1, 2024274N15266 (heatmap top and left on the map) and Trajectory 2, 2023247N20130 (heatmap bottom and right on the map).}
        \label{fig:hurrianes-angles-vcoeff}
    \end{minipage}
    \hfill
\end{figure}

To further verify this, additional statistical variables are reviewed from the feature importance results. Namely, \textit{angles kurtosis} and \textit{angles variation coefficient} are selected, calculated, and illustrated. Figure \ref{fig:hurrianes-angles-kurtosis} shows the angles kurtosis, while figure \ref{fig:hurrianes-angles-vcoeff} shows the variation coefficient.

\noindent From these additional statistical variables, it can also be discerned that angles were more significant in Trajectory 1 than in Trajectory 2. To understand the specific angles that mark such a difference between the two trajectories, it is possibly useful to review which unique angles are found. More specifically, reviewing the highest degree values for several statistical variables may point the current analysis towards the main descriptors for the geometric behavior in Trajectory 1 (hybrid behavior) and Trajectory 2 (curvature only). 

\begin{figure}[H]
    \centering
    \includegraphics[width=1.03\linewidth]{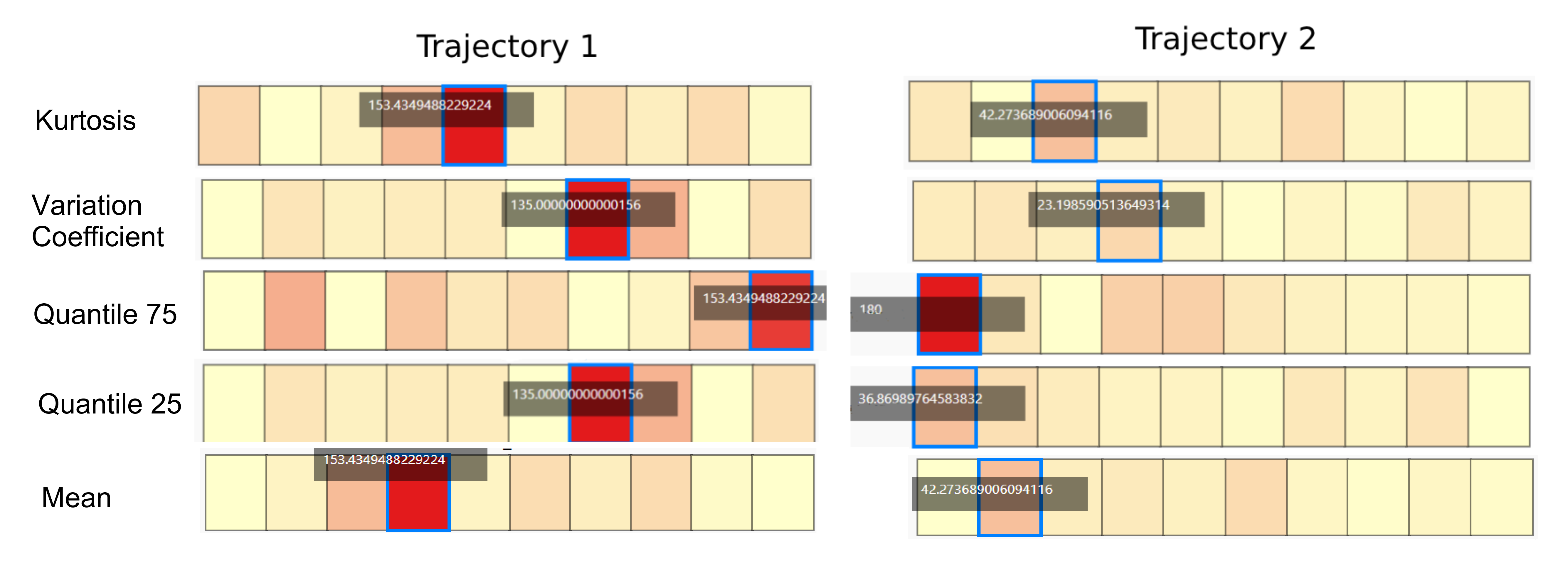}
    \caption{Angle trajectory samples (from top to bottom): kurtosis, variation coefficient, quantile 75, quantile 25, and mean.}
    \label{fig:hurricanes-stat-variables-angles}
\end{figure}

The figure \ref{fig:hurricanes-stat-variables-angles} compares several statistical variables side-by-side, highlighting the \textbf{highest angle degrees} found in each of the trajectory samples (i.e., a total of \textbf{10} trajectory points derived from the entire dataset), which was derived by finding the closest angle throughout the entire dataset matching the variable calculated. The five variables selected were kurtosis, variation coefficient, quantile 75, quantile 25, and mean. 

\noindent From reviewing peak angles in this manner, it becomes evident that Trajectory 1, which was labeled by the outlier detection algorithm as \textit{hybrid}, was correctly categorized by having both indentation and curvature behaviors. Although curvature values were lower than those seen in Trajectory 2, which was labeled as curvature-based behavior only, it still showed relevant levels of curvature. This further supports the findings and correlation between the outlier detection, the decision boundary visualization, and the feature importance results, which share categorical results.

Further analysis over the highest angles shown through all 5 statistical variables presented in Fig. \ref{fig:hurricanes-stat-variables-angles} may enable further understanding of the hurricanes' trajectories. Table \ref{table:hurricanes-stat-variables-angles} synthesizes the top angles (i.e., the highest angle values per sample) per trajectory, as for each statistical variable.

\begin{table}[ht]
    \begin{center}
        \begin{tabular}{|| c | c | c ||} 
             \hline
             \textbf{Statistical Variable} & \textbf{Trajectory 1 (angle degrees)} & \textbf{Trajectory 2 (angle degrees)} \\ [0.5ex] 
             \hline\hline
             Kurtosis & 153° & 42° \\ 
             \hline
             Variation Coefficient & 135° & 23° \\
             \hline
             Quantile 75 & 153° & 180° \\
             \hline
              Quantile 25 & 135° & 36° \\
             \hline
              Mean & 153° & 42° \\ [0.5ex] 
             \hline
        \end{tabular}
    \end{center}
    \caption{Significance for each of the zone-trajectory distributions based on the currently selected combination.}
    \label{table:hurricanes-stat-variables-angles}
\end{table}

\noindent Noticeably, two repeated angles with the \textbf{highest degrees} in trajectory 1  with different statistical variables can be found. Namely, kurtosis, quantile 75, and mean samples showed the highest angle of \textbf{153 degrees}, while variation coefficient and quantile 25 samples had \textbf{135 degrees} at their highest points. It must be noted that the \textbf{exact} angle values are found (i.e., 153.4349488229223° and 135.00000000000156°), which clearly indicates that the degree values come from the same sample in space and time. These points are of interest, since they represent the maximal degrees found in each of the statistical variables (i.e., 10-point samples for each statistical variable). It is important not to confuse these with the exact values of the angle kurtosis, v. coefficient, quantiles, and mean results, as these are found in the same samples (i.e., the neighboring points). This finding redirects further analysis towards the insight that certain unique points in space and time, during the hurricane in trajectory 1, had the \textbf{most significance} for the categorization within the indentation-based behavior, and possibly, to the trajectory's shape overall (i.e., its geometry).

\begin{figure}[H]
    \centering
    \includegraphics[width=1.1\linewidth]{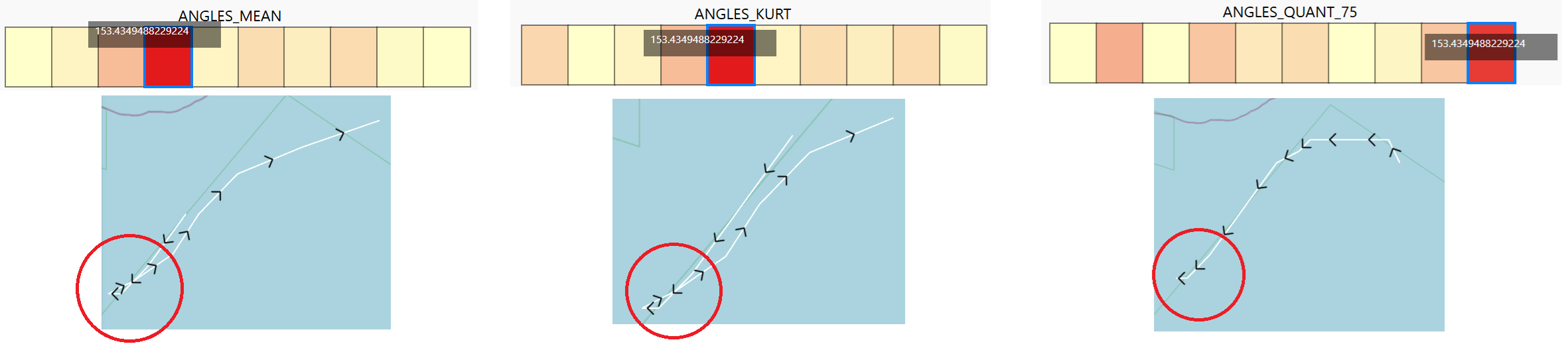}
    \caption{Trajectory 1, ID 2024274N15266. Angles mean, kurtosis, and quantile 75 (from left to right), highlighting the specific trajectory point that had an angle of 153°.}
    \label{fig:hurricanes-153-degrees-angle}
\end{figure}

With these new insights over the angles and their importance, one additional step is performed, which aims to visually link the highest degree angles with their point location on a map. Figure \ref{fig:hurricanes-153-degrees-angle} shows the three statistical variables mean, kurtosis, and quantile 95, which had the 153-degree angle point, highlighting such point in the 2D map view. 

\begin{figure}[H]
    \centering
    \includegraphics[width=1\linewidth]{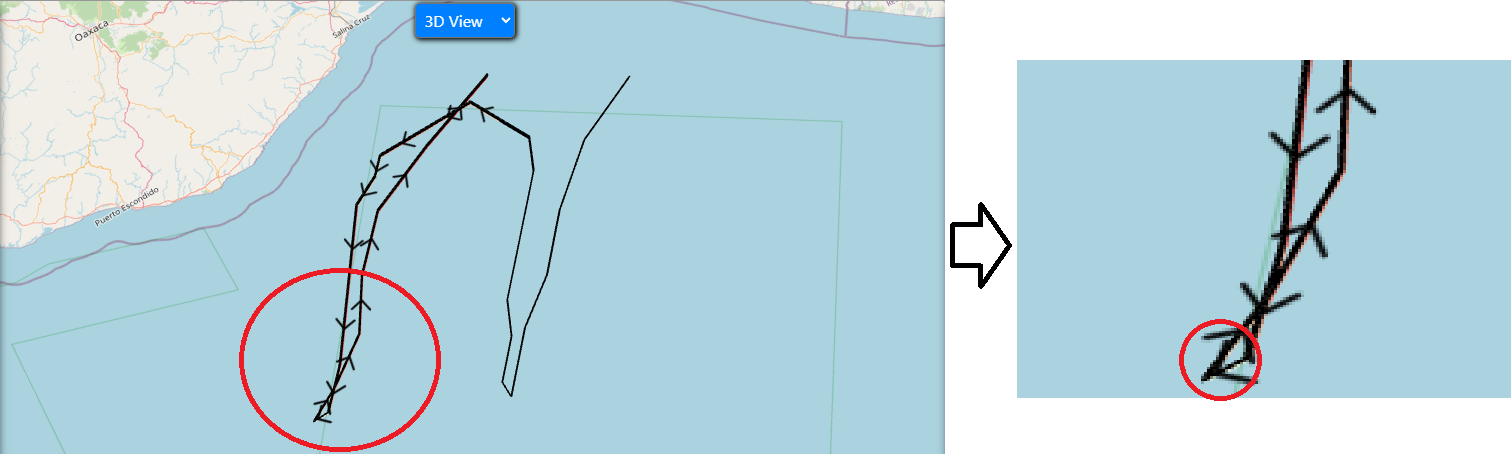}
    \caption{Full trajectory 1, ID 2024274N15266, highlighting the segment in signature 2\_2 that had a 153 degree angle (left) and zoomed view (right).}
    \label{fig:hurricane-signature-2-1-with-153-degrees}
\end{figure}

\noindent Viewing such point through the entire trajectory path (highlighting signature 2\_2), it can clearly be noted the pivot point of the hurricane with such degree, as displayed in Fig. \ref{fig:hurricane-signature-2-1-with-153-degrees}

\begin{figure}[H]
    \centering
    \includegraphics[width=1\linewidth]{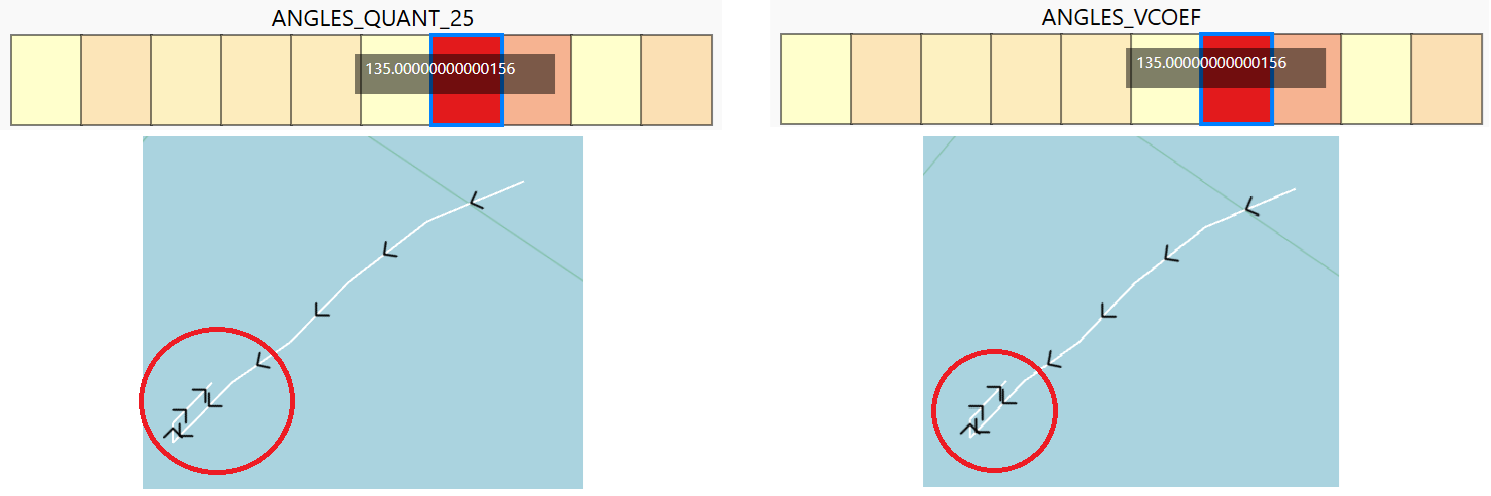}
    \caption{Trajectory 1, 2024274N15266.}
    \label{fig:hurricanes-135-degrees-angle}
\end{figure}

\noindent By repeating the process with the two statistical variables which had an angle of 135 degrees, namely, quantile 25 and variation coefficient, the figure \ref{fig:hurricanes-135-degrees-angle} was produced. Notice that for this sample, although the computation is clearly different for diverse statistical calculations, the same 10-point samples are derived for both statistical variables.

\begin{figure}[H]
    \centering
    \includegraphics[width=1\linewidth]{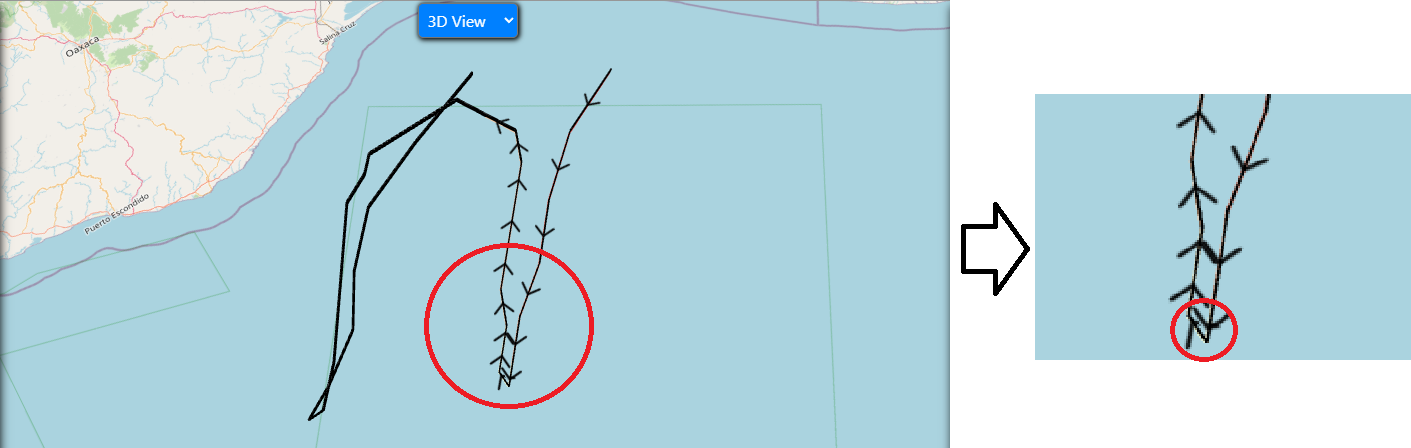}
    \caption{Full trajectory 1, ID 2024274N15266, highlighting the segment in signature 2\_1 that had a 135 degree angle (left) and zoomed view (right).}
    \label{fig:hurricanes-pivot-angle-135}
\end{figure}

\noindent Then, through the view of the entire trajectory, it is possible to pinpoint the exact pivot point (highlighting signature 2\_1), as shown in Fig. \ref{fig:hurricanes-pivot-angle-135}

\begin{figure}[H]
    \centering
    \includegraphics[width=1\linewidth]{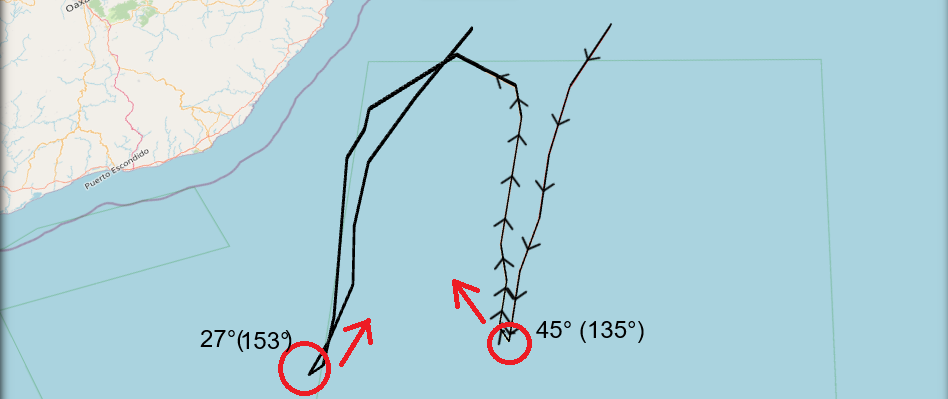}
    \caption{Full trajectory 1, ID 2024274N15266, in signature 2\_1 highlight, showing the two pivot points; 153° (left, supplementary 27°) and 135° (right, supplementary 45°).}
    \label{fig:pivot-points-together}
\end{figure}

\noindent The figure below \ref{fig:pivot-points-together}, shows the two key pivot points for the entire trajectory, outlining the angle and direction changes.

\begin{figure}[H]
    \centering
    \includegraphics[width=1\linewidth]{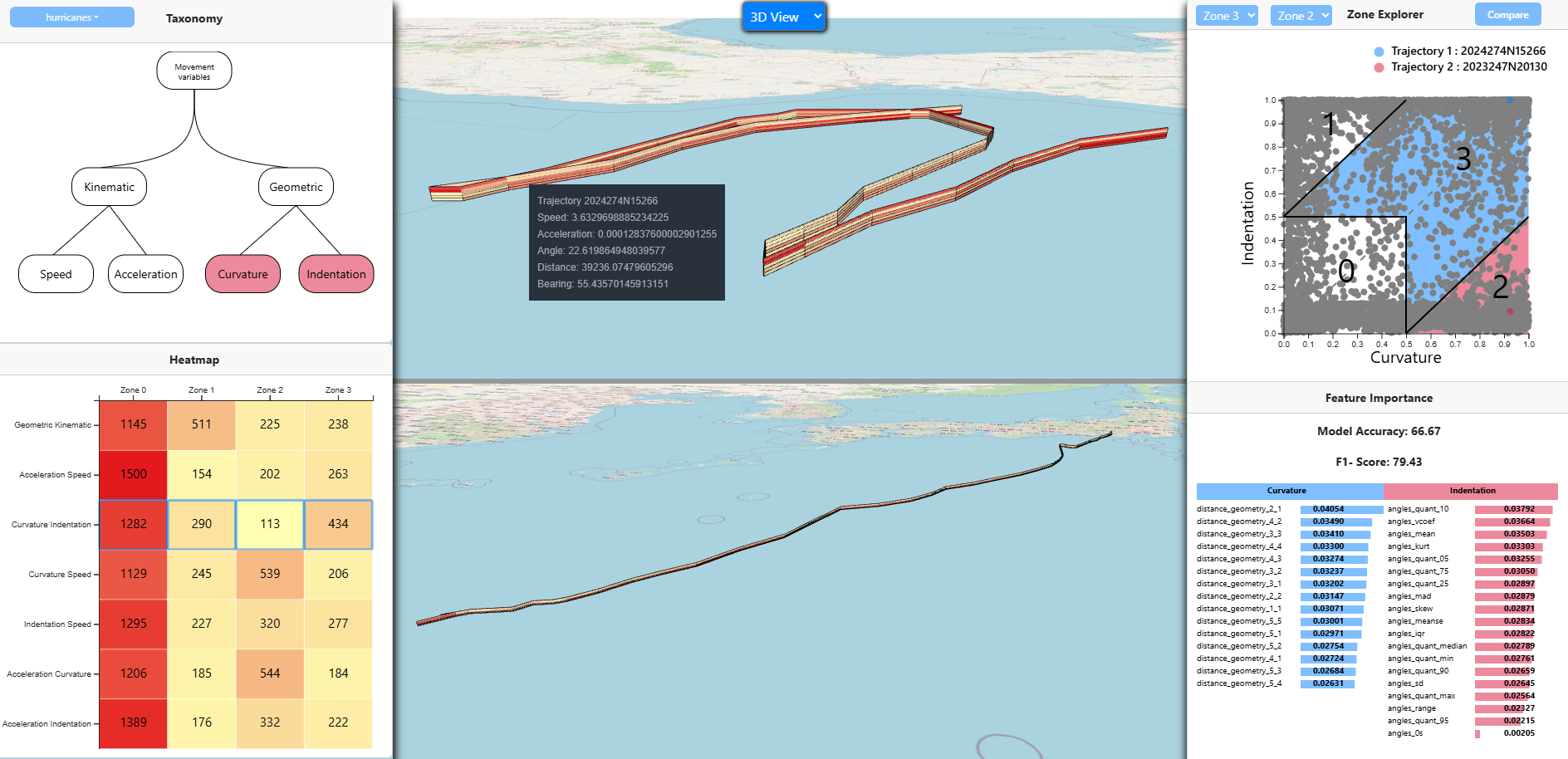}
    \caption{Full tool capture: Trajectory 1, ID 2024274N15266 (top) and Trajectory 2, id 2023247N20130 (bottom).}
    \label{fig:full-tool-case-study-2}
\end{figure}

\noindent With the final visual representation over the angles with the highest degrees on the map view, the intention is to present how the results in the machine learning algorithms, which include pre-processing, feature extraction, outlier detection, boundary analysis, and feature importance, produced several measurable values through statistical results, labeling the trajectory as \textit{hybrid}, with both curvature and indentation, in contrast to a trajectory that had only curvature-based behavior. The visualization on the map is meant to be self-evident, with the intent to prove that the details gathered from the analytical views, using algorithms, mathematics, and statistics, which found the trajectory to be highly characterized by indentation, are cohesive. Then, providing visual validity, and making a \textit{full cycle} between the low-level details, the taxonomy used, and high-level, easy-to-conceptualize, map visualizations. Figure \ref{fig:full-tool-case-study-2} displays the entire tool view, with the 3D map trajectory comparisons.

\subsubsection{Case Study 2 Final Results}
\noindent For the second case study, a dataset over tropical cyclones was analyzed, using trajectories between the years 2004 and 2024. The initial iteration of the analysis presented a comparison and investigation of the subdivision taxonomy groups \textit{Speed} and \textit{Curvature}.

The initial results for the first iteration showed that \textbf{11.56\%} of the hurricanes were grouped in \textbf{zone 1}, which meant that \textbf{245} trajectories had significant \textbf{curvature} behavior only, and \textbf{25.44\%} of the total trajectories had significant \textbf{speed} behavior, as \textbf{539} points were grouped in \textbf{zone 2}. A \textbf{53.28\% }of the total hurricanes had \textbf{no} curvature, speed, or hybrid behavior, as they were grouped in \textbf{zone 0}, leaving \textbf{206} trajectories in \textbf{zone 3}, which accounted for \textbf{9.72\% }of the hurricanes, presenting \textbf{\textit{hybrid}} behavior (i.e., speed and curvature together).

With the intent to further understand the zone distribution, zone 0 was compared against zone 3. Two individual trajectories were also selected from each of the zones, namely, hurricane ID 1959013S20042 (from zone 0, no significant speed/curvature behavior) and hurricane ID 2020299N18277 (from zone 3, hybrid behavior). That comparison produced a feature importance bar chart table, which showed that, \textbf{for speed-based behavior}, the statistical variables \textbf{median absolute deviation, variance coefficient, quantile median, interquartile range, and standard error of mean} had the highest impact over their categorization. On the other hand, \textbf{curvature-based behavior} was characterized by 5 distance geometry signatures: \textbf{2\_2, 3\_1, 5\_4, 5\_3, and 4\_1}.

Additional visual comparison was then performed, selecting \textit{speed median absolute deviation} and visualizing on a map the trajectories previously selected for individual analysis. The first iteration showed rather easy-to-discern details over the trajectories, as the trajectory with hybrid behavior clearly reached a higher speed over time, with \textbf{7.55 meter per second}, showing \textbf{higher straightness}, in comparison to the trajectory without significant speed and curvature behavior, which reached a maximum speed of \textbf{5.23 meters per second}. Nonetheless, the visual comparison in contrast to the statistical variables and the decision boundary distribution from the outlier scores presented a new point of interest, guiding a second iteration on the analysis. The results until then showed that curvature did not necessarily represent the shape of the first trajectory selected (i.e., 1959013S20042, from zone 0, without significant speed/curvature), however, it was possible to convey that indentation had an effect on the overall geometry and shape of the trajectory, possibly indicating that angles could provide better insights and representativeness when analyzing cyclones. Therefore, a second iteration comparing curvature and indentation was performed.

The second iteration of the analysis presented the selection of \textbf{Curvature} and \textbf{Indentation}. The analysis over the heatmap frequencies results showed \textbf{60.50\%} of the trajectories to be grouped \textbf{zone 0}, with \textbf{no} significant curvature/indentation behavior, \textbf{13.69\% in zone 1} representing 290 hurricanes with significant \textbf{indentation-based} behavior, \textbf{5.33\% in zone 2} with significant \textbf{curvature-based} behavior, and \textbf{20.48\%} of the hurricanes showing \textbf{both, curvature and indentation-based behavior in zone 3} (i.e., hybrid behavior). With the intent to understand curvature and indentation from both the individual and the combined characterizations, the zones selected for comparison were zone 3 (i.e., curvature \textbf{and} indentation \textbf{together}) and zone 2 (i.e., \textbf{curvature-based behavior only}). Individual trajectories were selected from each of the decision boundary zones. Namely, ID 2024274N15266 from zone 3, referred to as \textbf{\textit{trajectory 1}}, and ID 2023247N20130 from zone 2, referred as \textbf{\textit{trajectory 2}}. The feature importance results showed the distance geometry variables (i.e., the mathematical tool used for representing curvature) had a rather even distribution of importance, with \textbf{signatures 2\_1, 4\_2, 3\_3, 4\_4, and 4\_3} as the top 5 most significant variables, nonetheless, with the 10th signature (i.e., 5\_5) having a difference of only \textbf{0.01053}. The indentation variables also showed an even distribution of importance, with the top statistical indicators being angle mean, kurtosis, variation coefficient, quantile 75, and quantile 25.

Visual analysis over the entirety of the trajectories' paths was then performed, specifically highlighting the directions for the two previously selected trajectories. Since signature 2\_1 was the feature with the highest measurable importance from the random forest algorithm results, 2\_1 and 2\_2 were investigated. The \textbf{visual analysis clearly showed} that \textit{trajectory 1} (i.e., hybrid) had significant representation in regards to \textbf{both curvature and angles}, with often changes in the hurricane's direction, while \textit{trajectory 2} (i.e., curvature only) showed \textbf{higher straightness and less angle shifting}. The following step focused on reviewing specific distance geometry results over time, as per each of the trajectories under analysis. The first comparison for signature 2\_1 showed that trajectory 2 was indeed characterized by curvature-based behavior mainly, by presenting higher straightness (which was visually discerned by the \textit{deeper-red} heatmap views, representing higher distance geometry scores), while trajectory 1 showed lower distance geometry scores, due to the changes on shape between the starting point and end point of the signature segment. This relation and labeling were further verified by analyzing the second \textit{part} for each of the trajectories, through signatures 3\_2, 4\_3, and 4\_4, which showed once again that trajectory 1 (i.e., hybrid) had a lower distance geometry representation than trajectory 2 (i.e., curvature only).

The analysis then continued by studying the indentation of the two same trajectories under review. Calculating the mean angle value for both trajectories led to the visual analysis of the hurricane trajectories \textit{side-by-side}, which showed that trajectory 1 (i.e., hybrid), had a number of key points were the angle values were high, in comparison to trajectory 2, which had lower angle values overall. This was further analyzed by first reviewing two additional statistical variables, namely, \textit{angle kurtosis} and \textit{variation coefficient}, and then furthermore with 2 quantile variables; \textit{quantile 75} and \textit{quantile 25}. Results showed that trajectory 1 (i.e., hybrid) had two specific high-angle values, which had the highest impact and contributed the most to its behavioral categorization: one angle with 153 degrees, and a second angle with 135 degrees. Trajectory 1 (i.e., curvature only) showed indeed low angle values, which supports even more all results, representation, and labeling of trajectory 2 from zone 2 being highly characterized by curvature, and not indentation.

The final step of the second iteration led to connecting the final findings between low-level details and results, with high-level visual representations. The two angles, 153° and 135°, were found to match two pivot points on the trajectory. These two points can be clearly seen from the visualization shown in Fig. \ref{fig:pivot-points-together}, which arguably make up for the \textbf{biggest two indication points} that differentiate the \textbf{hurricane's trajectory 2} from trajectory 1, and \textbf{characterizes the overall shape}, and therefore \textbf{geometry}, of this specific hurricane.  

The tropical cyclones case study showed several findings derived from a combination of data pre-processing, machine learning techniques, statistical representations, mathematical and visual-interactive analyses. These include outlier detection, outlier score analysis, decision boundary representation, score-to-group correlation, taxonomy selection, labeling, subdivision overview, \textbf{several} visual analytic steps, linkage of low-level and high-level details, and overall representation of the specific hurricanes under analysis. The tool and analytical steps successfully derived, grouped, labeled, and separated hurricanes and their behaviors, showcasing specific trajectories that had curvature and indentation behavior together, and curvature-based behavior only. The findings from the analytics tool provided specific statistical indicators that showed how distance geometry was more representative for one trajectory, while a second trajectory had a hybrid behavioral representation. The system's output was able to show not only the two specific angles significant for the entirety of the trajectory's shape and geometry, but also under which statistical variable indicators, group categorization, and visual locations on a map were critical to properly contextualize the findings.

\newpage
	
\section{Discussion}


In this thesis project, a tool was developed to facilitate the explanation and comprehensive analysis of spatio-temporal data by integrating statistical representation with interactive visualizations. 
In order to present, analyze, and discuss the tool and the thesis project proposal, two case studies were conducted.

The first case study, using a dataset of Arctic Foxes' movement, demonstrated the cyclic approach for the entire tool, with two key iterations. 
The final results presented the importance, differences, and insights over Kinematic and Geometric behaviors, and then between Acceleration and Curvature behaviors, as their respective group subdivisions. 
More specifically, the case study showed the key statistical variables, their importance, the zone distribution, and impact on the trajectories and trajectory groups, using several visualizations to represent such findings. 
The outlier scores, visually separated into zones through the boundary analysis axes, correctly identified the foxes to be represented with Kinematic, Geometric, Acceleration, and Curvature-based behavior, as supported by the statistical variables derived from feature importance bar charts, then visually represented in 2D and 3D map views over time. 
Through deeper analysis of individual foxes in regards to Acceleration, a pattern was found, which showed that foxes labeled as having Acceleration-based behavior had constant and steady acceleration, while those foxes identified as Curvature-based behavior had several peaks of acceleration and deceleration through short periods of time. 
The findings were supported by the low-level details presenting statistical indicators and outlier scores, as well as high-level views, from 2D and 3D map visualizations, with heatmap patterns clearly visible when doing \textit{side-by-side} comparisons.

The second case study was performed by analyzing tropical cyclones' trajectories. 
The method and the visual analytics tool again showed that they could correctly identify and contextualize different hurricanes into specific taxonomy groups and subgroups, through a cyclic approach. 
The specific subgroup divisions analyzed were Speed, Curvature, and Indentation, with specific investigation over Geometric combinations, comparing Curvature and Indentation hybrid behavior against Curvature-based behavior only. 
The top statistical variables were correctly identified, as well as the zone distribution from the outlier score results. 
Further analysis of Distance Geometry and Angles presented the comparison of several signatures, effectively isolating the different sections of the trajectories over time, when comparing individual hurricanes. 
The visual analysis demonstrated the output over labeling and grouping of the trajectories to be accurate, as the trajectory with hybrid behavior was characterized by both Curvature and Indentation. 
Furthermore, the unique angles that represented the hurricane's hybrid behavior were identified and presented, validating the statistical variable findings and machine learning model results onto map visualizations, proving results that consistently aligned and supported such findings with simple, easy-to-comprehend, trajectory visualizations in 2D. 

The overall results were positive, as the utilization of the tool and the pre-processed dataset were the only sources used in order to find such patterns, anomalies, and compare the trajectories. 
The multi-level approach was key in order to understand and redirect the analytic process during the investigation of the datasets, as all relevant indicators found in the trajectories were only discovered after several view levels, which also showcased how numerical and visual-interactive analyses support each other for such discoveries to happen. 
Furthermore, the cohesive linkage between the different tool components significantly facilitated the understanding of the multiple views, rather rapidly.

Two of the important aspects of data analysis in both exploratory and pattern-finding processes are contextualization and meaning. The tool proposed is, intentionally, limited to only using spatio-temporal datasets in order to reach any findings. 
This means that the tool aims to generally allow the identification of such patterns and anomalies through the same process. 
This has both advantages and disadvantages. 
The main advantage is that a researcher can perform comprehensive analyses over several spatio-temporal datasets, from different objects, and diverse subject areas, and consistently unveil insights, patterns, and find meaningful representations of such datasets and moving objects, regardless of their nature. 
A second key advantage is that an analyst may follow the exact same process, with the same exact steps, for a number of different datasets and be able to discern diverse insights. 
A disadvantage is the lack of specification or uniqueness towards the dataset under analysis. 
This, however, is expected for a tool that aims to analyze different datasets, from different moving objects, through the same methodological approach. 
The tool being open-sourced, and this thesis project being public, provides the necessary knowledge to further reuse them, increasing the specification to a single application area. For instance, the tool could be modified to specifically analyze some of the following areas: tropical cyclones, arctic foxes' movement, unmanned aerial vehicles (i.e., UAVs), or even the spreading of unique diseases, such as COVID-19. This specification point significantly relates to the \textit{semantic interpretation and domain-specific meaning} mentioned in the first chapter, scope and limitations, subsection \ref{Scope and Limitations}, but also with the related work discussed throughout section \ref{Related Work}. 
For semantic interpretation to take place, extensive additional data and information are required. 
As an example, in the second case study, two pivot point angles were found to be critical for the shape and geometry of a cyclone. 
Discovering how and why these angles took place, could require an analyst to review the specific location's temperature, wind speed, general wind direction, air pressure, geographic plane outline, terrain shape, storm surge, landfall damage, air outflow, ocean temperature, cyclone hight, duration, atmospheric circulation, among many others. 
Noticeably, the information, background, and indicators needed to explain further, for instance, Arctic foxes instead, would be drastically different, disregarding the details needed to explain cyclone angles. 
From the related work discussed, semantic interpretation is often possible to achieve \cite{theoretical_framework_activity_groups_profiles_police}, however, such analysis was outside of this thesis' scope, due to the objective of generalizing the method and tool's utility for different datasets, with a centered approach on the analysis of spatio-temporal data with a taxonomy. 
In contrast, Tavakoli et al.'s \cite{Yashar} taxonomy framework was shown to be critical during the trajectory labeling process. Such a taxonomy was effective and validated by analyzing several trajectories, through the multi-level tool components, which showed how the labeling from the machine learning techniques was supported by statistical results and the visual-interactive map views.

\subsection{(RQ1) How can high-dimensional and unlabeled movement data be explored and better understood from a visual analytics tool using a taxonomy?}
The visual analytics tool proposed in this study presented a multi-level representation with a taxonomy structure, which facilitated group and individual trajectory comparisons. Such comparisons show several ways in which high-dimensional and unlabeled movement data can be explored and better understood:
\begin{itemize}
    \item For group comparisons, by following a taxonomical approach to group trajectories with similar behavior, and allowing a label-based comparison among them (i.e., into Kinematic, Geometric, and their respective subdivisions). This allows for exploring all combination groups based on frequency (in this study, with a frequency heatmap visualization) and decision boundaries representing group distributions. Furthermore, additional group and sub-group comparisons (in this study, with \textit{zone vs zone} comparisons) with visual analysis of low-level details within a given taxonomy, reveal insights on statistical data based on a selected taxonomical approach, which would not be available otherwise, effectively increasing data exploration, and overall, enhancing the understanding of such complex data.
    
    \item From individual comparisons, by incorporating \textit{side-by-side} selection and visualizations for taxonomy labels, allowing to explore unique and smaller samples of the trajectories (i.e., in this study, with 2D and 3D views), highlighting key statistical variables with respect to the taxonomy combination and trajectories selected, leading to additional iterations, each time refining the exploratory approach, and ultimately increasing the analyst's understanding of the spatio-temporal data under review.
\end{itemize}

\subsection{(RQ2) How can a multi-level approach assist in finding patterns and representing movement data?}
The multi-level approach showed several ways in which movement data can be represented, and patterns can be found:
\begin{itemize}
    \item Bridging low and high-level details: By having statistical information reflected onto the trajectories’ paths over a map, the representation of such low-level details gets easily reflected, providing better contextualization, and hence, increases the representation of the data. By repeatedly comparing groups, zones, and individual trajectories from both high and low-level views, patterns are revealed.
    
    \item Visual-interactive relation levels: By having points of interaction at each level, the analyst is presented with several possible views of the groups generated and the trajectories under analysis. Such visual-interactive changes provide several ways to represent the different groups and trajectories for each level of detail. When these interactions repeatedly lead to the same visual and statistical results at multiple levels, for several trajectories, patterns are unveiled.

    \item Cyclic-based approach: By starting from high-level views (e.g., taxonomy tree), moving onto middle-level views (e.g., frequency heatmap, decision boundary), then onto low-level views (e.g., feature importance and statistical analysis), back onto middle and high-level views (e.g., 2D and 3D map views, with point feature visual representations over time) the same data and samples are inspected through several analytical scopes, which improve representation at each level, and when performed multiple times, leads to patterns that would not be easy to discern otherwise.
\end{itemize}

\noindent The result demonstrated the capabilities of this tool and manifested its effectiveness in transforming raw movement data into interpretable statistical and visual formats, allowing analysts to conduct more thorough investigations and perform in-depth analyses. 
These findings suggest that, despite the inherent complexity of movement data, it can still be explained and analyzed after undertaking a series of steps and following certain procedures, including pre-processing prior to its analytical use, with the possibility of finding patterns, anomalies, or simply explaining and visualizing such information.

\newpage
		
\section{Conclusion and Future Works}

This thesis study presented a visual analytics tool for spatio-temporal data exploration and discovery of trajectory patterns using a movement taxonomy. 
The methodology proposed several tool components, which proved to be key for the visual-interactive needs in spatio-temporal data analysis, often being difficult to process and interpret due to their high dimensionality.

Through two case studies, one analyzing Arctic foxes' trajectories and a second one using tropical cyclones' paths, the tool was shown to be effective at explaining their movement and finding patterns. 
The first case study results showed that they successfully explained the foxes' movement, finding a pattern, where the foxes with acceleration-based behavior had a constant and steady acceleration pattern, while those represented by curvature-based behavior had several peaks in acceleration and deceleration. 
The second case study results revealed the geometric significance of the hurricanes analyzed, showcasing the importance and impact of indentation over the trajectories, effectively finding the key angles with the highest significance for the overall path's shape. 
Both case studies were supported by all tool components, including taxonomy selection, frequency heatmap representation, outlier detection scores, boundary analysis visual analysis, feature importance, statistical variables, as well as 2D and 3D map visualizations with detailed sample representations.

As such, data analysts and researchers in the computer science field may benefit from two valuable contributions. First, the research methodology and case studies showcased here support further research and development within the areas of EDA, data analysis, and data visualization. 
Secondly, by using the tool itself for the analysis of different spatio-temporal datasets.

Furthermore, researchers from other fields may also benefit from this study by analyzing movement objects from varied application areas such as aerospace and transportation engineers, animal biologists, and even epidemiologists. 
More often than not, if movement data is available, and variables such as speed, acceleration, angle, distance, and bearing are extracted, it becomes possible to apply the proposed methodology and analytics tool to further analyze such data.

With all this considered, improvement could still be made in the current project. Representation could still be increased by considering further components, such as a time series, which may enable analysts to capture the temporal context of each data point, preserving the sequence of movement data and enhancing the quality of the comparisons across different trajectories. 
In addition, normalization and pre-processing techniques could be performed in the data pipeline, for every dataset under analysis, to reduce the noisy trajectories, which are present in most spatio-temporal datasets and are often unavoidable. 
For instance, incorporating Hampel filtering could possibly reduce the outliers for each dataset loaded into the tool, effectively reducing the time required during pre-processing. 

Future work may aim to incorporate the components and features discussed in this section, either for the proposed tool in this thesis project or for a new method, aggregating them into a framework focusing on spatio-temporal data analysis. 
Researchers working on specific application areas that require spatio-temporal data analysis may consider incorporating and adapting the proposed thesis project to individual subject areas, which may then enable the necessity of unique analytical components, techniques, and functionalities for each individual application.

\newpage

%
\newpage

\hypersetup{urlcolor=black}
\bibliographystyle{IEEEtran}
\bibliography{references}
\newpage





\end{document}